\pdfoutput=1

\documentclass[11pt,twoside,a4paper,cmspaper,final,collab]{cms-tdr}
\def\svnVersion{0d3ee56-D}\def\svnDate{2019/06/28}\def\cmsCernNoTag{CERN-EP-2019-012}\def\cmsCernDate{\today}\def\cmsMessage{"Published in the Journal of Instrumentation as \href{http://dx.doi.org/10.1088/1748-0221/14/06/P06032}{\doi{10.1088/1748-0221/14/06/P06032}.}"}

\begin{document}\cmsNoteHeader{TAU-18-001}

\hyphenation{had-ron-i-za-tion}
\hyphenation{cal-or-i-me-ter}
\hyphenation{de-vices}
\RCS$HeadURL$
\RCS$Id$

\newcommand{\Wjets}{\ensuremath{\PW}{+}\text{jets}\xspace}
\newcommand{\ZTT}{\mbox{\ensuremath{\PZ\to\Pgt\Pgt}\xspace}}
\newcommand{\ZMM}{\mbox{\ensuremath{\PZ\to\Pgm\Pgm}\xspace}}
\newcommand{\ZEE}{\mbox{\ensuremath{\PZ\to\Pe\Pe}\xspace}}
\newcommand{\Pgth}{\Pgt_{\text{h}}\xspace}
\newcommand{\Irelem}{\ensuremath{I_{\text{rel}}^{\Pe(\PGm)}}\xspace}
\newcommand{\Irele}{\ensuremath{I_{\text{rel}}^{\Pe}}\xspace}
\newcommand{\Irelm}{\ensuremath{I_{\text{rel}}^{\PGm}}\xspace}
\newcommand{\mmumu}{\ensuremath{m_{\mu\mu}}\xspace}
\newcommand{\mjj}{\ensuremath{m_{\text{jj}}}\xspace}
\newcommand{\mvis}{\ensuremath{m_{\text{vis}}}\xspace}
\newcommand{\emu}{\ensuremath{\Pe\PGm}\xspace}
\newcommand{\ee}{\ensuremath{\Pe\Pe}\xspace}
\newcommand{\mumu}{\ensuremath{\PGm\PGm}\xspace}
\newcommand{\mutau}{\ensuremath{\PGm\Pgth}\xspace}
\newcommand{\etau}{\ensuremath{\Pe\Pgth}\xspace}
\newcommand{\tautau}{\ensuremath{\Pgth\Pgth}\xspace}
\newcommand{\ptem}{\ensuremath{\pt^{\Pe(\PGm)}}\xspace}
\newcommand{\mTe}{\ensuremath{m_{\text{T}}^{\Pe}}\xspace}
\newcommand{\mTm}{\ensuremath{m_{\text{T}}^{\PGm}}\xspace}
\newcommand{\mTem}{\ensuremath{m_{\text{T}}^{\Pe(\PGm)}}\xspace}
\newcommand{\Dzeta}{\ensuremath{D_{\zeta}}\xspace}
\newlength\cmsTabSkip\setlength{\cmsTabSkip}{1ex}

\cmsNoteHeader{TAU-18-001}
\title{An embedding technique to determine $\Pgt\Pgt$ backgrounds in proton-proton collision data}

\date{\today}

\abstract{
	An embedding technique is presented to estimate standard model $\Pgt\Pgt$ backgrounds from data with minimal simulation input. In the data, the muons are removed from reconstructed \mumu events and replaced with simulated tau leptons with the same kinematic properties. In this way, a set of hybrid events is obtained that does not rely on simulation except for the decay of the tau leptons. The challenges in describing the underlying event or the production of associated jets in the simulation are avoided. The technique described in this paper was developed for CMS. Its validation and the inherent uncertainties are also discussed. The demonstration of the performance of the technique is based on a sample of proton-proton collisions collected by CMS in 2017 at $\sqrt{s}=13\TeV$ corresponding to an integrated luminosity of 41.5\fbinv.
}

\hypersetup{
	pdfauthor={CMS Collaboration},
	pdftitle={An embedding technique to determine tau tau backgrounds in proton-proton collision data},
	pdfsubject={CMS},
	pdfkeywords={CMS, physics,  performance of high energy physics detectors}}

\maketitle
\section{Introduction}

An important background for many measurements at the CERN LHC is the decay of $\PZ$ bosons into
pairs of tau leptons ($\ZTT$). Among those measurements are studies of Higgs boson events in the
$\Pgt\Pgt$~\cite{Aad:2012mea,Aad:2015vsa,Chatrchyan:2012vp,Chatrchyan:2012xdj,Chatrchyan:2014nva}
and $\PW\PW$~\cite{Chatrchyan:2013lba,Chatrchyan:2013iaa} decay channels, and searches for additional
supersymmetric and charged Higgs bosons~\cite{Aad:2012cfr,Chatrchyan:2012vp,Chatrchyan:2011nx,
Khachatryan:2014wca,Khachatryan:2015tha,Aad:2014kga,Khachatryan:2015qxa}. This background can be
estimated from observed events, using selected $\PZ$ boson events in the $\Pgm\Pgm$ final state
($\ZMM$). Initially, the method was only used to model events originating from $\ZTT$ decays, 
which are the most prominent source of $\Pgt\Pgt$ background events at the LHC. However, all 
statements made throughout this paper are equally true for other standard model (SM) background 
processes that decay into two tau leptons. The aim of this method is to model all such processes.

In the embedding technique, all energy deposits of the recorded muons are removed from the $\ZMM$
events collected by CMS and replaced by the energy deposits of simulated tau lepton decays with the
same kinematic properties for the tau leptons as for the removed muons. In this way, a hybrid event
is created, comprised of information from both observed and simulated events. The parts of an event
that are challenging to describe in the simulation, such as the underlying event or the production
of additional jets, are taken directly from observed data. Only the tau lepton decay, which is well
understood, relies on the simulation. In Higgs boson analyses, the small coupling strength of the muon
with respect to the tau lepton guarantees a negligible contamination by signal events. The $\ZMM$
selection thus serves as a sideband region for those analyses that rely on this technique, referred
to as target analyses in the following. In this picture, the simulation of the tau leptons in place
of the removed muons corresponds to the extrapolation into the signal region.

The method itself can be studied by applying the embedding technique to a reference sample of simulated
$\ZMM$ events and comparing the result to an independent validation sample of simulated $\PZ\to\ell\ell$
events, where $\ell=\Pe,\,\Pgm,\,\Pgt$ stands for the embedded lepton flavor. All lepton flavors are
embedded for the validation of the technique. The corresponding application is referred to as $\Pe$-,
$\Pgm$-, or $\Pgt$-embedding throughout the text. The $\Pgm$-embedding holds the special role of validating
the technique itself. The $\Pe$-embedding serves to validate the sophisticated electron identification
in CMS, which relies on many detector quantities. Reconstruction efficiencies are determined from each
application, using the ``tag-and-probe'' method, as described in Ref.~\cite{Khachatryan:2010xn}. This
monitors the level of understanding of the reconstruction of each lepton flavor, and allows us to derive
residual correction factors for final use in the target analyses. Since these correction factors are
derived for the simulated leptons that have been embedded into the event, they are expected to be
similar to the correction factors obtained without the embedding technique. The branching fractions
for $\ZEE$, $\ZMM$, and $\ZTT$ are equal so the normalizations for all the decays are equal.

The embedding technique was implemented successfully for the first time by the CMS Collaboration in
the search and analysis of Higgs boson events in the context of the SM and its minimal supersymmetric
extension (MSSM) based on the data set obtained during the first operational run of the LHC between
2009 and 2013 (Run-1)~\cite{Chatrchyan:2012vp,Chatrchyan:2012xdj,Chatrchyan:2013lba,Chatrchyan:2014nva,
Chatrchyan:2011nx,Khachatryan:2014wca}. The technique has been upgraded since then to cope with the
new challenges of the most recent LHC data-taking periods that are related to the increased proton-proton
$(\Pp\Pp)$ collision rate. Further developments of the method include (i) the inclusion of other 
processes than $\ZTT$; (ii) the estimate of the normalization of the corresponding background processes 
from data; (iii) and an improved description of the electron identification. The upgraded embedding 
technique served as a cross-check of the estimate of the $\ZTT$ background events from simulation in 
the first CMS search for additional Higgs bosons in the $\Pgt\Pgt$ final state at 13\TeV, in the 
context of the MSSM~\cite{Sirunyan:2018zut}. A similar technique was used during the LHC Run 1 
data-taking period by the ATLAS Collaboration~\cite{Aad:2012mea,Aad:2015vsa,Aad:2012cfr} and is 
described in Ref.~\cite{Aad:2015kxa}.

In this paper, the methodology, validation, and application of the embedding technique developed for
the CMS experiment are described. The data sample used for the demonstration of the technique has been
recorded in 2017 and corresponds to an integrated luminosity of 41.5\fbinv. The validation of the
method is based on event samples that have been simulated for the same run period.

In Sections~\ref{sec:detector} and~\ref{sec:reconstruction} the CMS detector and event reconstruction
are introduced. The production of simulated events used for the validation of the technique is described
in Section~\ref{sec:simulation}. In Sections~\ref{sec:embedding} and~\ref{sec:validation} the technique
itself and its validation are discussed. Section~\ref{sec:application} contains a demonstration of the
performance of the technique, when applied to data, for the selection and analysis of $\PZ$ or Higgs
boson events in the $\Pgt\Pgt$ final state. The paper is concluded with a brief summary in
Section~\ref{sec:summary}.

\section{The CMS detector}
\label{sec:detector}

The central feature of the CMS apparatus is a superconducting solenoid of 6\unit{m} internal diameter,
providing a magnetic field of 3.8\unit{T}. Within the solenoid volume are a silicon pixel and strip
tracker, a lead tungstate crystal electromagnetic calorimeter (ECAL), and a brass and scintillator
hadron calorimeter (HCAL), each composed of a barrel and two endcap sections. Forward calorimeters
extend the pseudorapidity coverage provided by the barrel and endcap detectors. Muons are detected
in gas-ionization chambers embedded in the steel flux-return yoke outside the solenoid.

The silicon tracker measures charged particles within the pseudorapidity range $\abs{\eta} < 2.5$.
It consists of 1440 silicon pixel and $15\,148$ silicon strip detector modules. For nonisolated particles
with a transverse momentum of $1 < \pt < 10\GeV$ and $\abs{\eta} < 1.4$, the track resolutions
are typically 1.5\% in $\pt$ and 25--90 (45--150)\unit{\mum} in the transverse (longitudinal) impact
parameter~\cite{TRK-11-001}. The electron momentum is estimated by combining the energy measurement
in the ECAL with the momentum measurement in the tracker. The momentum resolution for electrons with
$\pt \approx 45\GeV$ from $\PZ\to\Pe\Pe$ decays ranges from 1.7\% for nonshowering electrons in the
barrel region to 4.5\% for showering electrons in the endcaps~\cite{Khachatryan:2015hwa}. Matching
muons to tracks measured in the silicon tracker results in a relative transverse momentum resolution,
for muons with $\pt$ up to $100\GeV$, of 1\% in the barrel and 3\% in the endcaps. The $\pt$ resolution
in the barrel is better than 7\% for muons with $\pt$ up to 1\TeV~\cite{Sirunyan:2018fpa}. In the
barrel section of the ECAL, an energy resolution of about 1\% is achieved for unconverted or
late-converting photons in the tens of GeV energy range. The remaining barrel photons have
a resolution of better than 2.5\% for $\abs{\eta} \leq 1.4$. In the endcaps, the resolution of
unconverted or late-converting photons is about 2.5\%, while the remaining endcap photons have a
resolution between 3 and 4\%~\cite{CMS:EGM-14-001}. When combining information from the entire
detector, the jet energy resolution typically amounts to 15\% at 10\GeV, 8\% at 100\GeV, and 4\% at
1\TeV, to be compared to about 40, 12, and 5\% obtained when the ECAL and HCAL calorimeters alone
are used.

Events of interest are selected using a two-tiered trigger system~\cite{Khachatryan:2016bia}. The
first level, composed of custom hardware processors, uses information from the calorimeters and
muon detectors to select events at a rate of around 100\unit{kHz} within a time interval of less than
4\unit{\mus}. The second level, known as the high-level trigger, consists of a large array of processors
running a version of the full event reconstruction software optimized for fast processing, and reduces
the event rate to around 1\unit{kHz} before data storage.

A more detailed description of the CMS detector, together with a definition of the coordinate system
used and the relevant kinematic variables, can be found in Ref.~\cite{Chatrchyan:2008zzk}.

\section{Event reconstruction}
\label{sec:reconstruction}

The reconstruction of the $\Pp\Pp$ collision products is based on the particle-flow (PF) algorithm
described in Ref.~\cite{Sirunyan:2017ulk}, which combines the available information from all CMS
subdetectors to reconstruct an unambiguous set of individual particle candidates. The particle
candidates are categorized into electrons, photons, muons, and charged and neutral hadrons. A good
understanding of the CMS lepton reconstruction is an important prerequisite for the assessment of
the embedding technique. Therefore the reconstruction of electrons, muons, and decays of tau leptons
to hadrons ($\Pgth$) from charged and neutral PF candidates is discussed in more detail in this
section.

In 2017, the CMS experiment operated with a varying instantaneous luminosity with, on average, between
28 and 47 $\Pp\Pp$ collisions per bunch crossing. Collision vertices are obtained from reconstructed
tracks using a deterministic annealing algorithm~\cite{726788}. The reconstructed vertex with the
largest value of summed physics-object $\pt^{2}$ is the primary collision vertex (PV). The physics
objects for this purpose are the jets, clustered using the anti-$\kt$ jet finding
algorithm~\cite{Cacciari:2008gp,Cacciari:2011ma}, as described below, with the tracks assigned to the
vertex as inputs, and the associated missing transverse momentum calculated as the negative vector $\pt$ sum of
those jets. Any other collision vertices in the event are associated with additional soft inelastic
$\Pp\Pp$ collisions called pileup (PU).

Electrons are reconstructed by combining energy deposits in the ECAL with tracks obtained from hits
in the tracker~\cite{Khachatryan:2015hwa}. Due to the strong curvature of the trajectory of charged
particles in the magnetic field and the significant amount of intervening material, an average fraction
of 33\% (at $\eta\approx 0$) to 86\% (at $\abs{\eta}\approx 1.4$) of the electron energy is radiated via
bremsstrahlung before the electron reaches the ECAL. All energy deposits above noise thresholds are
combined into clusters, using different algorithms for the ECAL barrel and endcap sections. The clusters
are further grouped into superclusters in a narrow window in $\eta$ and an extended window in the
azimuthal angle $\phi$ (measured in radians). The energy and position of the superclusters are
obtained from the sum of the energies and the energy-weighted mean of the positions of the building
clusters. This way of clustering is complemented by an alternative clustering algorithm, based on the
PF-reconstruction algorithm~\cite{Sirunyan:2017ulk}, resulting in an independent collection of PF
clusters.

Hits in the tracker are combined into tracks, using an iterative tracking procedure as described in
Ref.~\cite{Sirunyan:2017ulk}. To be efficient for the reconstruction of electrons, the track finding
must include the additional bending of the particle trajectory due to the bremsstrahlung emissions.
This is achieved by a dedicated Gaussian-sum filter algorithm~\cite{Adam:2005}. Since this method of
track reconstruction can be time consuming, it is initiated only on a selected set of electron track
seeds, which are likely to correspond to electron trajectories. Two approaches are followed to determine
these seeds. In the first approach, starting from the ECAL, the energy and position of the superclusters
are used to extrapolate the electron trajectory to its origin. The intersections of this extrapolation
with the innermost tracker layers or discs are matched to hits in the corresponding detectors. In the
second approach, starting from the tracker, reconstructed tracks obtained from a less efficient, but
also less CPU intensive, algorithm are extrapolated to the ECAL surface and matched to PF clusters.
The seeds of both approaches are combined to initiate the final electron track finding with an efficiency
of $\gtrsim$95\% for electrons from $\PZ$ boson decays.

The combination of the electron tracks with the ECAL clusters is achieved via a matching of the track
extrapolated to the ECAL surface with the supercluster in $\eta$-$\phi$ space with an efficiency of
${\approx}$93\% for electrons from $\PZ$ boson decays. Alternatively, the electron track is matched
to a PF cluster, while at each intersection with a layer or disc of the tracker a straight line is
extrapolated to the ECAL surface, tangent to the electron trajectory, to identify further PF clusters
due to bremsstrahlung emission. This approach improves the reconstruction for low $\pt$ electrons and
electrons in jets. To increase their purity, the reconstructed electrons are required to pass a
multivariate electron identification discriminant~\cite{Khachatryan:2015hwa}, which combines information
on the quality of the differently reconstructed tracks, shower shape, and kinematic quantities. In
the target analyses, for which the embedding technique is primarily foreseen, working points of this
discriminant with an efficiency between 80 and 90\% are used to identify electrons.

Two main approaches are also pursued to reconstruct muons with the CMS detector~\cite{Sirunyan:2018fpa}:
in the initial steps tracks are reconstructed independently in the inner silicon tracker and the outer
track detectors of the muon system. In the first approach inner and outer tracks are matched by comparing
their parameters propagated to a common surface. If a match is found, a global-muon track is fitted
combining the hits from both tracks. In a second approach, tracks from the inner tracker are extrapolated
to the muon system taking into account the magnetic field, the average expected energy losses, and
multiple Coulomb scattering in the detector material. If at least one muon segment (\ie, a short track
stub made of drift tube or cathode strip chamber hits) matches the extrapolation, the corresponding
track is identified as a muon track. The second approach improves the reconstruction efficiency for
muons with $\pt\leq 5$\GeV, which are unlikely to traverse the entire muon system. For muons
within the geometrical acceptance and with sufficiently high $\pt$ to reach the muon system, the
reconstruction efficiency reaches up to 99\%. It is supplemented by specialized algorithms for muons
with a $\pt$ of several hundreds of GeV. The presence of hits in the muon chambers already leads to
a strong suppression of particles misidentified as muons. Additional identification requirements on
the track fit quality and the compatibility of individual track segments with the fitted track can
reduce the misidentification rate further. In the analyses for which the embedding technique is
primarily foreseen, muon identification requirements with an efficiency of about 99\% are chosen.

The contribution from nonprompt leptons to the electron (muon) selection is further reduced by
requiring the selected leptons to be isolated from any hadronic activity in the detector. This property
is quantified by a relative isolation variable
\begin{linenomath}
  \begin{equation}
    \Irelem = \frac{1}{\pt^{\Pe(\Pgm)}} \left[\sum p_{\text{T}, i}^{\text{charged, PV}} +
    \max\left(0,\sum E_{\text{T}, i}^{\text{neutral}} - E_{\text{T}}^{\text{neutral, PU}}\right)\right],
\label{eq:iso_rel}
\end{equation}
\end{linenomath}
which uses the sum of the $\pt$ of all charged and transverse energy of all neutral
particles in a cone of radius $\Delta R = \sqrt{\smash[b]{\left(\Delta\eta\right)^{2}+\left(
\Delta\phi\right)^{2}}}$ around the lepton direction at the PV, where $\Delta\eta$ and $\Delta\phi$
correspond to the angular distance of the particle to the lepton in the $\eta$ and $\phi$ directions.
 The chosen cone sizes are $\Delta R=0.3$ and 0.4 for electrons and muons, respectively. The lepton
  itself is not included in this calculation. To mitigate any distortions from PU, only those charged
   particles whose tracks are associated with the PV are included in the sum. The presence of neutral
particles from PU around muons is estimated by summing the $\pt$ of charged particles in the isolation
cone whose tracks have been associated with PU vertices and multiplying this quantity by a factor of
0.5 to account for the approximate ratio of neutral to charged hadron production, such that $E_{\text{T}
}^{\text{neutral, PU}} = 0.5 \, \sum p_{\text{T}, i}^{\text{charged, PU}}$. For electrons, the
\FASTJET technique~\cite{Cacciari:2007fd,Cacciari:2008gn} is applied as described
in Ref.~\cite{Khachatryan:2015hwa}. The energy of neutral particles from PU is estimated as $E_{\text{T}
}^{\text{neutral,PU}} = \rho A_{\text{eff}}$, where $\rho$ is the median of the energy density
distribution per area in the $\eta$-$\phi$ plane around any jet in the event and $A_{\text{eff}}$ is
an effective area in $\eta$ and $\phi$. The value obtained is subtracted from the transverse energy
sum, and the result set to zero in the case of negative values. Finally, the result is divided by
the $\pt$ of the lepton to result in $\Irelem$.
	
	For further characterization of the event, all reconstructed PF candidates are clustered into jets using
	the anti-$\kt$ jet clustering algorithm as implemented in \FASTJET~\cite{Cacciari:2008gp,
		Cacciari:2011ma} with a distance parameter of 0.4. To identify jets resulting from the hadronization
	of \PQb quarks (\PQb jets), a reoptimized version of the combined secondary vertex \PQb tagging algorithm is used
	that exploits information from the decay vertices of long-lived hadrons and the impact parameters of
	charged-particle tracks in a combined discriminant~\cite{Sirunyan:2017ezt}. A typical working point for
        analyses for which the embedding technique is foreseen corresponds to a \PQb jet
	identification efficiency of ${\approx}$70\% and a misidentification rate for jets induced by light quarks and
	gluons of 1\%. For the validation of the embedding technique, jets with $\pt> 20$\GeV and $|\eta|<4.7$ 
        and $\PQb$ jets with $\pt> 20$\GeV and $|\eta|<2.5$ are used, unless otherwise indicated.
	
	Jets are also used as seeds for the reconstruction of $\Pgth$ candidates. The $\Pgth$ reconstruction
	is performed by further exploiting the substructure of the jets, using the hadrons-plus-strips algorithm
        described in Refs.~\cite{Khachatryan:2015dfa,Sirunyan:2018pgf}. The
	decay into three charged hadrons, and the decay into a single charged hadron, accompanied by up to two
	neutral pions with $\pt>2.5\GeV$, are used for the target analyses. The neutral pions are reconstructed as \emph{strips}, \ie,
	clusters of electron or photon constituents of the seeding jet with stretched energy deposits along the azimuthal
	direction. The strip size varies as a function of the $\pt$ of the electron or photon candidate. The
	$\Pgth$ decay mode is then obtained by combining the charged hadrons with the strips. High-$\pt$
	tau leptons are expected to be isolated from any hadronic activity in the event, as are high-$\pt$
	electrons and muons. Furthermore, in accordance with its finite lifetime, the charged decay products
	of the tau lepton are expected to be slightly displaced from the PV. To distinguish $\Pgth$ decays
	from jets originating from the hadronization of quarks or gluons, a multivariate $\Pgth$ identification
	discriminant is used~\cite{Sirunyan:2018pgf}. It combines information on the hadronic activity in
	the detector in the vicinity of the $\Pgth$ candidate with the reconstructed properties related to
	the lifetime of the tau lepton. Of the predefined working points given in Ref.~\cite{Sirunyan:2018pgf},
	the \text{tight}, \text{medium}, and \text{very loose} working points are used in the target analyses.
	These have efficiencies between 27\% (\text{tight}) and 71\% (\text{very loose}) for genuine tau
	leptons, \eg, from $\ZTT$ decays, for quark/gluon misidentification rates of less than $4.4\times
	10^{-4}$ (tight), and $1.3\times 10^{-2}$ (very loose). Finally, additional discriminants are imposed 
        to reduce the misidentification probability for electrons and muons as $\Pgth$ candidates, using 
        predefined working points from Ref.~~\cite{Sirunyan:2018pgf}. For the discrimination against electrons 
        these working points have identification efficiencies for genuine tau leptons ranging from 65\% 
        (\text{tight}) to 94\% (\text{very loose}) for misidentification rates between $6.2\times10^{-4}$ 
        (\text{tight}) and $2.4\times10^{-2}$ (\text{very loose}). For the discrimination against muons the 
        typical $\Pgth$ identification efficiency is 99\% for a misidentification rate of $\mathcal{O}(10^{-3})$.
	
	The missing transverse momentum vector $\ptvecmiss$, defined as the negative vector $\pt$ sum of all
	reconstructed PF objects, is also used to characterize the events. Its magnitude is referred
	to as $\ptmiss$. It enters the target analyses via selection criteria and via the calculation
	of the final discriminating variable used for the statistical analysis, which is usually correlated
	with the invariant mass of the $\Pgt\Pgt$ system.
	
	\section{Simulation}
	\label{sec:simulation}
	
	For the validation of the embedding technique and to demonstrate its performance, simulated events
	are used to model the most important processes contributing after the event selections described in
	Sections~\ref{sec:embedding} and~\ref{sec:application}. The Drell--Yan production in the $\Pe\Pe$,
	$\Pgm\Pgm$, and $\Pgt\Pgt$ final states, and the production of $\PW$ bosons in association with jets
	($\Wjets$) are generated at leading order (LO) precision~\cite{Alwall:2011uj} in the strong coupling
	constant $\alpS$, using the \MGvATNLO 2.2.2 event generator~\cite{Alwall:2014hca}. To increase
	the number of simulated events in phase space regions with high jet multiplicity, supplementary samples
	are generated with up to four outgoing partons in the hard interaction. For diboson production
	\MGvATNLO is used at next-to-leading order (NLO) precision. For $\ttbar$ and single t quark production
	samples are generated at NLO precision using \POWHEG v2~\cite{Nason:2004rx,Frixione:2007vw,Alioli:2008tz,
        Alioli:2009je,Alioli:2010xd,Alioli:2010xa,Bagnaschi:2011tu}. For the generation of all processes the 
        NNPDF3.0 parton distribution functions~\cite{Ball:2014uwa} are used. The simulation
	of the underlying event is parametrized according to the CUETP8M1 tune~\cite{Khachatryan:2015pea}.
	Hadronic showering and hadronization, as well as the $\Pgt$ decays, are modeled using \PYTHIA
	8.212~\cite{Sjostrand:2014zea}. For all generated events the effect of the PU is included by generating
	additional inclusive inelastic $\Pp\Pp$ collisions with \PYTHIA and adding them
	to the simulated events according to the expected PU distribution profile in data. Differences between
	this expectation and the observed PU profile are mitigated by reweighting the simulated events.
	All events generated are passed through a \GEANTfour-based~\cite{Agostinelli:2002hh} simulation of
	the CMS detector and reconstructed using the same version of the CMS event reconstruction software as
	used for the data.
	
	\section{Embedding procedure}
	\label{sec:embedding}
	
\begin{figure}[htbp!]
		\centering
		\includegraphics[width=\textwidth]{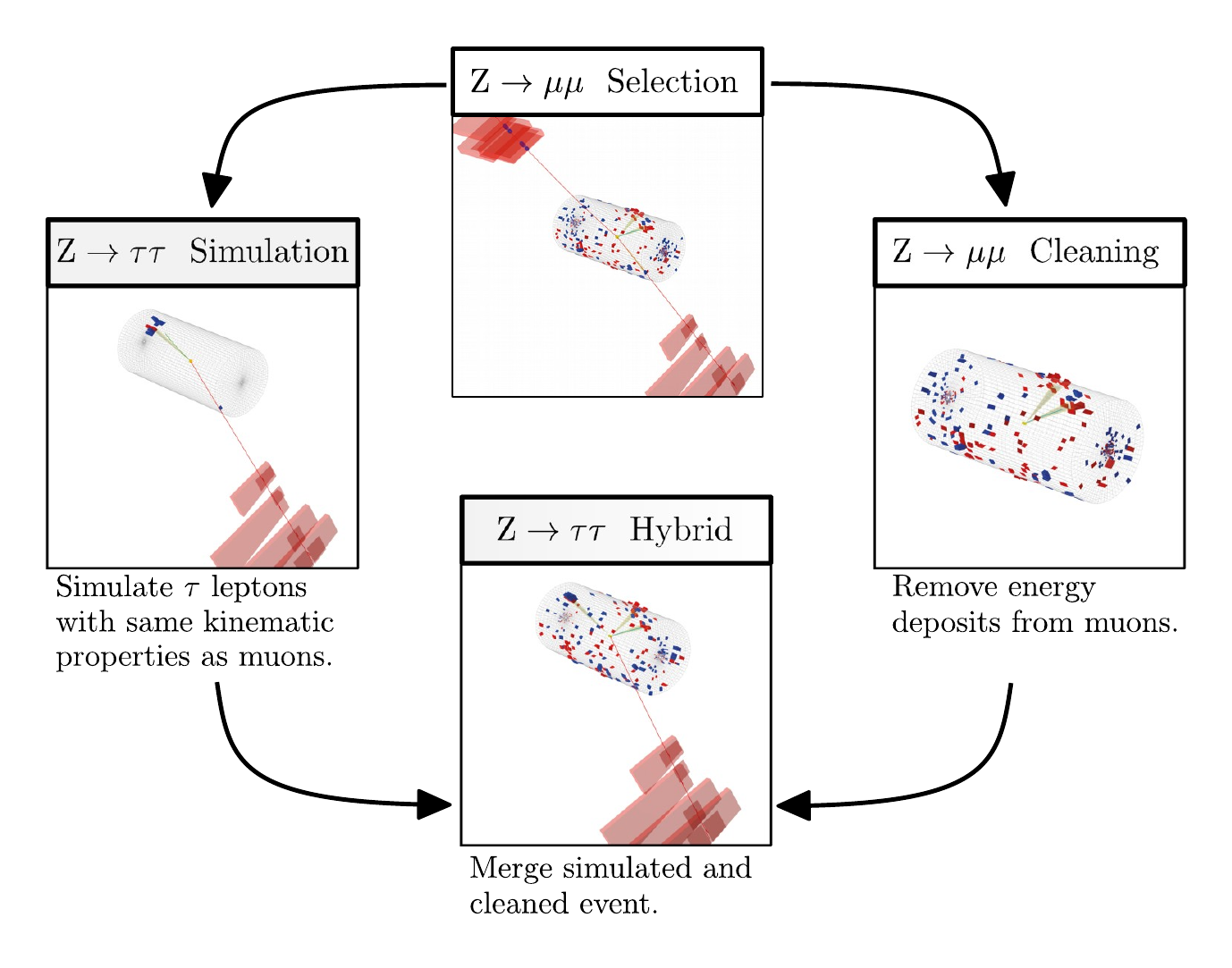}
	  \caption {
	  	Schematic view of the four main steps of the $\Pgt$-embedding technique, as described in
	  	Section~\ref{sec:embedding}. A $\ZMM$ candidate event is selected in data (``$\ZMM$ Selection''),
	  	all energy deposits associated with the muons are removed from the event record (``$\ZMM$ Cleaning''),
	  	and two tau lepton decays are simulated in an otherwise empty detector (``$\ZTT$ Simulation'').
	  	Finally all energy deposits of the simulated tau lepton decays are combined with the original
	  	reconstructed event record (``$\ZTT$ Hybrid''). In the example, one of the simulated tau leptons
	  	decays into a muon and the other one into hadrons.
	  }
		\label{fig:embedding-overview}
	\end{figure}

	The embedding procedure can be split into four steps: 
	\begin{itemize}
	\item the selection of $\Pgm\Pgm$ events from data
	(Section~\ref{sec:Zmm-treatment}), 
	\item the removal of tracks and energy deposits of the selected muons
	from the reconstructed event record (Section~\ref{sec:Zmm-cleaning}), 
	\item the simulation of two $\Pgt$
	leptons with the same kinematic properties as the removed muons in an otherwise empty detector
	(Section~\ref{sec:Ztt-simulation}), and
	\item the combination of the energy deposits of the simulated
	tau lepton decays with the original reconstructed event record (Section~\ref{sec:Ztt-merging}).
	\end{itemize}
	For validation purposes, electrons or muons can also be injected into the simulation to form an
	embedded $\Pe\Pe$ or $\Pgm\Pgm$ event, referred to as an $\Pe$- or $\Pgm$-embedded event. A schematic
	view of the procedure is given in Fig.~\ref{fig:embedding-overview}.
	
	\subsection{Selection of \texorpdfstring{$\Pgm\Pgm$}{mumu} events}
	\label{sec:Zmm-treatment}
	
	In the first step of the embedding procedure, $\Pgm\Pgm$ events are selected from data. Although the
	selected muons might not necessarily originate from $\PZ$ boson decays, $\ZMM$ events are a natural
	target of this selection, which helps to identify genuine $\Pgm\Pgm$ events. The selection should be
	tight enough to ensure a high purity of genuine $\Pgm\Pgm$ events and at the same time loose enough
	to minimize biases of the embedded event samples. The selection of the muons defines the minimal
        selection requirements to be used in
	the target analyses that are discussed in more detail in Section~\ref{sec:Ztt-simulation}.
	Inefficiencies of the reconstruction and selection of the muons due to the geometrical acceptance of
	the detector are estimated, giving correction factors which are applied to the final distributions.
	
	While strict isolation requirements help to increase the purity of prompt
	muons, \eg, from $\ZMM$ decays, in the selection, they introduce a bias towards less hadronic
	activity in the vicinities of the embedded leptons that will appear more isolated than expected
	in data. To minimize this kind of bias, which cannot be corrected by a scale factor, isolation
        requirements are omitted as much as possible. At the
	same time the selected phase space is desired to be as inclusive as possible for the embedded event
	samples to be applicable for a variety of target analyses. The loose selection in turn leads to an
	admixture of other processes in addition to $\ZMM$. This admixture and the consequences for the
	embedded event samples are carefully checked and assessed.

	\subsubsection{Selection requirements}
	
	\begin{table}[b]
		\topcaption{
			Expected event composition after the selection of two muons, as described in
			Section~\ref{sec:Zmm-treatment}. The label ``QCD'' refers to SM events composed 
                        exclusively of jets produced via the strong interaction.
                        The compositions after adding selections on $\mmumu>70
			$\GeV or on the number of \PQb jets in the event
			are shown in column 3 and 4 respectively. In the second column the fraction of
			events where the corresponding process has two genuine muons in the final state is given in
			parentheses. For $\Wjets$ events the second muon originates from additional heavy 
                        flavor production.
		}
		\label{tab:event-composition}
		\centering
		\begin{tabular}{lcccc}
			& \multicolumn{4}{c}{Fraction (\%)} \\
			\multicolumn{1}{c}{Process}
			& \multicolumn{2}{c}{Inclusive} & $\mmumu>70$\GeV & $N(\text{\PQb jet})>0$ \\
			\hline
			\hspace{0.3cm}$\ZMM$
			& $97.36$ & $(97.36)$ & $99.11$ & $69.25$ \\[\cmsTabSkip]
			\hspace{0.3cm}QCD
			& $\hphantom{0}0.84$ & $\dagger$ & $\hphantom{0}0.10$ & $\hphantom{0}2.08$ \\[\cmsTabSkip]
			$\hspace{0.3cm}\ttbar$
			& $\hphantom{0}0.78 $ & $(\hphantom{0}0.60)$ & $\hphantom{0}0.55$ &            $25.61$ \\[\cmsTabSkip]
			$\hspace{0.3cm}\ZTT$
			& $\hphantom{0}0.74$ & $(\hphantom{0}0.71)$ & $\hphantom{0}0.05$ & $\hphantom{0}0.57$ \\[\cmsTabSkip]

			\hspace{0.3cm}Diboson, single t
			& $\hphantom{0}0.20$ & $(\hphantom{0}0.17)$ & $\hphantom{0}0.17$ & $\hphantom{0}2.35$ \\[\cmsTabSkip]
			\hspace{0.3cm}$\Wjets$
			& $\hphantom{0}0.08$ & $(\hphantom{0}0.01)$ & $\hphantom{0}0.02$ & $\hphantom{0}0.14$ \\
			\hline
			\multicolumn{5}{l}{$\dagger$ \small{Data-driven estimate, information not available.}} \\
		\end{tabular}
	\end{table}
	
	At the trigger level, the events are required to be selected by at least one of a set of $\Pgm\Pgm$
	trigger paths, with a minimum requirement between 3.8 and 8.0\GeV on the invariant mass of the two muons, $\mmumu$. All
	trigger paths require $\pt>17\,(8)$\GeV for the leading (trailing) muon, very loose isolation
	in the tracker, and a loose association of the muon track with the PV. Offline, the reconstructed muons
	are required to match the objects at the trigger level, their distance extrapolated to the PV is required
	to be $\left|d_{z}\right|<0.2$\unit{cm} along the beam axis, and both muons are required to have $\left|\eta\right|<2.4$.
	Their transverse momentum is required to be $\pt>
	17\,(8)$\GeV for the leading (trailing) muon to match the online selection requirements. No
	additional selection requirements are imposed on the isolation of the muons to minimize any bias of
	the embedded event samples in this respect.
	
\begin{figure}[htbp]
		\centering
		\includegraphics[width=0.48\textwidth]{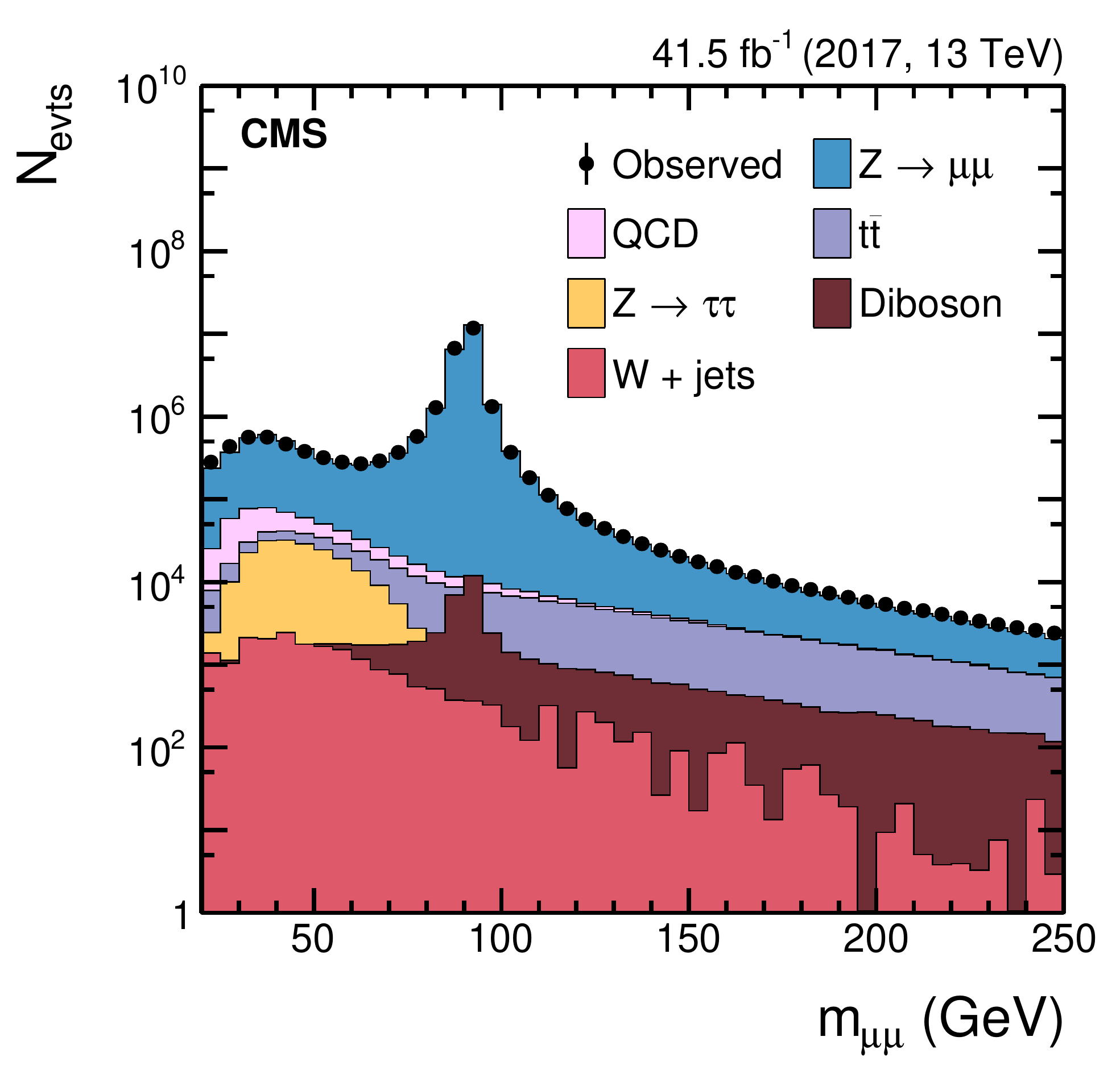}
		\includegraphics[width=0.48\textwidth]{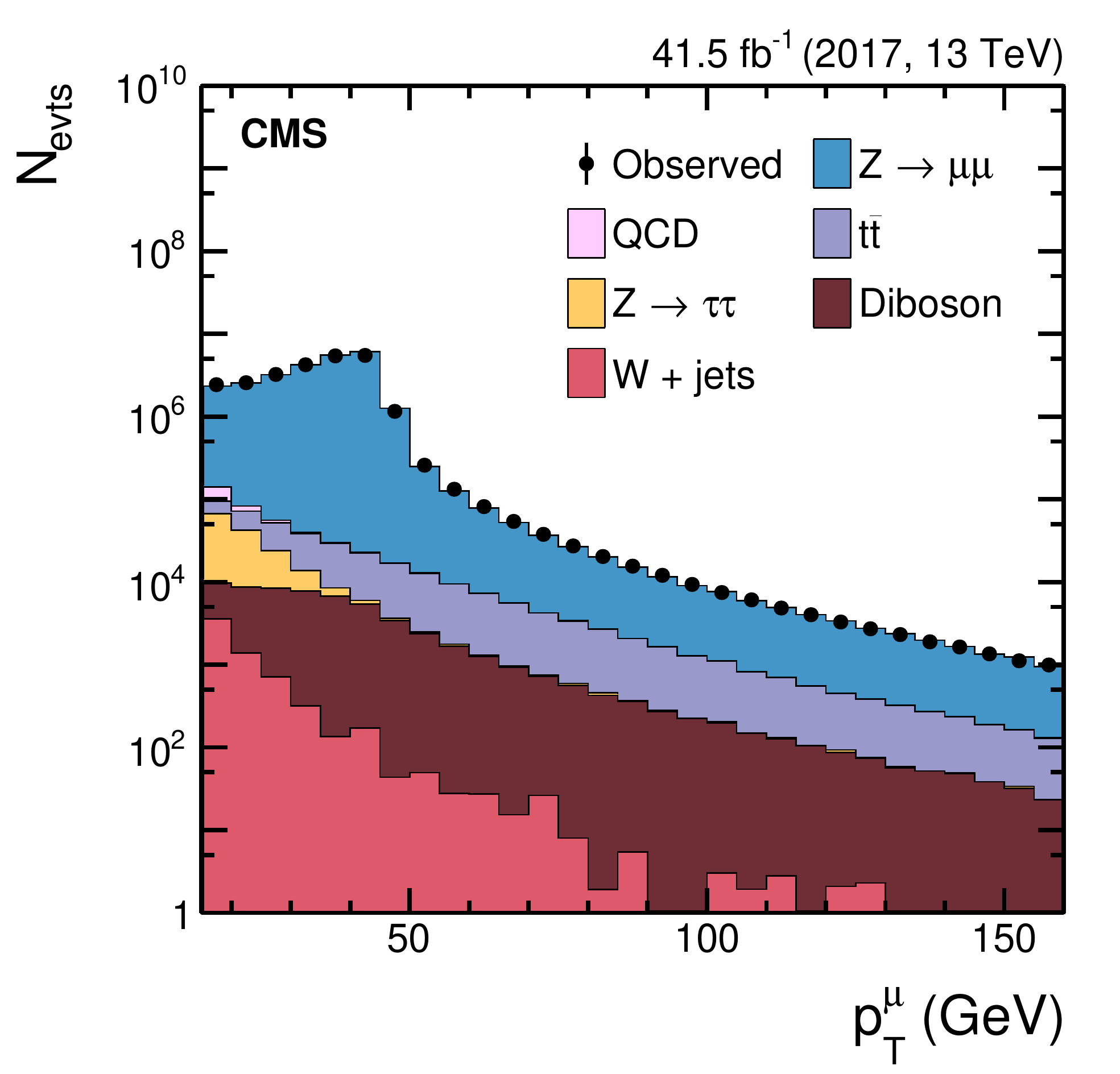}
		\caption {
			(Left) invariant mass, $\mmumu$, of the selected dimuon $\PZ$ boson candidates and (right) $\pt$ of the trailing muon after
			the event selection, as described in Section~\ref{sec:Zmm-treatment}.
		}
		\label{fig:dimuon-selection}
	\end{figure}

	To form a $\PZ$ boson candidate, each muon is required to originate from a global-muon track. The
        muons are required to be of opposite charge with an
        invariant mass of $\mmumu>20$\GeV. If more than one
	$\PZ$ boson candidate is found in the event, the one with the value of $\mmumu$ closest to the nominal
	$\PZ$ boson mass is chosen. This selection results in a total of more than 65 million events, with
	an average rate of about 1.5 million events per 1\fbinv of collected data. The expected event
	composition after these and several further selection requirements that will be specified in the
	following discussion is given in Table~\ref{tab:event-composition}. SM events composed exclusively of jets produced via the
	strong interaction are referred to as quantum chromodynamics (QCD) multijet production. Throughout the paper
	this contribution is estimated from data using a background estimation method described in
	Ref.~\cite{Sirunyan:2018zut}. The distributions of $\mmumu$ and $\pt$ of the trailing muon for all
	selected events are shown in Fig.~\ref{fig:dimuon-selection}. Also shown are the contributing processes
	estimated by the simulation, to illustrate their kinematic distributions.
	
	\subsubsection{Expected sample composition}
	\label{sec:sample-composition}
	
	In Table~\ref{tab:event-composition}, a relaxed selection of two muons compatible with
	the properties of a $\PZ$ boson candidate already results in a sample of $\ZMM$ events with an
	expected purity of more than 97\%. Smaller contributions are expected from
	$\ZTT$ events, mostly where both tau leptons subsequently decay into muons, and from QCD multijet,
        $\ttbar$, and diboson production.
	
	Without further correction, the presence of QCD multijet and $\ZTT$ events in the selected event sample
	leads to an
	overestimate of the $\ZMM$ event yield and a bias of the $m_{\ell\ell}$ and $\pt$ distributions
	of the embedded leptons towards lower values. This can be inferred from Fig.~\ref{fig:dimuon-selection},
        where the accumulation of these events is visible for $\mmumu<70$\GeV and $\pt^{\mu}<20$\GeV.
	The fraction of QCD multijet and $\ZTT$ events can be significantly suppressed by raising the requirement
        on $\mmumu$ to
	be higher than 70\GeV, at the cost of a loss of ${\approx}13\%$ of selected $\ZMM$ events.
        However, because of the low transverse momentum of the selected muons, these events have a low
	probability to end up in the final sample of $\Pgt$-embedded events, see
	Section~\ref{sec:Ztt-simulation}.
	
	The contribution from $\ttbar$ and diboson events is distributed over the whole range of $\mmumu$.
        Its relative contribution is larger at high values of $m_{\ell\ell}$, where the overall event yield is
        small, and in
	event selections with \PQb jets, as shown in the last column of Table~\ref{tab:event-composition}.
	These conditions are met, \eg, in searches for additional Higgs bosons in models beyond the
	SM~\cite{Sirunyan:2018zut}. A large fraction of this contribution originates from events where the
	$\PW$ bosons \eg, from both t quark decays subsequently decay into a muon and neutrino ($\ttbar(\Pgm
	\Pgm)$). The contribution from $\ttbar$ and diboson production in all other modes is below the current
	accuracy requirements of the method. The substitution of the muons
	by tau leptons provides an additional estimate for $\ttbar$ and diboson production with two tau leptons
	in the final state from data. This class of events needs to be removed from simulation in the target analyses
	to prevent double counting. For simplicity, all further discussion of the embedding technique will refer to the estimate
	of all genuine $\Pgt\Pgt$ events from either $\ZTT$, $\ttbar$, or diboson production, unless explicitly stated
	otherwise.
	
	\subsubsection{Correction for the detector acceptance}
	
	As discussed above, inefficiencies in the reconstruction and selection of the $\Pgm\Pgm$ events lead to
	kinematic biases in the embedded event samples because of the limited detector acceptance. The global
	efficiency of the trigger selection in the kinematic regime where embedded event samples can be applied amounts to
	about 80\%, the combined reconstruction and identification efficiency lies well above 95\%. Both
	efficiencies are estimated differentially in a fine grid in muon $\eta$ and $\pt$, using the ``tag-and-probe''
	method. They are then used to correct for the effects of the detector
	acceptance.

        As a consequence, not only the kinematic distributions but also the yield of the estimated $\Pgt\Pgt$
        events can be obtained directly via the embedding technique, assuming the same branching fraction of
        the $\PZ$ boson into muons and tau leptons. This is achieved by correcting for the
        detector acceptance and
	selection efficiency of the $\Pgm\Pgm$ events and applying the reconstruction and selection
        efficiency from the
	$\Pgt$-embedded event sample. Residual corrections of these efficiencies with respect to the
	data, are discussed in Section~\ref{sec:correction-factors}. When applied to the data this estimate
	renders uncertainties in the production cross sections and integrated luminosity irrelevant for the
	involved processes, as will be further discussed in Section~\ref{sec:uncertainties}.
	
	\subsection{Removal of \texorpdfstring{$\Pgm$}{mu} energy deposits from
		the reconstructed event record}
	\label{sec:Zmm-cleaning}
	
	In the second step, all energy deposits of the selected muons are removed from the reconstructed event
	record. This is done at the level of hits in the inner tracker and muon systems, and clusters in the calorimeters.
	Hits in the tracker are identified by their association to the fitted global-muon track. Clusters in
	the calorimeters are identified by the intercept of the muon trajectory interpolated through the
	calorimeters, as discussed in Section~\ref{sec:reconstruction}. If an intercept matches with the position
	of a calorimeter cluster, an energy amount corresponding to a minimum ionizing particle is subtracted
	from the cluster. If the energy of the modified cluster drops below the noise
	threshold defined for the event reconstruction, the cluster is removed from the event record. By this
	procedure, all traces of the selected muons in the detector can be removed from the event reconstruction
	even in detector environments with additional hadronic activity in the vicinity of the selected muons.
	
\begin{figure}
		\centering
		\includegraphics[width=0.49\textwidth]{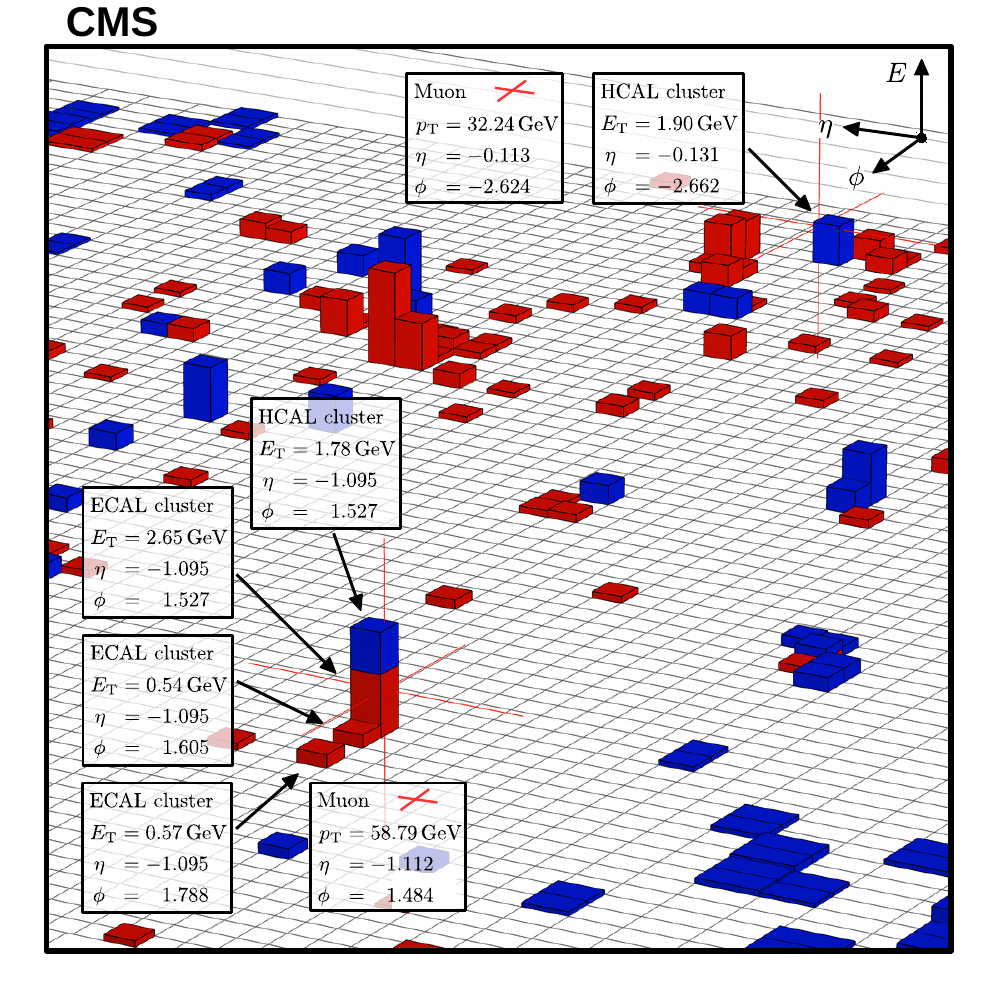}
		\includegraphics[width=0.49\textwidth]{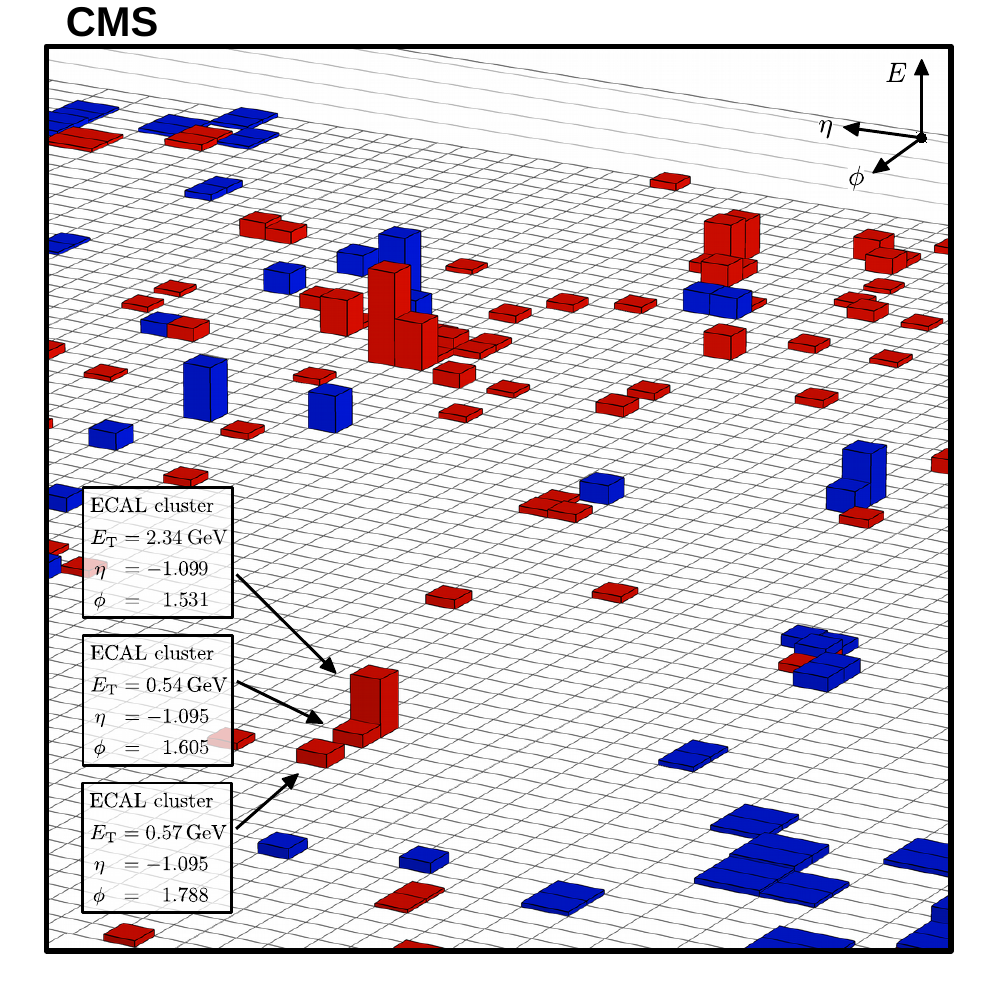}
		\caption {
			Display of a $\ZMM$ candidate event in the data set, in the $\eta$-$\phi$ plane at the surface
			of the calorimeters (left) before and (right) after the hits and energy deposits associated with the
			muons have been removed from the reconstructed event record. The red crosses indicate the intercepts
			of the reconstructed muon trajectories with the calorimeter surface. The red (blue) boxes correspond
			to clusters in the ECAL (HCAL).
		}
		\label{fig:cleaning}
	\end{figure}

	Effects of the removal of energy deposits in the calorimeters can arise in cases where the
	energy deposit of the muon is not completely removed or leads to the split of a geometrically extended
	cluster into more than one piece. Such a removal may lead to the reconstruction of spurious photon
	or neutral hadron candidates. These additionally reconstructed objects are usually of low energy and
	low reconstruction quality, and play a negligible role in the target analyses. The removal of the
	energy deposits of the muons from the detector is illustrated in Fig.~\ref{fig:cleaning}. In
	Fig.~\ref{fig:cleaning} (left), a selected $\ZMM$ candidate event in the data set is displayed in
	the $\eta$-$\phi$ plane of the calorimeters, with the intercepts of the reconstructed
	muons with the calorimeter surface and clusters in the ECAL (HCAL) shown. One
	muon (with $\pt=32$\GeV) in the upper and one muon (with $\pt=59$\GeV) in the lower parts
	of the figure are visible. Several clusters in the calorimeters have been associated with the incident
	muon trajectories. In Fig.~\ref{fig:cleaning} (right) the same detector area is shown after the hits
	and energy deposits associated with the muons have been removed from the reconstructed event record.
	The HCAL clusters associated with each corresponding muon have been completely removed, whereas the energy of the ECAL
	cluster associated with the muon in the lower part of the figure has been reduced. The
	remaining ECAL cluster is identified as low-energy photon in the subsequent reconstruction.
	
	\subsection{Simulation of tau lepton decays}
	\label{sec:Ztt-simulation}
	
	In the third step, the energy and momentum of the selected muons are either directly injected as electrons
	or muons into the detector simulation, for validation purposes, or used to seed the simulation of tau
	lepton decays via \PYTHIA, before entering the detector simulation. For this purpose an event record
	is prepared that contains only the information related to the kinematic properties of the two selected
	muons in an otherwise empty detector that is free of any other particles from additional jet production,
	underlying event, or PU. The invariant mass of the selected muons is fixed to the reconstructed value, as
	shown in Fig.~\ref{fig:dimuon-selection} (left). Polarization effects are neglected in embedded events, since they are below
the sensitivity of the target analyses.
	
	To account for the mass difference between the muon and the tau lepton or electron (referred to by
	$\ell=\Pe,\,\Pgt$), the four-momenta of the muons are boosted into the center-of-mass frame of the
	$\Pgm\Pgm$ pair, where the energy ($E^{*}_{\ell}$) and momentum ($\vec{p}^{\,*}_{\ell}$) of each lepton,
	with mass $m_{\ell}$, are determined from
	\begin{linenomath}
		\begin{equation}
		E^{*}_{\ell} = \frac{\mmumu}{2}\,;
		\qquad
		\left|\vec{p}^{\,*}_{\ell} \right| = \sqrt{E_{\ell}^{*\,2} - m_{\ell}^{2}}\,;
		\qquad
		\ell=\Pe,\,\Pgt\,.\\
		\label{eq:energy-momentum}
		\end{equation}
	\end{linenomath}
	The corrected values $\vec{p}^{\,*}_{\ell}$ and $E^{*}_{\ell}$ are then boosted back into the laboratory frame and used either for the electrons
	or to seed the tau lepton decays. The event vertex for the simulation of the embedded leptons
	is set to the PV of the initially reconstructed $\Pgm\Pgm$ event. Four distinct samples of $\Pgt$-embedded
	events are produced from the same $\Pgm\Pgm$ event sample, for use in the most important final states
	of the target analyses, namely $\emu$, $\etau$, $\mutau$, and $\tautau$. This is
	achieved by enforcing the subsequent decay of the injected $\Pgt$ lepton pair in the simulation,
	with a branching fraction of 100\%. It has been checked that the overlap of the resulting
        $\Pgt$-embedded event samples is small enough, such that even those distributions that are related
        to the part of the event that originates from the observed data, \eg like jet distributions, are
        fully uncorrelated.
	
	\subsubsection{Post-processing of the simulated tau lepton decays}
	
	A significant amount of the energy and momentum of the tau lepton is not transferred to the visible
	decay products, but carried away by the neutrino(s) in the decay. As a consequence, the visible products
	of the tau lepton decays are usually significantly lower in $\pt$ than that of the  originally selected
	muons. A restricted phase space of the selected muons results from the finite
	detector acceptance. For each set of $\Pgt$-embedded events, this
	translates into a final-state-dependent
	kinematic range, for later use in the
	target analyses. This range is further restricted by the acceptance requirements that have to be
	imposed in the target analyses. For example, the ability to create $\Pgt$-embedded events in the
	$\tautau$ final state, with reconstructed $\Pgth$ candidates with a $\pt^{\Pgth}$ as low as $20\GeV$
	each is useless for an analysis with a trigger threshold of $\pt^{\Pgth}>30\GeV$.
	To save computing time during the CPU-intensive detector simulation, a
	kinematic filtering is applied to the visible decay products, after the simulation of the tau lepton
	decay and before the detector simulation. The final-state-dependent thresholds of this filtering on
	the $\pt$ of the visible decay products (prior to the detector simulation) define the kinematic range
	of eligibility of the $\Pgt$-embedded event samples for later use in the target analyses. They are
	given in Table~\ref{tab:eligibility}.
	
	\begin{table}[b]
		\topcaption{
			Kinematic range of eligibility for each $\Pgt$-embedded event sample in the $\emu$, $\etau$,
			$\mutau$, and $\tautau$ final states. The expression ``First/Second object'' refers to the final
			state label used in the first column. Also given are the probability of the simulated tau lepton
			pair to pass the kinematic filtering ($\epsilon_{\text{kin}}$), described in the text, and the
			equivalent of the integrated luminosity $\mathcal{L}_{\text{int}}$, of the corresponding
			$\Pgt$-embedded event sample, in multiples of the data set, from which the embedded event sample
			has been created.
		}
		\label{tab:eligibility}
		\centering
		\begin{tabular}{lr@{$>$}l@{, }r@{$<$}l@{\hspace{1.cm}}r@{$>$}l@{, }r@{$<$}lcc}
			Final state
			& \multicolumn{4}{c}{First object}
			& \multicolumn{4}{c}{Second object}
			& $\epsilon_{\text{kin}}$ & $\left.\mathcal{L}_{\text{int}}/41.5\fbinv\right.$ \\
			\hline
			$\hspace{0.3cm}\emu$
			& \multicolumn{4}{c}{$\pt^{\Pe}>21\,(10)$\GeV}
			& \multicolumn{4}{c}{$\pt^{\Pgm}>10\,(21)$\GeV}
			& $0.58$ & $60$ \\[\cmsTabSkip]
			$\hspace{0.3cm}\etau$
			& $\pt^{\Pe\hphantom{\text{h}}}$&$~22$\GeV & $|\eta^{\Pe}|$&$~2.2$
			& $\pt^{\Pgth}$&$~18$\GeV & $|\eta^{\Pgth}|$&$~2.4$
			& $0.50$ & $14$ \\[\cmsTabSkip]
			$\hspace{0.3cm}\mutau$
			& $\pt^{\Pgm\hphantom{\text{h}}}$&$~18$\GeV & $|\eta^{\Pgm}|$&$~2.2$
			& $\pt^{\Pgth}$&$~18$\GeV & $|\eta^{\Pgth}|$&$~2.4$
			& $0.53$ & $15$ \\[\cmsTabSkip]
			$\hspace{0.3cm}\tautau$
			& $\pt^{\Pgth}$&$~33$\GeV & $|\eta^{\Pgth}|$&$~2.2$
			& $\pt^{\Pgth}$&$~33$\GeV & $|\eta^{\Pgth}|$&$~2.2$
			& $0.27$ & $\hphantom{0}5$ \\
			\hline
		\end{tabular}
	\end{table}
	
	To increase the number of $\Pgm\Pgm$ events that can be used in the target analyses, the decay is repeated
        1000 times for each tau lepton
	pair. This is done to give the decay products a higher probability
	to pass the eligibility requirements. Only the last trial that fulfills the kinematic requirements for the
	given final state is saved for the subsequent detector simulation. If at least one trial succeeds, the
	number of successful trials divided by 1000 times the branching fraction of the subsequent $\Pgt\Pgt$
	decay is saved as an additional weight factor to the event. These weights take values below the corresponding
	branching fraction and can be as low as $10^{-4}$ at the kinematic thresholds of eligibility.
	Depending on the $\Pgt\Pgt$ final state, the fraction of events that pass the kinematic filtering
	ranges between $\epsilon_{\text{kin}}=27\%$ (in the $\tautau$ final state) and 58\% (in the $\emu$
	final state). In the $\tautau$ final state this means that 73\% of the $\Pgt$-embedded
	events that could in principle be used, according to the acceptance restrictions of the originally
	selected $\Pgm\Pgm$ events, are usually not accessible due to the stricter acceptance requirements in
	the target analyses. 
	
	Overall this procedure allows for the production of final-state-specific $\Pgt$-embedded event
	samples of approximately 5 to 60 times the size of the event sample of selected tau lepton
	pairs in the target analyses, independent of the integrated luminosity corresponding to this event sample.
	The efficiency of the kinematic filtering and the size of each $\Pgt$-embedded event
	sample are given in Table~\ref{tab:eligibility}.
	
	In Section~\ref{sec:sample-composition}, $\ZTT$ events where both tau leptons subsequently decay into
	muons and the corresponding neutrinos are discussed as a potential source of bias of the $\Pgt
	$-embedded event samples. Of all $\ZTT$ events in this final state a fraction of less than 0.25\%
	is expected to end up in the $\Pgt$-embedded event samples, in the given eligibility ranges. This
	corresponds to less than 2.8\% of the events indicated by the $\ZTT$ contribution in
	Fig.~\ref{fig:dimuon-selection}, and a fraction far below the 1\% level in the initial event
	composition as given in Table~\ref{tab:event-composition}.
	
	\subsubsection{Discussion of additional reconstruction effects}
	
	\begin{figure}[b]
		\centering
		\includegraphics[width=0.48\textwidth]{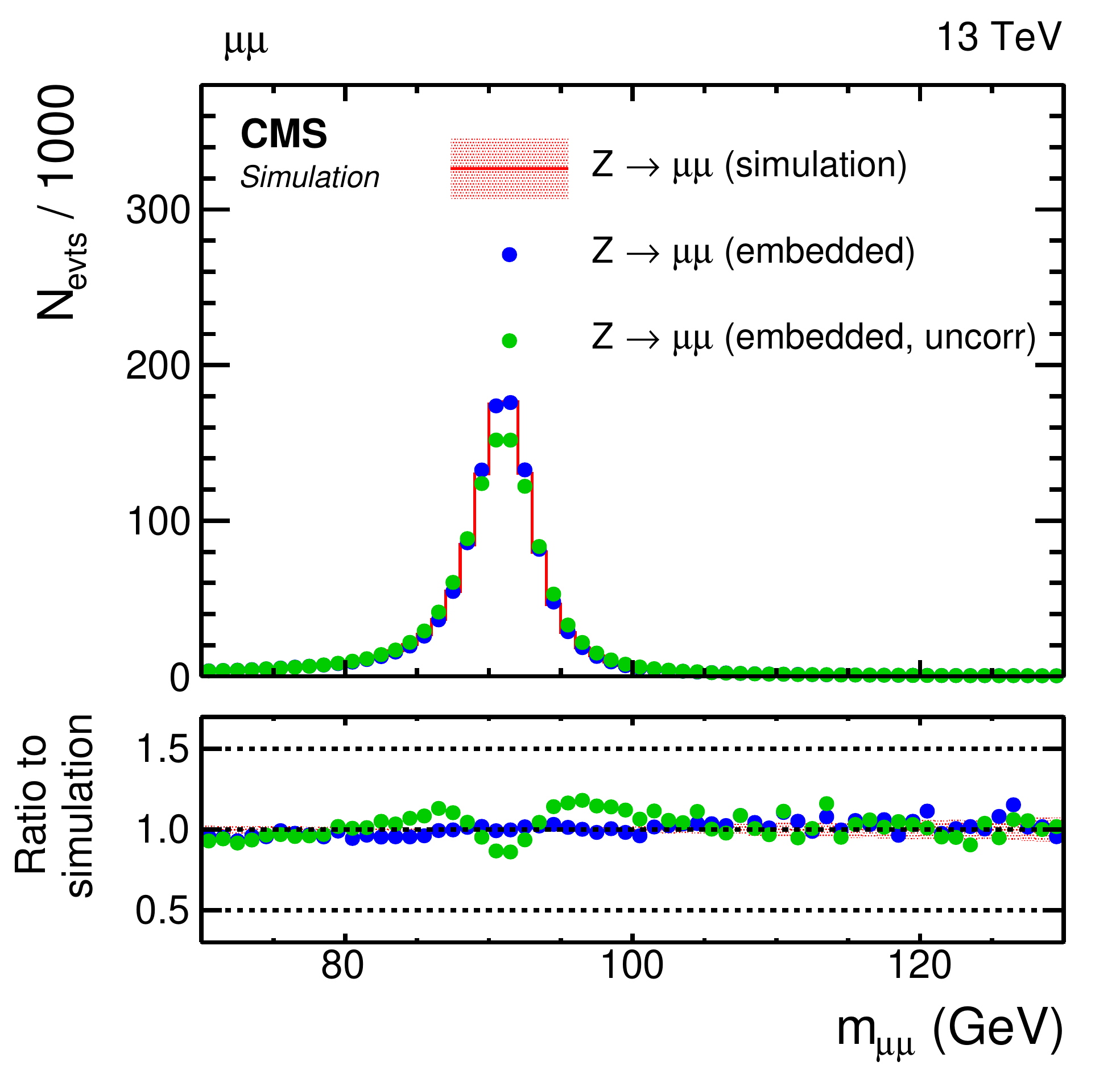}
		\includegraphics[width=0.48\textwidth]{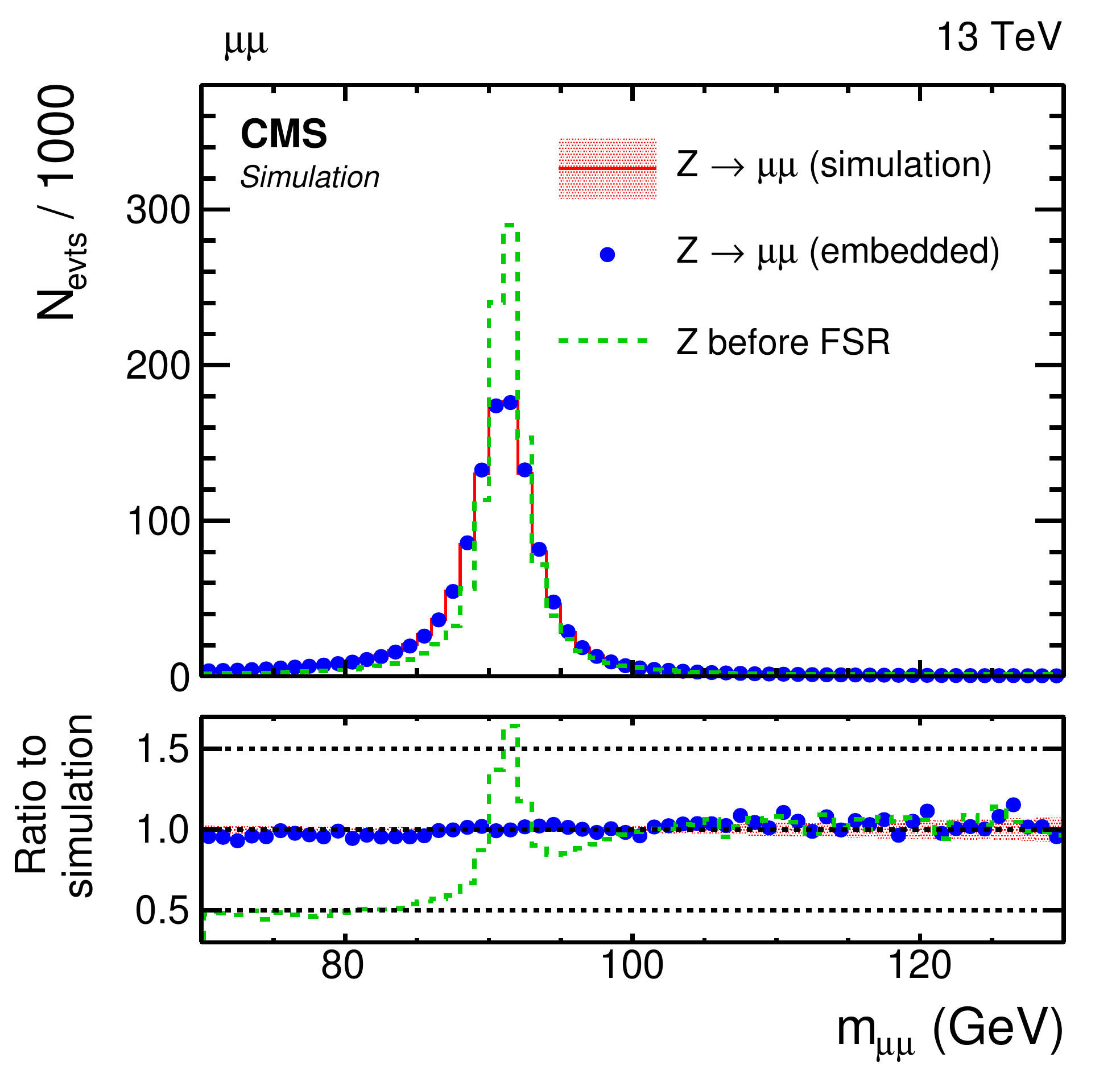}
		\caption {
			Comparison of the reconstructed invariant mass, $\mmumu$, of the selected muons from a simulated
			$\ZMM$ sample with the corresponding $\Pgm$-embedded event sample. On the left the (red
			histogram) simulated $\ZMM$ sample and the $\Pgm$-embedded event sample (blue dots) with and (green dots)
			without the correction for the effects of the finite detector resolution, as described in the
			text, are shown. On the right (green histogram) $\mmumu$ from the simulated $\ZMM$ sample before FSR
			is shown in addition, to illustrate the effect.
		}
		\label{fig:FSR-effects}
	\end{figure}
	
	Two more reconstruction effects arise in the discussion of the simulation
	step. First, the four-momenta of the selected muons correspond to already reconstructed objects, which
	are reinjected into the simulation of the detector response, effects due to the finite momentum
	resolution of the detector lead to a broadening, especially of the $\pt$ and $m_{\ell\ell}$ distributions
	of the embedded leptons. The distributions are corrected for this effect by an $\mmumu$-dependent
	rescaling of the energy and momentum of the selected muons on an event-by-event basis, before using
        them to generate the simulated leptons for embedding. A simulated $\ZMM$ sample is used to derive 
        this $\mmumu$-dependent rescaling. Figure~\ref{fig:FSR-effects} (left) shows the $\mmumu$ distribution
        from a sample of simulated $\ZMM$ events as well as the corresponding $\Pgm$-embedded event sample
        before and after the correction. In the lower panel of the
	figure, the ratio is given with respect to the simulated $\ZMM$ sample. The $\Pgm$-embedded
        event sample
	without the correction reveals a slight broadening with respect to the simulated $\ZMM$ sample, which is
	compensated by the correction.
	
	A second effect can be attributed to the emission of photons from the initially selected muons, referred
	to as final-state radiation (FSR) in the following. When missed in the reconstruction, FSR leads to an
	additional broadening of the kinematic distributions and a systematic shift to lower values of the energy and momentum
	of the initially selected muons. This shift is subsequently transferred to the
	embedded leptons. Figure~\ref{fig:FSR-effects} (right) shows the $\mmumu$ distribution of the $\ZMM$ simulation
	sample for muons before and after FSR, to illustrate the effect. For the validation of $\Pgm
	$-embedded events, this effect can be eliminated by executing the simulation step of the embedding
	procedure without FSR. The $\ZMM$ simulation sample and the corresponding $\Pgm$-embedded event data sample are
	then subjected to the same FSR effects during the initial simulation. For $\Pe$-embedded events the
	effects of FSR are underestimated; for $\Pgt$-embedded events they are overestimated.
	
	In the case of $\Pgt$-embedding, both effects that were discussed in this section are
	negligible compared to the energy and momentum fluctuations introduced by the undetected neutrinos
	in the decay, which already lead to a significant broadening of the related kinematic distributions.
	A more detailed discussion is given in Section~\ref{sec:validation}.
	
	\subsection{Hybrid event creation}
	\label{sec:Ztt-merging}
	
	In a fourth and final step of the procedure, all energy deposits of the simulated electrons,
	muons, or tau lepton decays are combined with the original reconstructed event record, from which
	the energy deposits of the initially selected muons had been removed, to form a hybrid event that is
	mostly obtained from data and only relies on the simulation for the embedded lepton pair. This is
	done at the earliest possible reconstruction step to guarantee that all subsequent quantities for
	the lepton identification are based on the full event information and not only on parts of the event.
	The ideal way is to combine the reconstructed object collections at the level of tracker hits and
	energy deposits in the calorimeter crystals. However, in practice, the information is combined at
	the level of reconstructed objects (tracks, calorimeter clusters, and muons) rather than at the level
	of individual hits. This is to avoid complications with residual small differences between the simulation geometry and the real detector. The tracks of the
	embedded leptons are reconstructed based on the geometry used for the simulation, in the otherwise
	empty detector, of the simulation step. Since the detector in the simulation step is free from other particles,
	jet production, underlying event, or PU there may be a biased track reconstruction efficiency that
	must be checked and possibly corrected. Residual effects are discussed in Section~\ref{sec:validation}.
	
	\begin{figure}[htbp]
		\centering
		\includegraphics[width=0.45\textwidth]{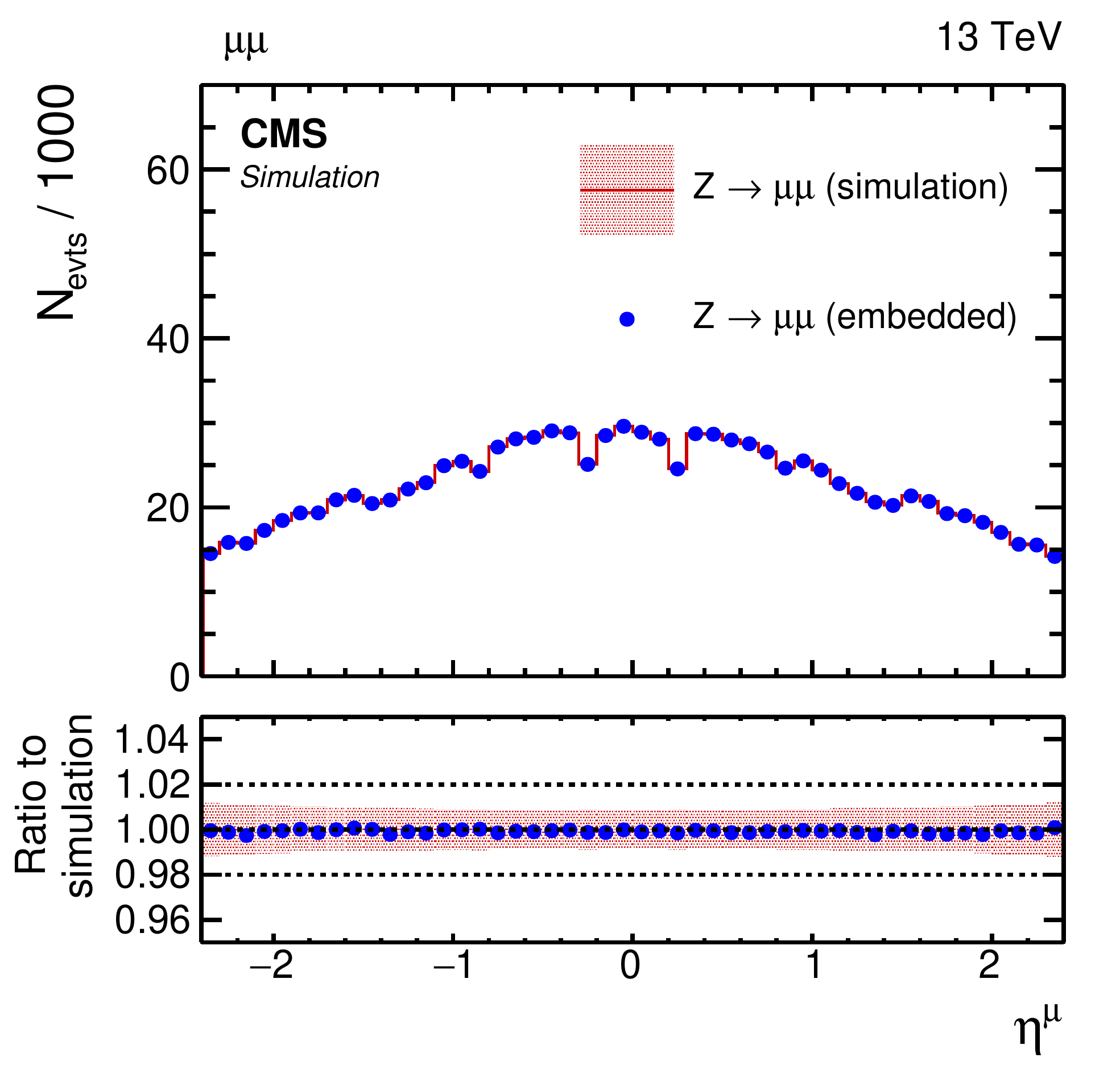}
		\includegraphics[width=0.45\textwidth]{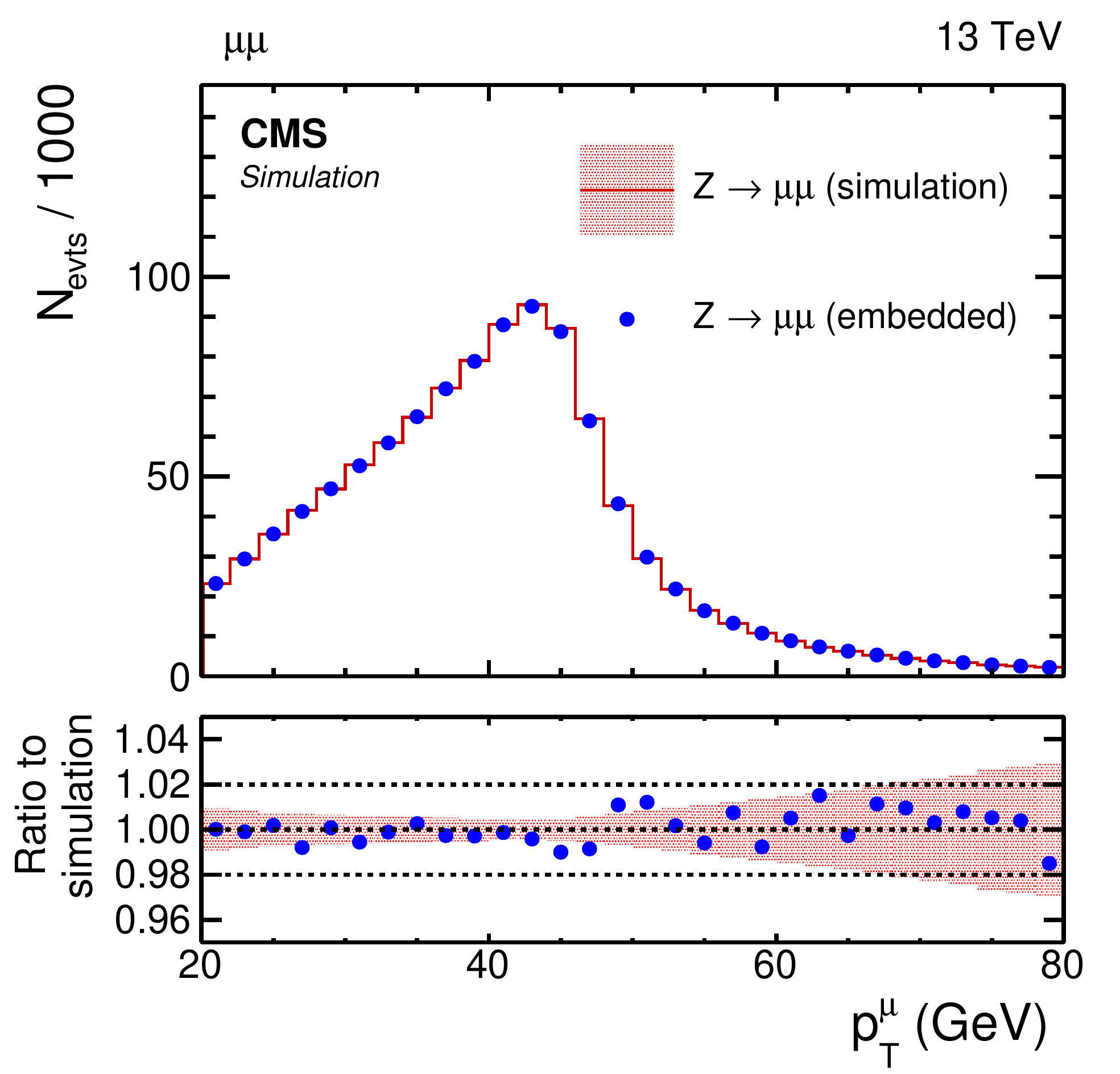}
		\includegraphics[width=0.45\textwidth]{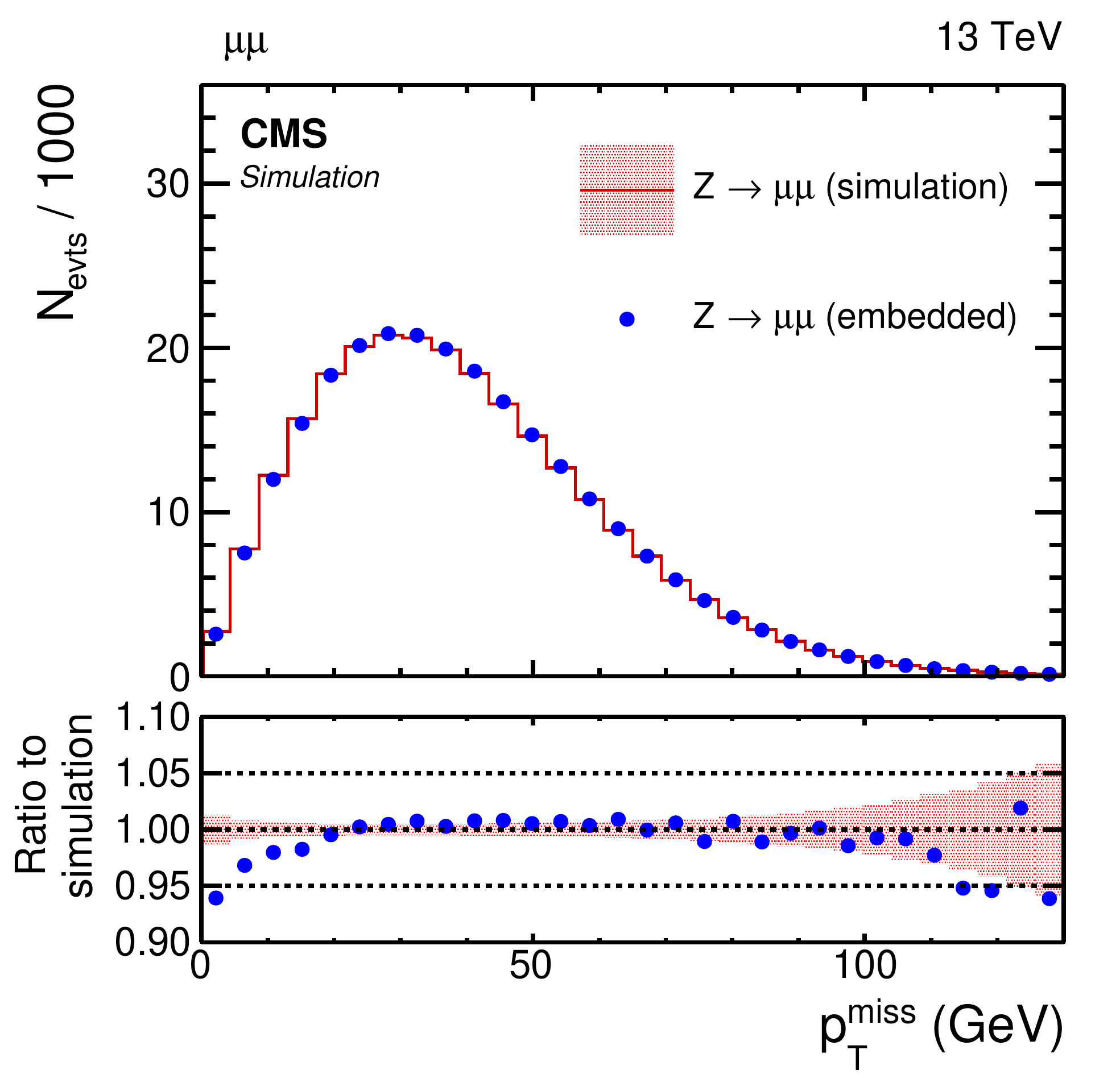}
		\includegraphics[width=0.45\textwidth]{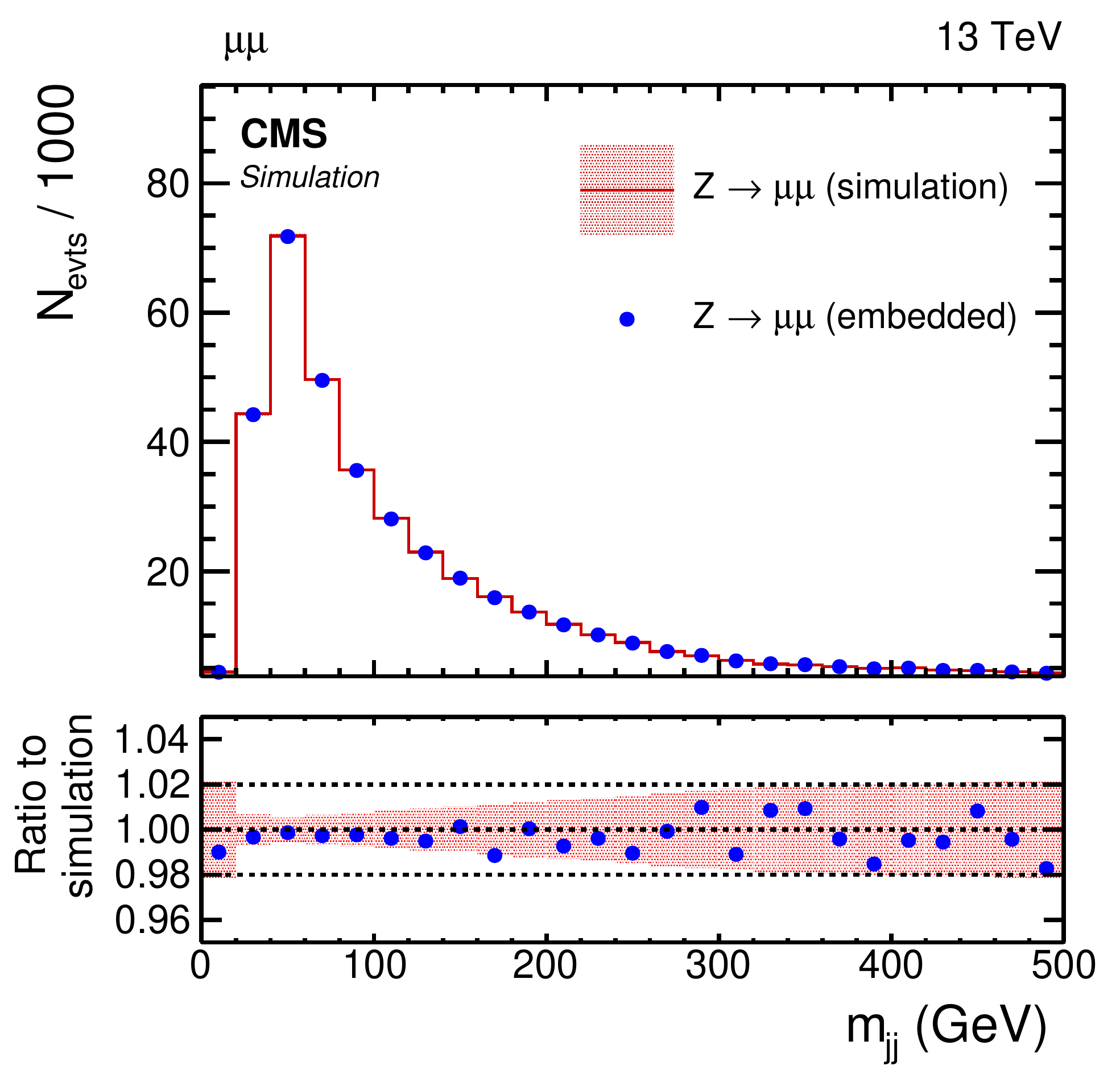}
		\includegraphics[width=0.45\textwidth]{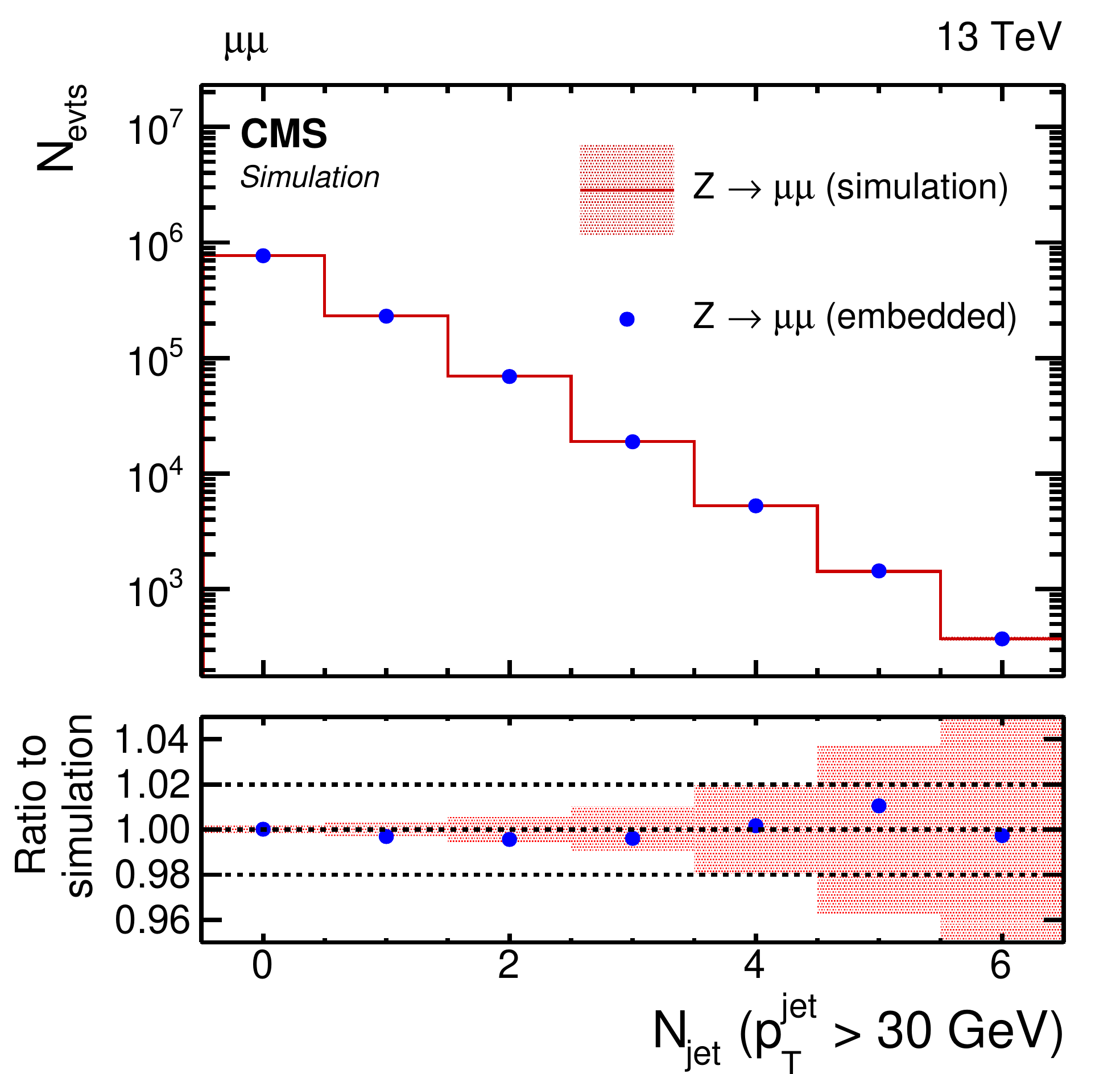}
		\includegraphics[width=0.45\textwidth]{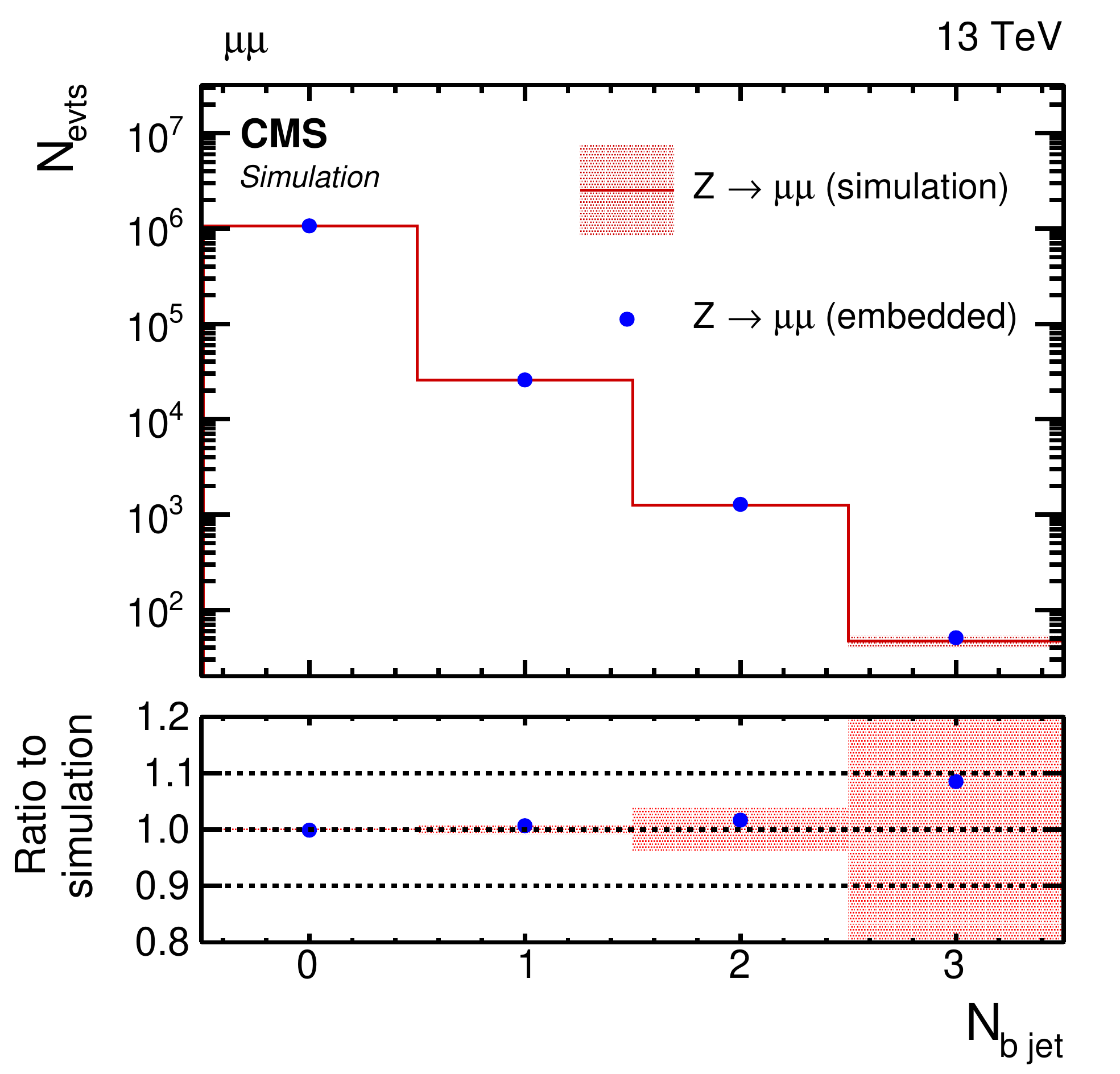}
		\caption {
			Comparison of $\Pgm$-embedded events with exactly the same $\ZMM$ events from simulation. Shown
			are the (upper left) $\eta$ and (upper right) $\pt$ distributions of the leading muon in $\pt$,
			(middle left) $\ptmiss$, (middle right) $\mjj$, (lower left) jet and, (lower right) \PQb jet
			multiplicities, as described in the text.
		}
		\label{fig:mm-embedding-validation}
	\end{figure}
	
	\begin{figure}[htbp!]
		\centering
		\includegraphics[width=0.45\textwidth]{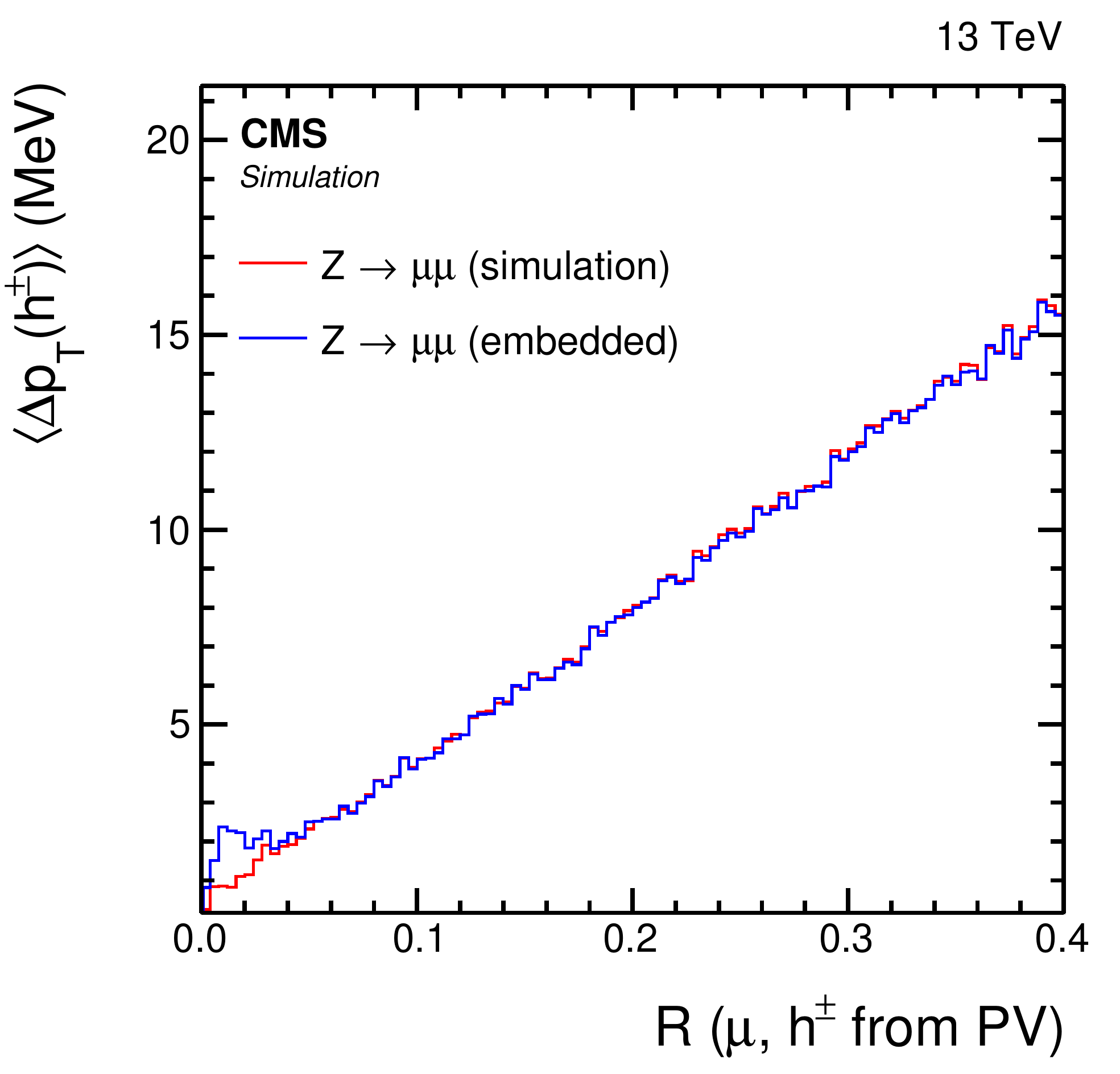}
		\includegraphics[width=0.45\textwidth]{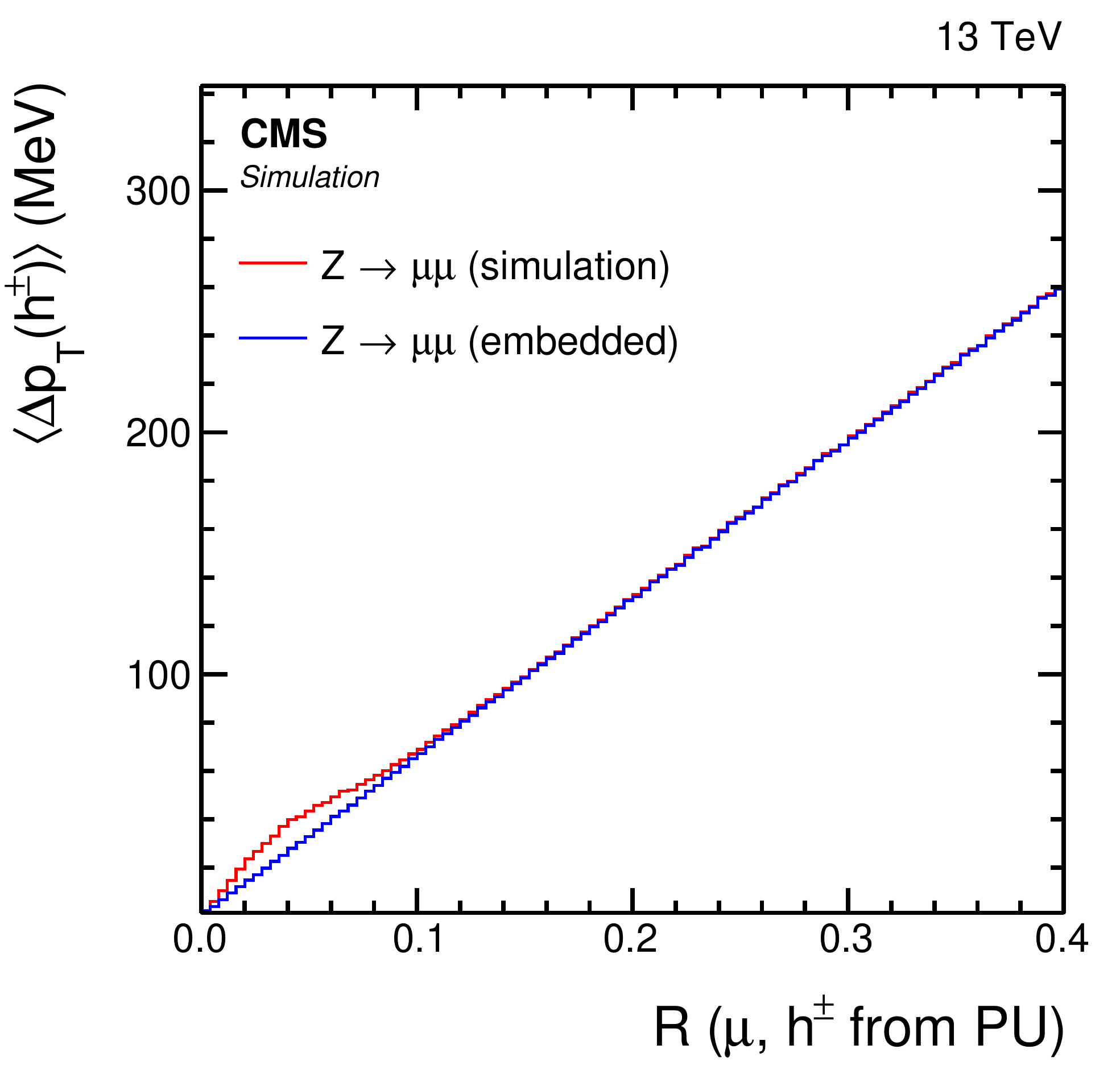}
		\includegraphics[width=0.45\textwidth]{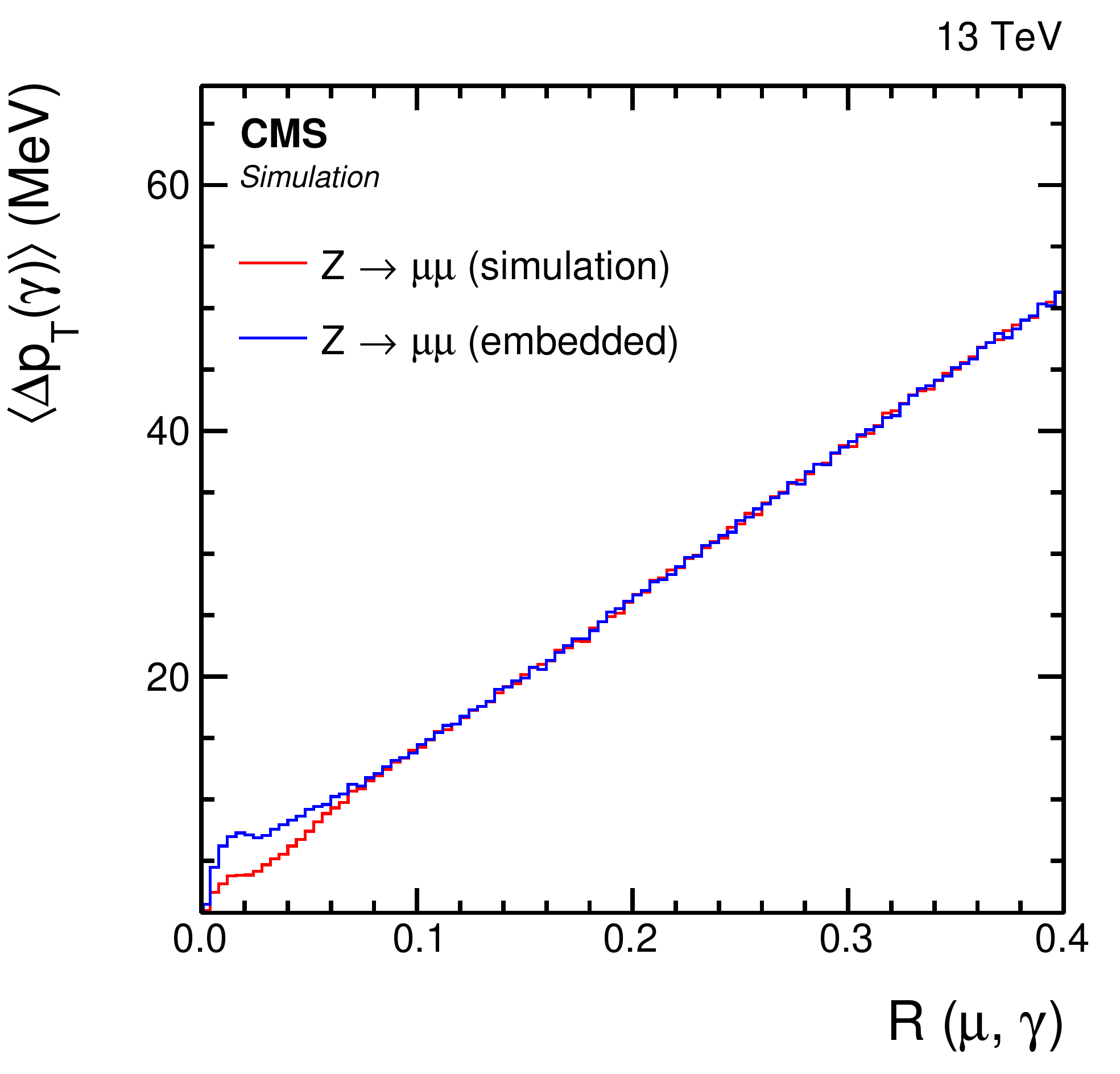}
		\includegraphics[width=0.45\textwidth]{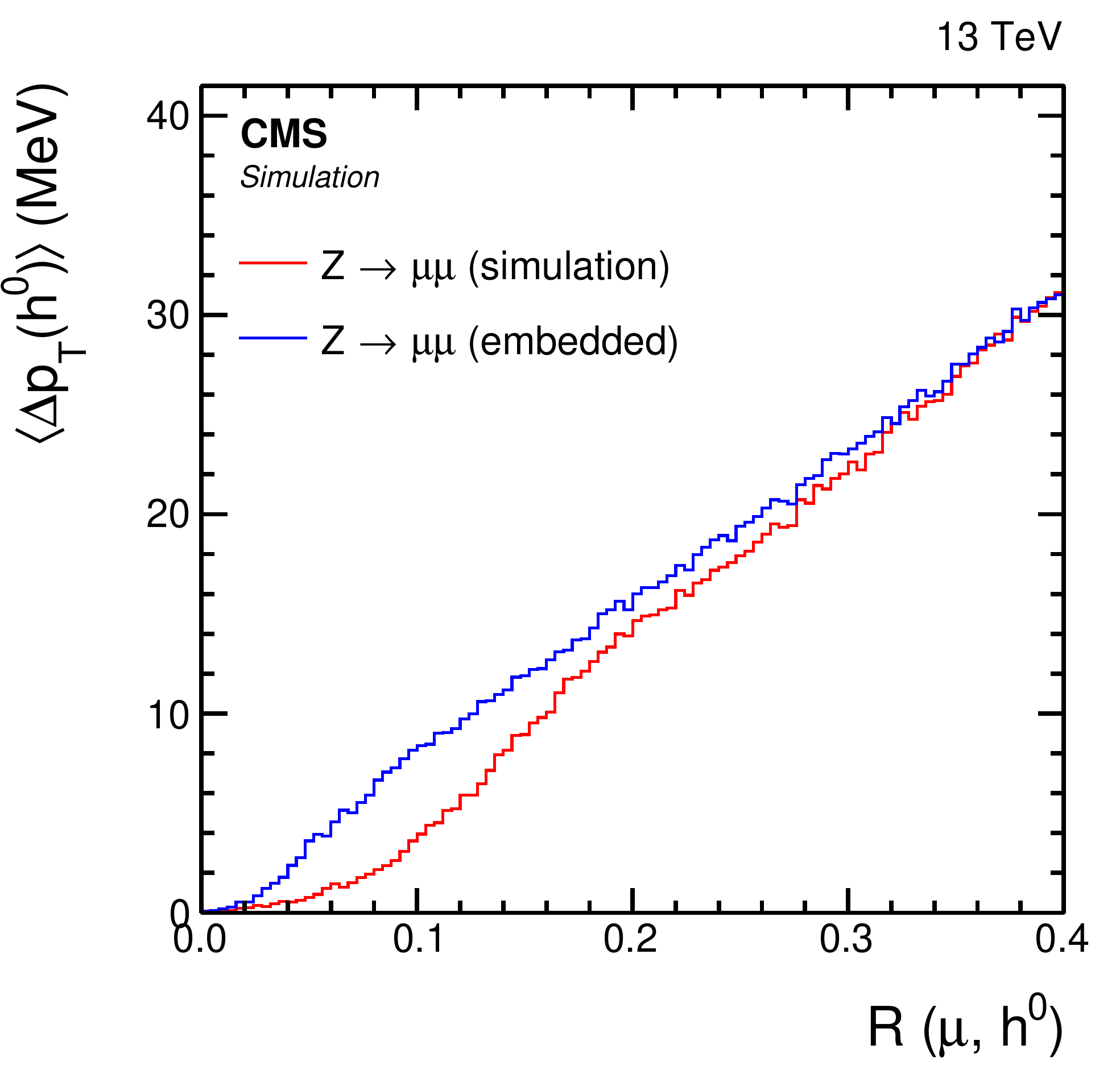}
		\caption {
			Comparison of $\Pgm$-embedded events with exactly the same $\ZMM$ events from simulation. Shown
			is the mean transverse momentum (energy) flux per muon, from all reconstructed particles with the
			distance $R$ from the muon, split by (upper left) charged hadrons from the PV and (upper right)
			PU vertices, (lower left) photons, and (lower right) neutral hadrons. The distributions are shown
			for the $\PGmm$ and for events with $\mmumu$ close to the nominal $\PZ$ boson mass.
		}
		\label{fig:mm-embedding-ptflow}
	\end{figure}
	
	\section{Validation of the method}
	\label{sec:validation}
	
	Simulation-based closure tests are performed to test the validity of the embedding method. For this purpose,
	a validation sample for embedded events is created from simulated $\ZMM$ events, in which the embedding technique is
	applied in the same way as in the observed data: the selected muons are removed from the
	reconstructed event record and replaced with electrons, muons, or tau leptons. The embedded event data samples created in this way
	are compared to simulated events in the same final states. For $\Pe$- and $\Pgt$-embedded events,
	this comparison is performed on statistically independent event samples. For $\Pgm$-embedded events,
	the comparison is performed on exactly the same simulated events, such that only the effects of the
	removal of energy deposits of the initially selected muons, and the reconstruction of the reinjected muons are tested.
	
	For $\Pe$- and $\Pgt$-embedded events, the normalization of the distributions is obtained
	from the yield of selected $\ZMM$ events in the first step of the procedure, as described in
	Section~\ref{sec:Zmm-treatment}. For the $\Pgt$-embedded events, the yield of selected
	$\Pgt\Pgt$ events matches the yield of the simulated $\ZMM$ sample within 1\% with a statistical
	uncertainty of 0.5\%. For the $\Pe$-embedded events a similar agreement is achieved.
	
	\subsection{Validation using the \texorpdfstring{$\Pgm$}{mu}-embedding
		technique}
	\label{sec:validation-mu}
	
	The muon plays a special role
	in validating the embedding procedure itself. The broadening of the kinematic distributions of the
	embedded muons, due to the repeated reconstruction and the finite angular and $\pt$ resolution of the
	detector, and the effects of FSR, have already been discussed in Section~\ref{sec:Ztt-simulation}. For
	the following discussion, the simulation of FSR is switched off in the simulation step of the embedding
	procedure. In this way FSR is simulated only once, during the initial simulation of the validation
	sample, and all FSR effects are the same for the simulated and the embedded event.
	
	Figure~\ref{fig:mm-embedding-validation} shows the $\eta$ and $\pt$ distributions of the leading muon in
	$\pt$, the $\ptmiss$, the invariant mass of the two leading jets in $\pt$, $\mjj$, the number of jets
	with $\pt>30$\GeV and $\abs{\eta}<4.7$, and the number of \PQb jets with $\pt>20$\GeV and $\abs{\eta}<2.5$. The blue dots correspond to the $\Pgm$-embedded event sample and the red histogram
	to the original simulation. The red-shaded bands represent the statistical uncertainty of the simulated event
	sample that is a reference for the comparison. All distributions are based on exactly
	the same events, so that the observed differences can exclusively be attributed to the removal and
	repeated simulation and reconstruction of the embedded muons. The uncertainty bands are added to facilitate the
	assessment of the observed differences between the compared samples. These differences are considered
	acceptable if they are compatible with the statistical uncertainty of the validation sample, which is
        chosen with 10 times more events than the expected number of events in the target analyses.
	
	The kinematic distributions of the muons and jets, and the jet multiplicities are well reproduced.
	The structure in the distributions of the muon $\eta$ follows the geometry of the detector. The
	Jacobian peak corresponding to the $\PZ$ boson decay is clearly visible in the $\pt$ distribution
	of the muon. A 5\% effect in the ratio is visible for low values of $\ptmiss$, which is caused by
	the finite angular and $\pt$ resolution of the detector that can lead to small residual values
	of $\ptmiss$ for events with little or no $\ptmiss$. Corrections due to the finite momentum resolution
	of the detector, as described in Section~\ref{sec:Ztt-simulation}, are not propagated to the
	$\ptmiss$. For $\Pgt$-embedded events this effect is negligible compared to the kinematic fluctuations
	related to the neutrinos involved in the decays, as will be discussed in Section~\ref{sec:validation-tautau}.
	Another 5\% effect in the ratio for $\ptmiss>100$\GeV is explained by rare reconstruction
	effects, where muons of high $\pt$ may create additional track segments, \eg, due to multiple scattering
	in the outer tracker, which are not associated with the initially reconstructed global muon track. After
	the cleaning step of the embedding procedure, such track segments may be picked up in a different way
	and thus lead to a different assignment of $\ptmiss$. Since the validation is based on simulated $\ZMM$
	events, without genuine $\ptmiss$, it is clear that such events point to a poor reconstruction of the
	original event. The fact that this is a 5\% effect only for a small fraction of events, and that the size
	of the effect is small compared to the statistical uncertainty of the validation sample, indicates that it is
	subdominant to the effect at low $\ptmiss$.
	
	\begin{figure}[htbp]
		\centering
		\includegraphics[width=0.45\textwidth]{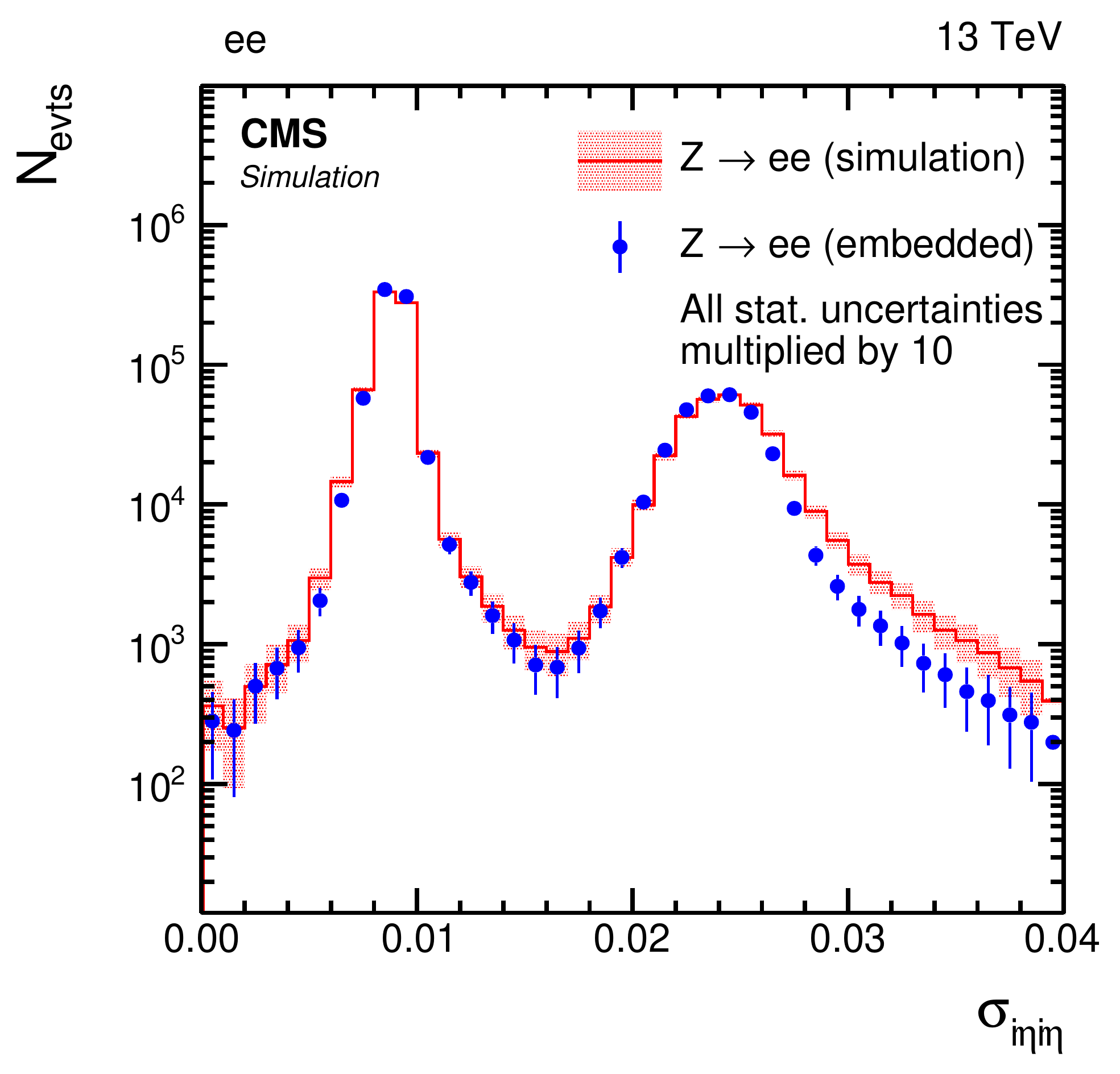}
		\includegraphics[width=0.45\textwidth]{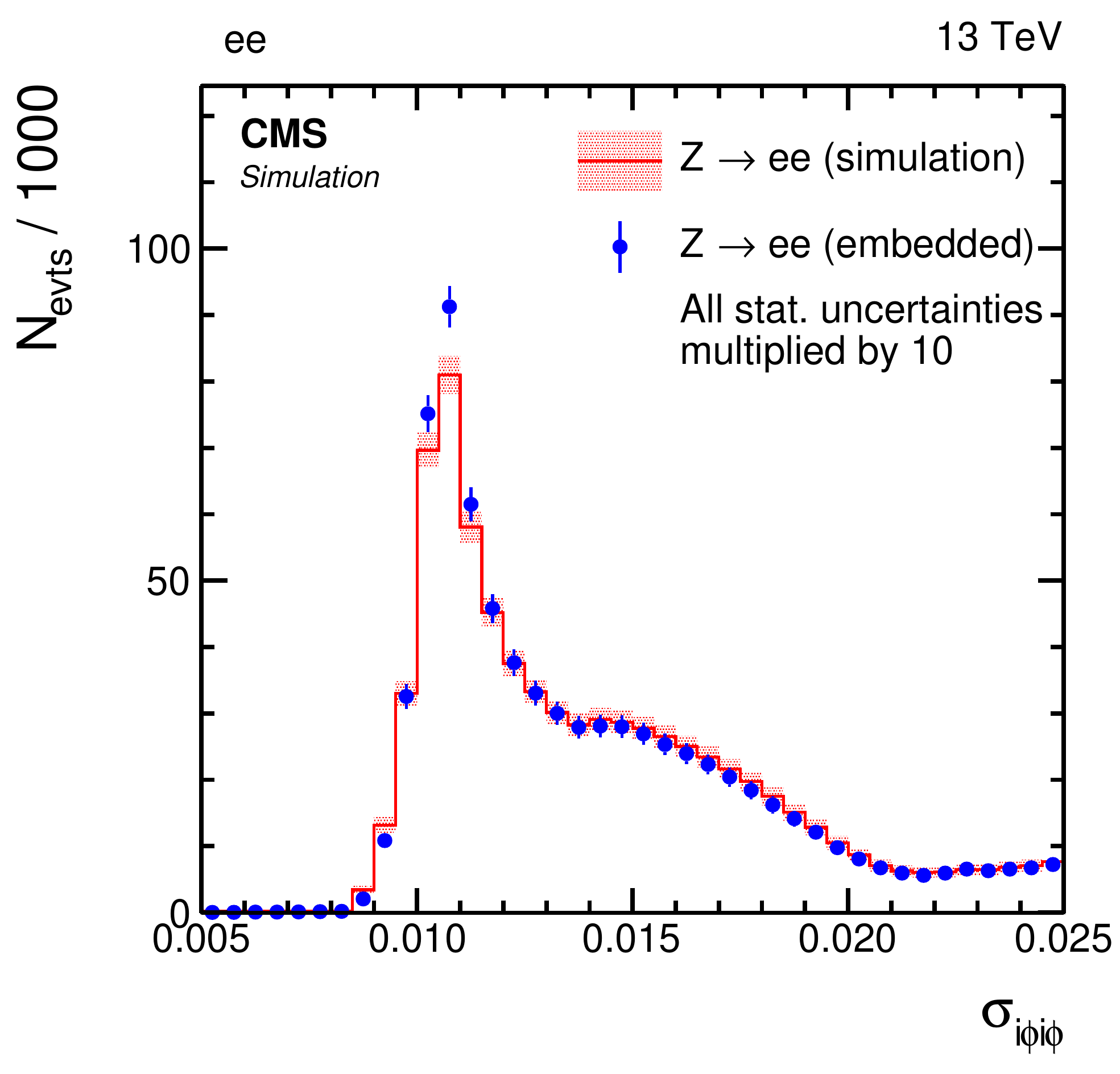}
		\includegraphics[width=0.45\textwidth]{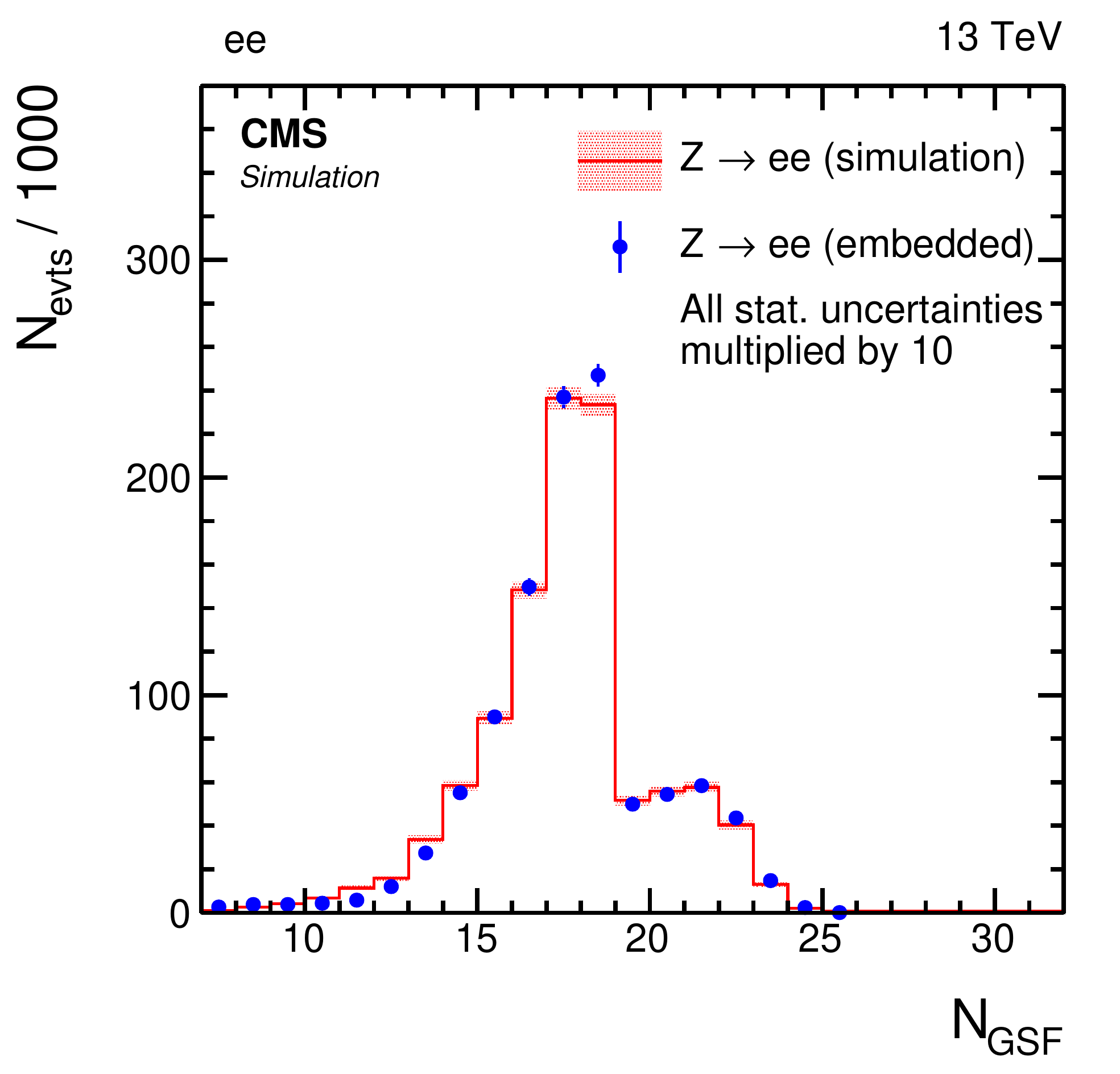}
		\includegraphics[width=0.45\textwidth]{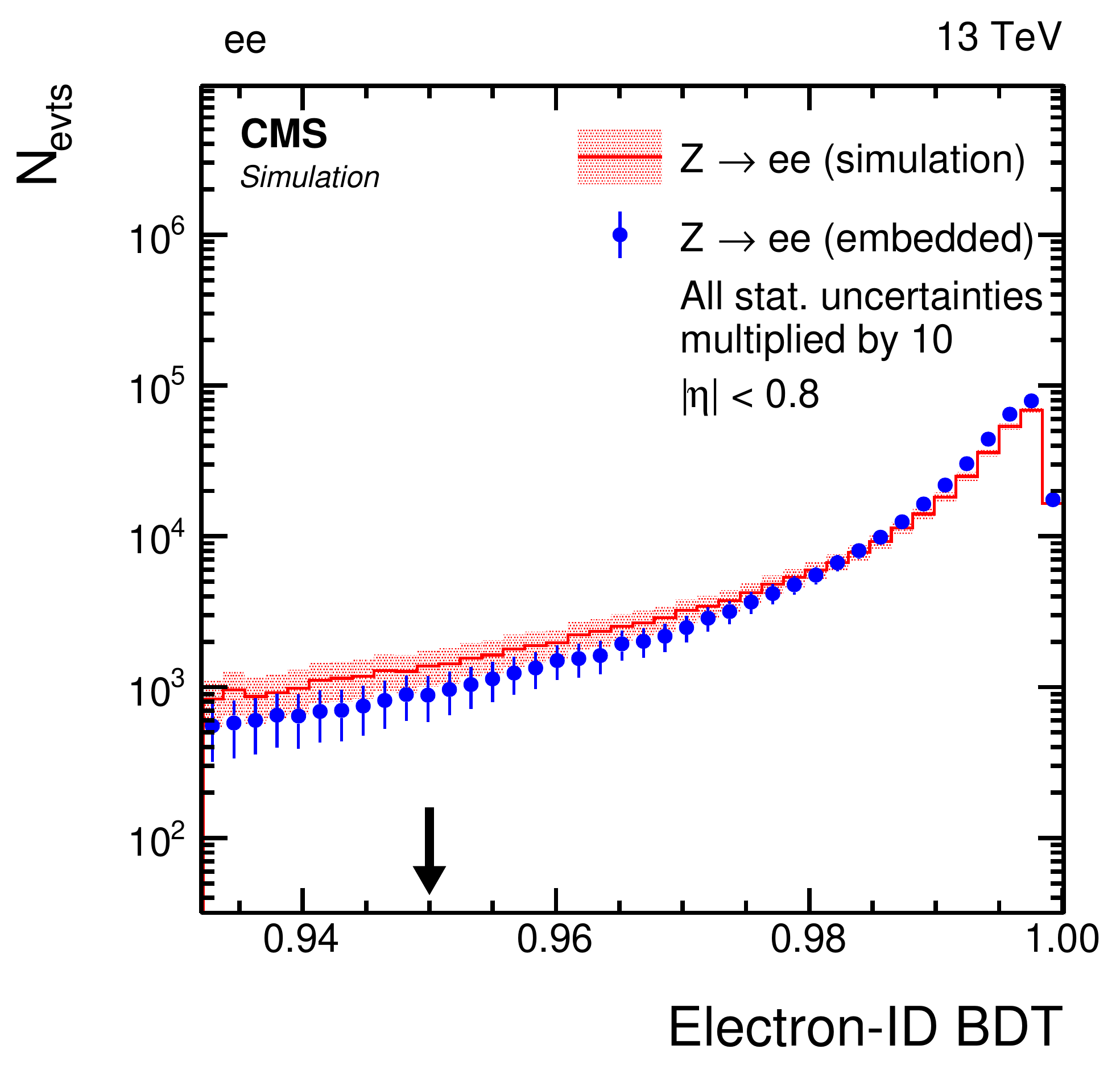}
		\caption {
			Comparison of $\Pe$-embedded events with a statistically independent sample of simulated
			$\ZEE$ events. Shown are distributions of the energy-weighted standard deviations of a $5\times
			5$ crystal array in (upper left) $\eta$, $\sigma_{i\eta i\eta}$, and (upper right) $\phi$,
			$\sigma_{i\phi i\phi}$, as described in the text, (lower left) the number $N_{\mathrm{GSF}}$ of
			detector hits, used for the Gaussian Sum Filter algorithm~\cite{Adam:2005} as described in
			Section~\ref{sec:reconstruction}, and (lower right) the multivariate discriminator for the
			identification of electrons (electron-ID BDT). The black arrow, shown in addition
			to the electron-ID BDT distribution, indicates the working point with 80\% efficiency in the
			displayed electron $\eta$ region. For better visibility, the statistical uncertainties of
			both samples, red-shaded band for simulated
			$\ZEE$ events, and blue vertical bars for $\Pe$-embedded events, are multiplied by
                        10 for the figures.
		}
		\label{fig:ee-embedding-validation_a}
	\end{figure}
	
	Figure~\ref{fig:mm-embedding-ptflow} shows the mean transverse momentum flux per muon, $\langle\Delta\pt
	\rangle$, from all reconstructed particles within the distance $R$ from the muon, split by
	charged hadrons originating from the PV and PU vertices, photons, and neutral hadrons. It is defined
	as the average sum of the $\pt$ (transverse energy in case of neutral particles) of all corresponding particles
	between two cones with radii $R$ and $R{+}\Delta R$ in the distance $R$ from the muon, where $\Delta
	R$ corresponds to the widths of the histogram bins. All distributions are shown for the $\PGmm$
	for events with $\mmumu$ close to the nominal $\PZ$ boson mass.
	
	The figures indicate that in most cases no other particles are reconstructed in the spatial
	vicinity of the muon. For a uniform $\pt$ flux distribution, $\langle\Delta\pt\rangle$
	is expected to increase linearly, because of the increasing area of the ring segments. This trend
	is roughly observed for all reconstructed particle types with a slope of 32\,(550)\MeV per
	unit of $R$ for $\langle\Delta\pt\rangle$ from charged hadrons originating from the PV (PU vertices),
	110\MeV for photons, and 66\MeV for neutral hadrons. The larger slope for charged
	hadrons from PU vertices, photons, and neutral hadrons is related to the simulated PU profile and
	may vary in data. The displayed distributions are shown for the simulated PU profile between
	40 and 70 additional inelastic $\Pp\Pp$ collisions. For charged hadrons and photons, the progression
	from the simulation is well reproduced, apart from small regions close to the muon, which show a small
	excess in $\langle\Delta\pt\rangle$ for charged hadrons from the PV and photons, and a small deficit
	in $\langle\Delta\pt\rangle$ for charged hadrons from PU vertices. A larger difference is observed
	for neutral hadrons, which is due to an incomplete removal of energy deposits of the muon
	in the HCAL, as discussed in Section~\ref{sec:Zmm-cleaning}. When integrated over $R$, and all
	reconstructed particle types, the additional hadronic energy in the predefined isolation cone adds
	up to less than 200\MeV.

	\subsection{Validation using the \texorpdfstring{$\Pe$}{e}-embedding
		technique}
	\label{sec:validation-e}
	
\begin{figure}[t]
		\centering
		\includegraphics[width=0.45\textwidth]{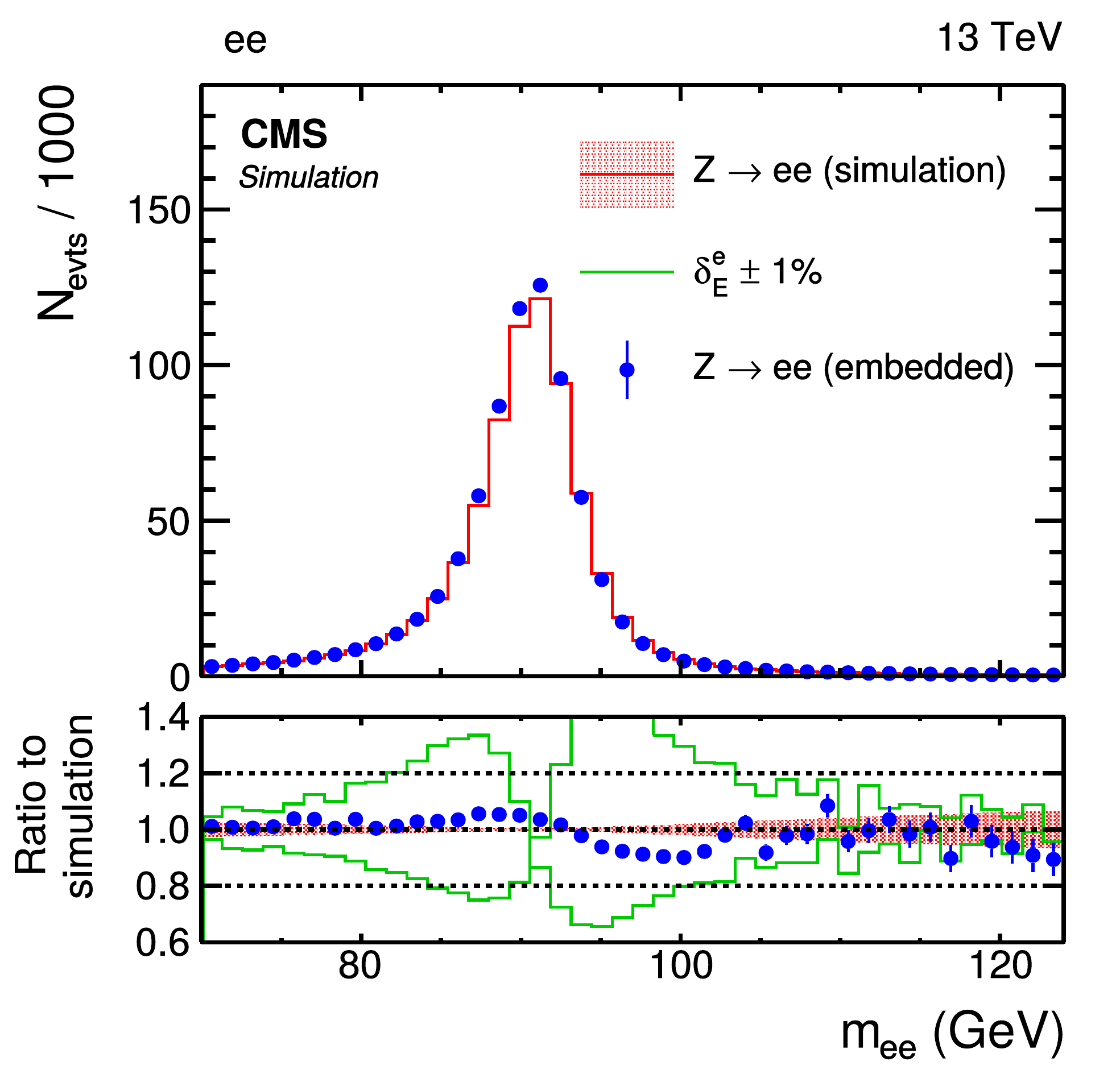}
		\includegraphics[width=0.45\textwidth]{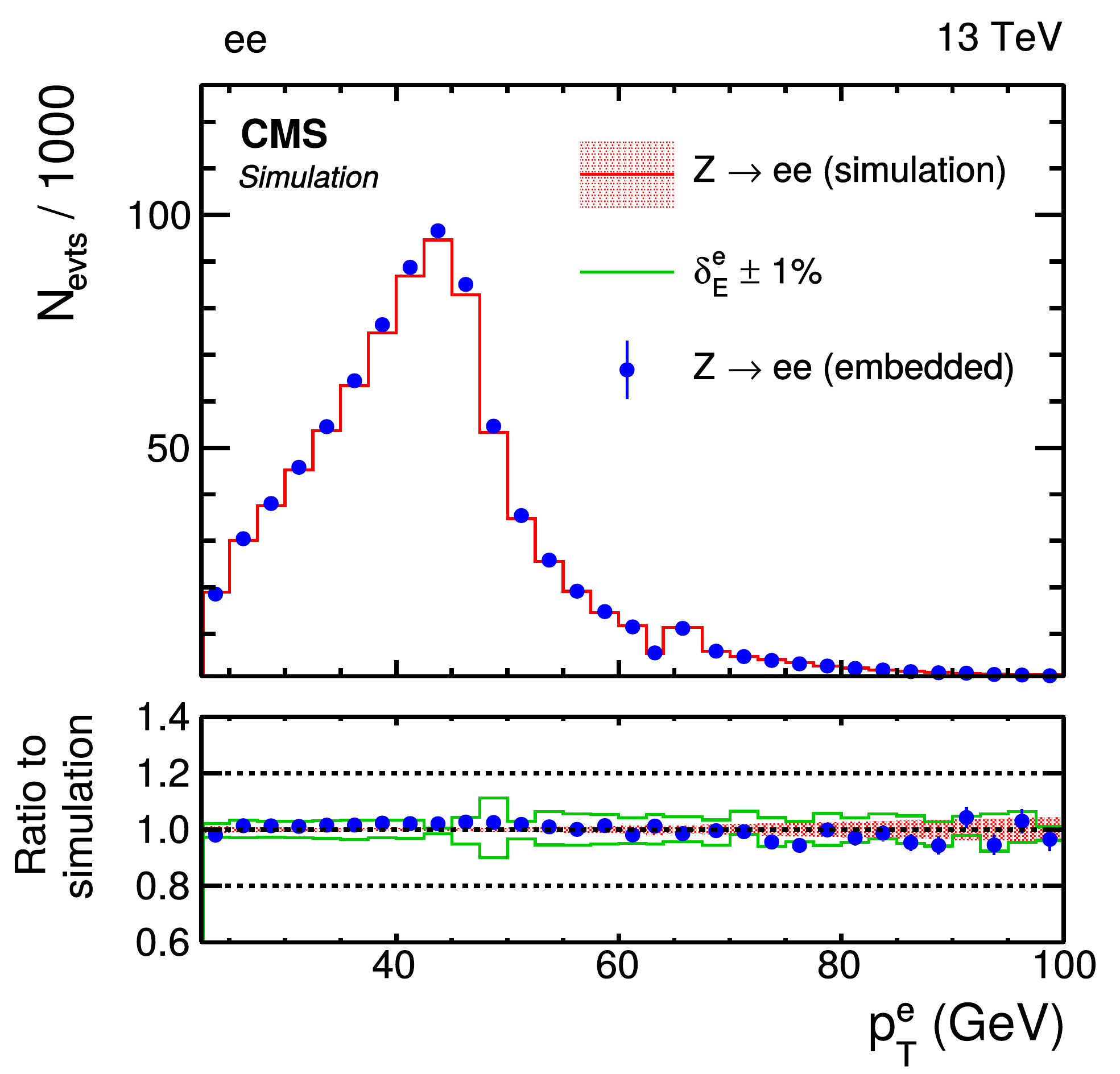}
		\caption {
			Comparison of the $\Pe$-embedded events with a statistically independent sample of simulated
			$\ZEE$ events. Shown are the distributions of (left) $m_{\ee}$ and (right) $\pt$ of the
			leading electron in $\pt$. The blue vertical bars and red-shaded bands correspond to the statistical
			uncertainty of each sample. The effect of a variation of the electron energy scale of $\pm$1\%
			is also shown by the green lines.
		}
		\label{fig:ee-embedding-validation_b}
	\end{figure}

	The identification of electrons in CMS is based on $\mathcal{O}(20)$ closely related detector
	variables that are combined into a multivariate discriminator~\cite{Khachatryan:2015hwa}. As discussed
	in Sections~\ref{sec:Ztt-simulation} and~\ref{sec:Ztt-merging} the simulation of the embedded
	lepton pair takes place in an otherwise empty detector with no other particles from PU, underlying event,
	or additional jet production. The tight relation of the electron reconstruction and identification
	to closely related detector quantities poses an extra challenge to the embedding technique
	for this lepton flavor, which therefore requires a unique validation procedure. To monitor the success in
	simulating the distribution of this discriminator and its inputs, $\Pe$-embedded events are created
	and compared to a statistically independent sample of simulated $\ZEE$ events. Figure~\ref{fig:ee-embedding-validation_a}
	shows, for the leading electron in $\pt$, the energy-weighted standard deviation of the position of a
	$5{\times}5$ ECAL crystal array in $\eta$ ($\sigma_{i\eta i\eta}$) and $\phi$ ($\sigma_{i\phi i\phi}$), 
        and $N_{\text{GSF}}$, the number of detector hits used for the Gaussian Sum Filter algorithm~\cite{Adam:2005} 
        that is introduced in Section~\ref{sec:reconstruction}. The quantities $i\eta$ and $i\phi$ are
	measured in integer crystal units, such that in a $5{\times}5$ array a peripheral crystal can
	be one or two units away from the central crystal in the array. All quantities are in reasonable
	agreement given their high sensitivity to the exact geometry, intercalibration, and level of noise
	suppression of the detector. Also shown is the multivariate discriminator itself (output of the electron-ID boosted decision tree (BDT)),
	which, among others, has the discussed quantities as input. The vertical arrow added to
        Figure~\ref{fig:ee-embedding-validation_a} (lower right) corresponds to the 80\% working point for
        the electron identification. Residual differences in the distributions of the electron-ID BDT
	are comparable to the differences between data and simulation. Correction factors for these differences
	are derived and applied to the $\Pgt$-embedded event samples, and are described in
	Section~\ref{sec:correction-factors}. In Fig.~\ref{fig:ee-embedding-validation_b}, the distributions
	of $m_{\text{ee}}$ and the $\pt$ of the leading electron are shown. The observed differences are explained by
        differences in FSR, as discussed in Section~\ref{sec:Ztt-simulation}. Also shown is the effect of a variation
        of the electron energy scale by $\pm$1\%, which is usually applied to the target analyses and fully covers
        the effect.
	
	\begin{figure}[htbp]
		\centering
		\includegraphics[width=0.45\textwidth]{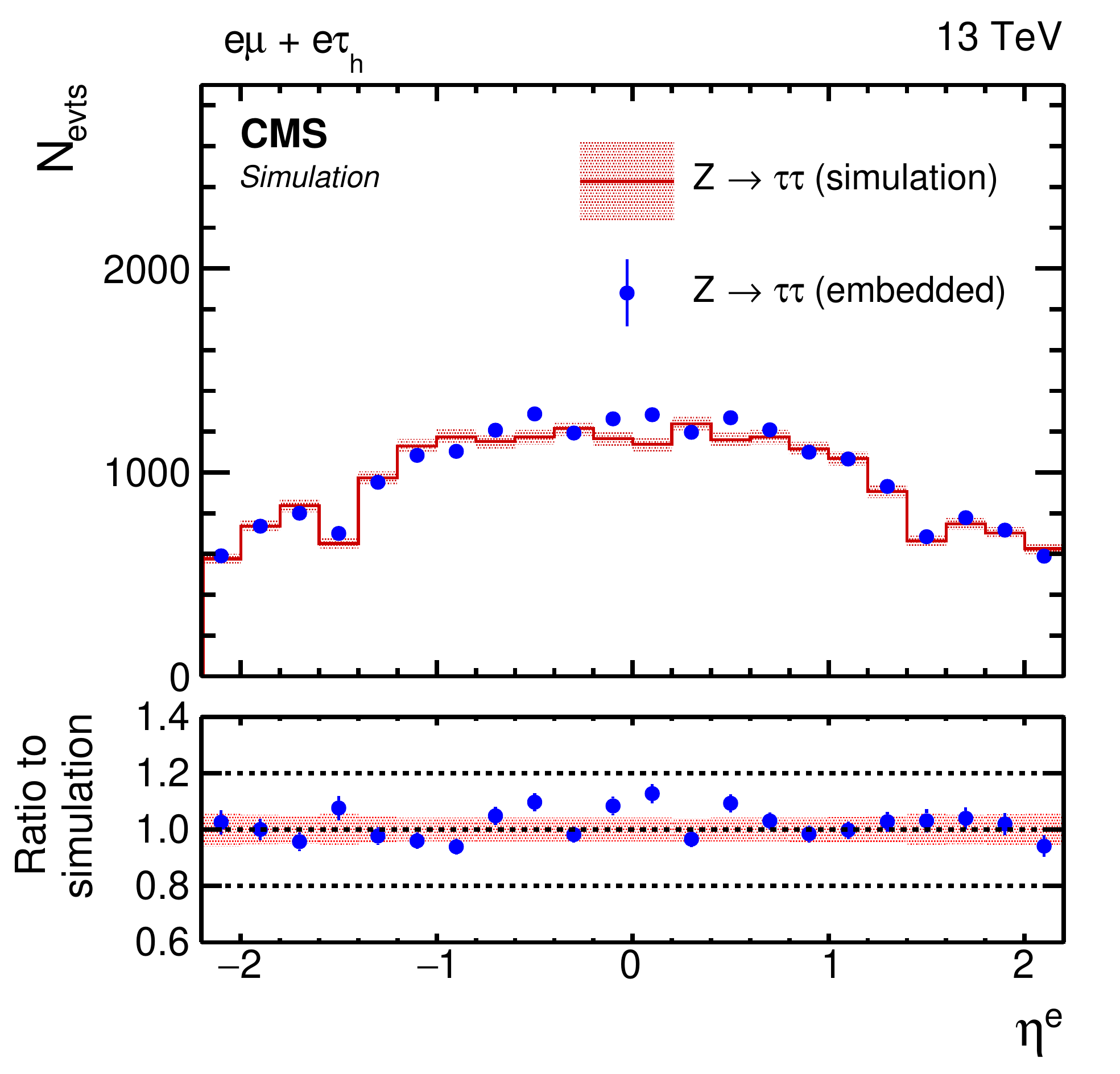}
		\includegraphics[width=0.45\textwidth]{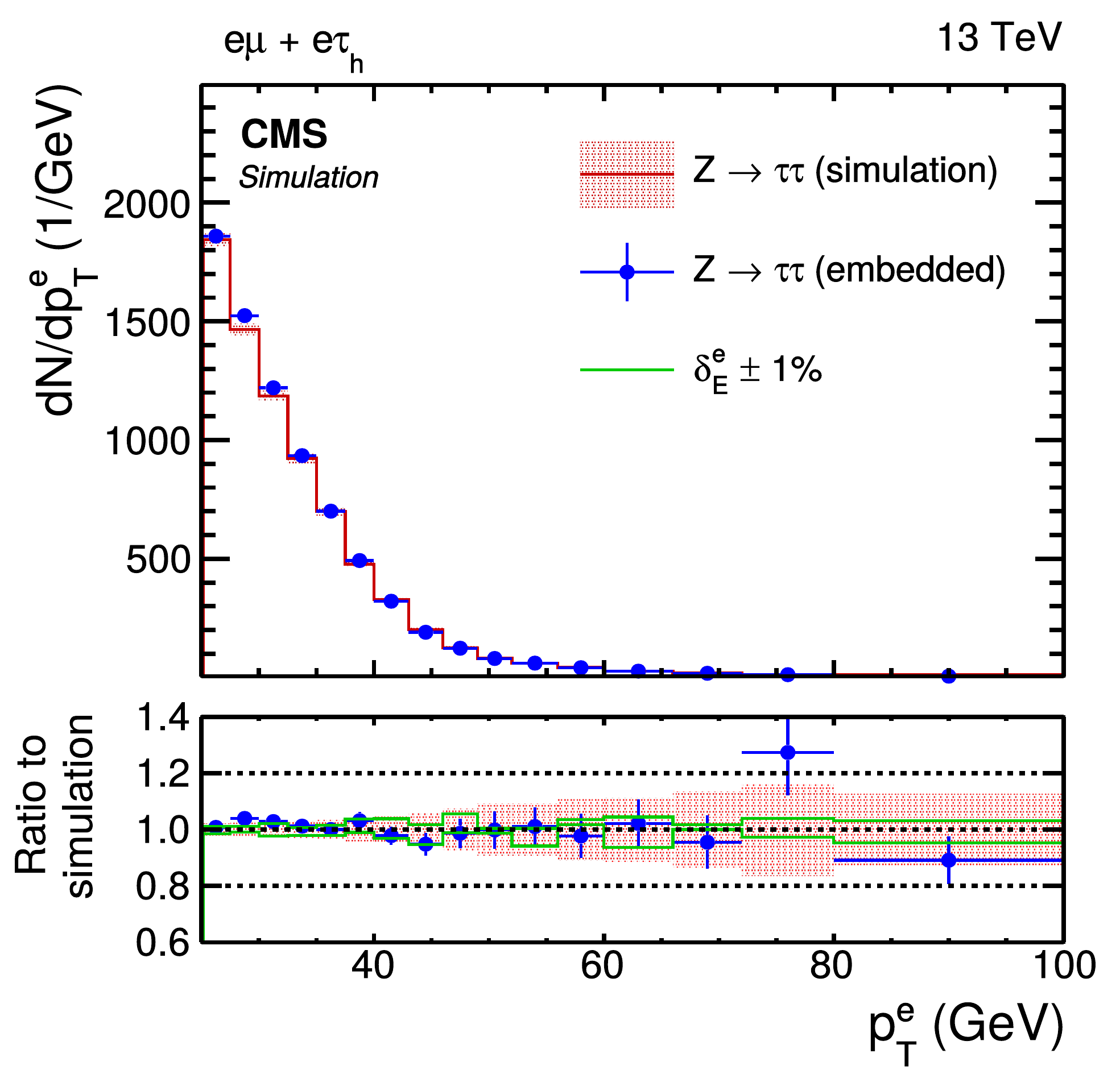}
		\includegraphics[width=0.45\textwidth]{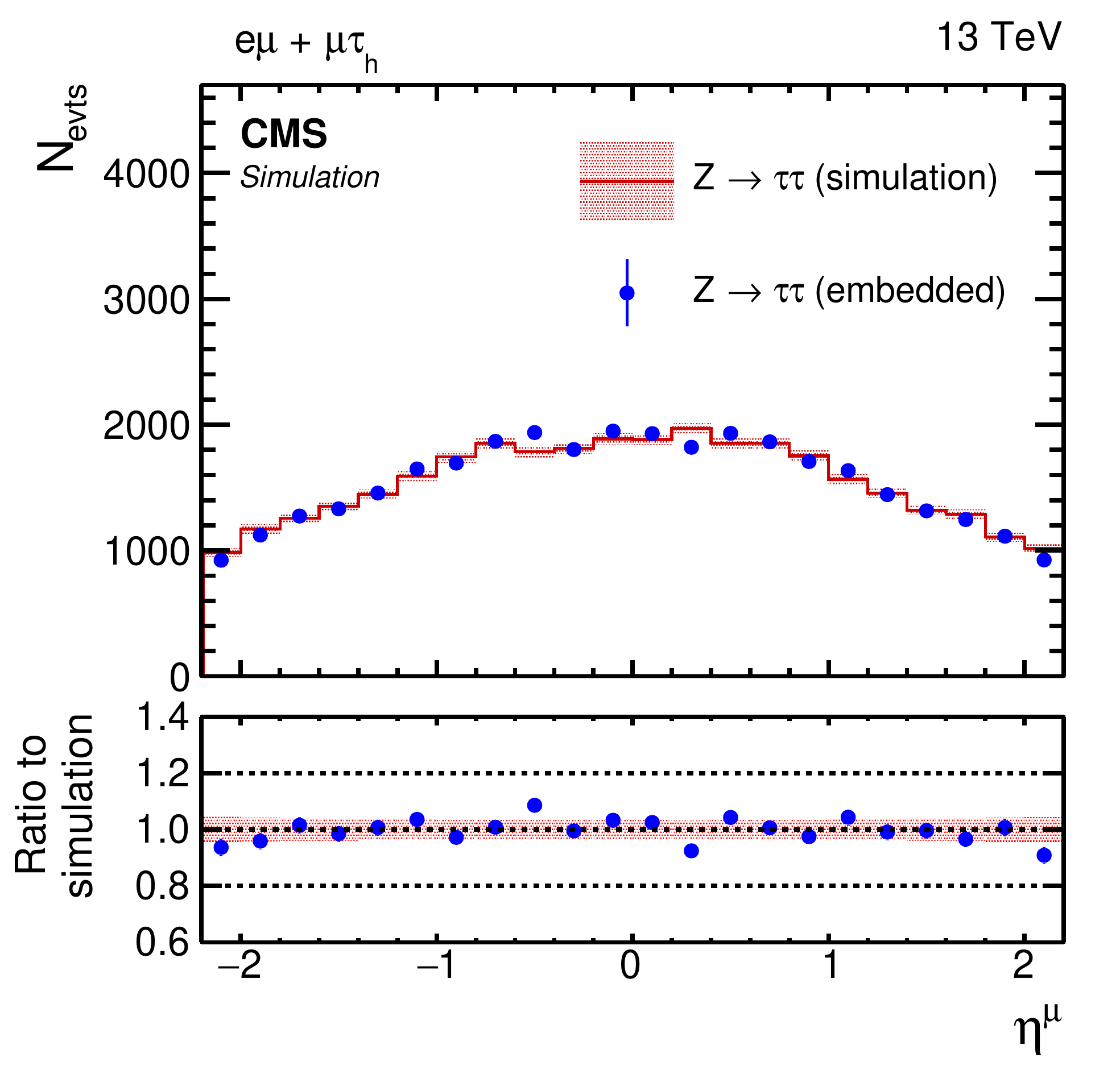}
		\includegraphics[width=0.45\textwidth]{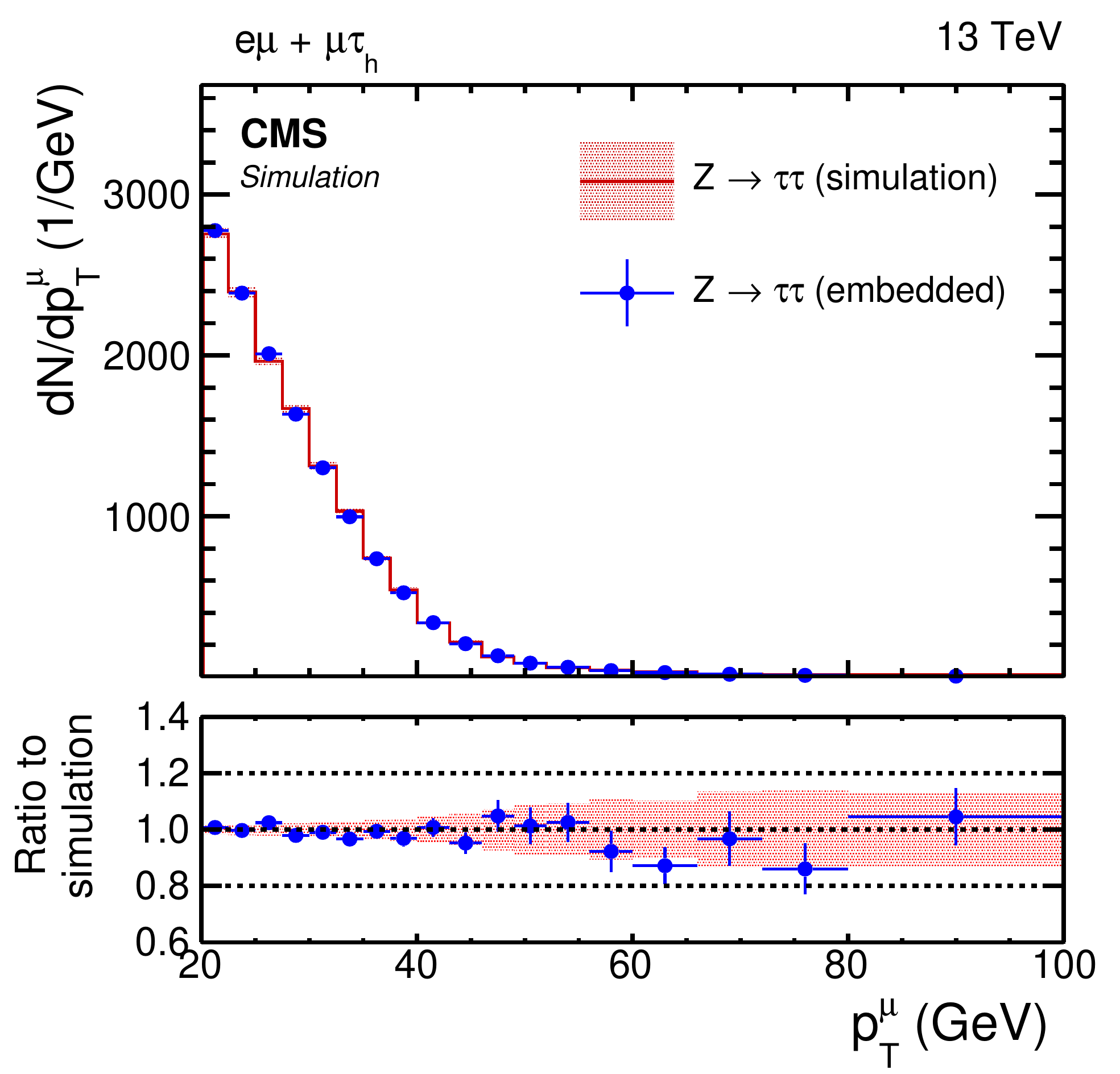}
		\includegraphics[width=0.45\textwidth]{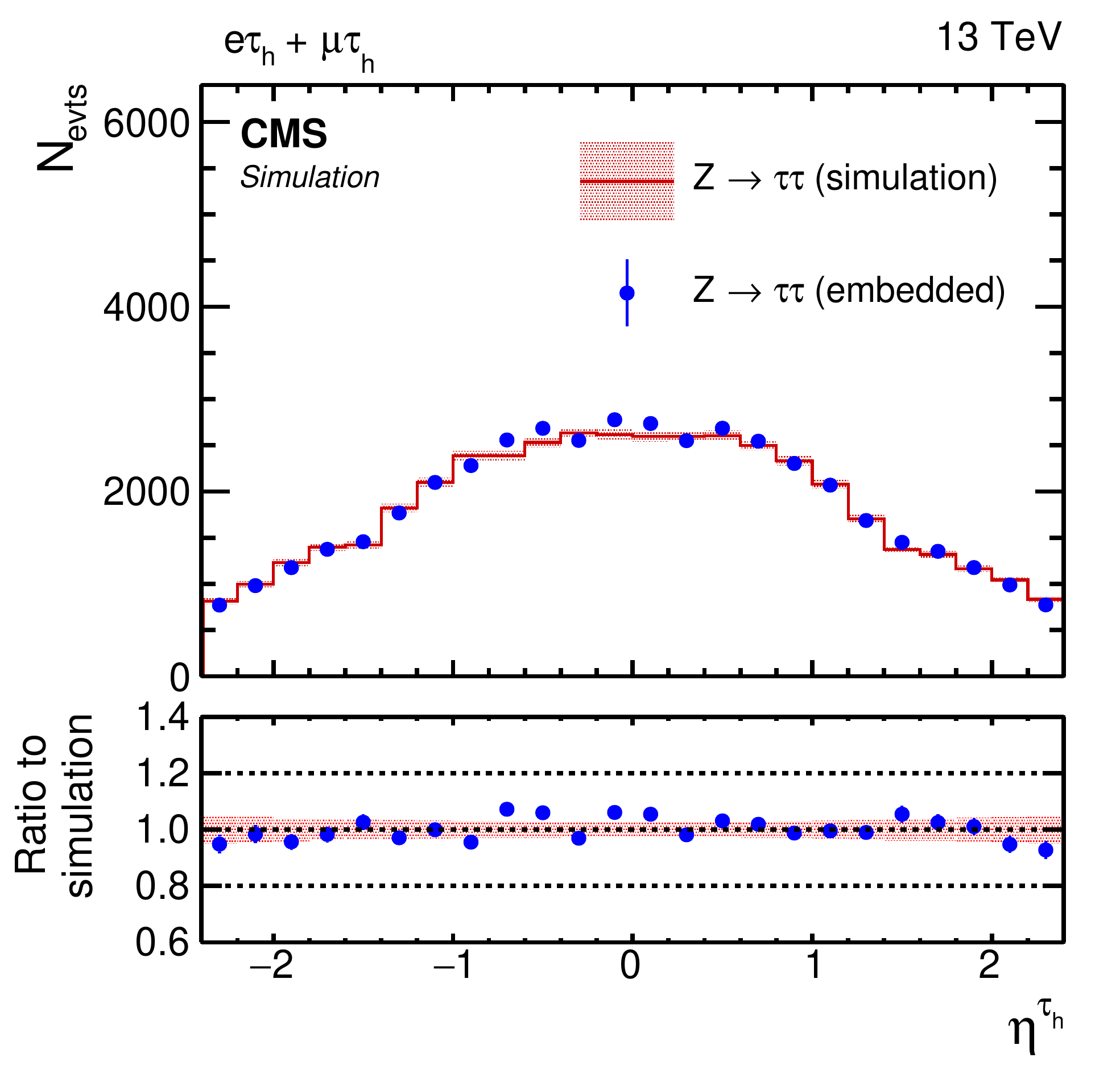}
		\includegraphics[width=0.45\textwidth]{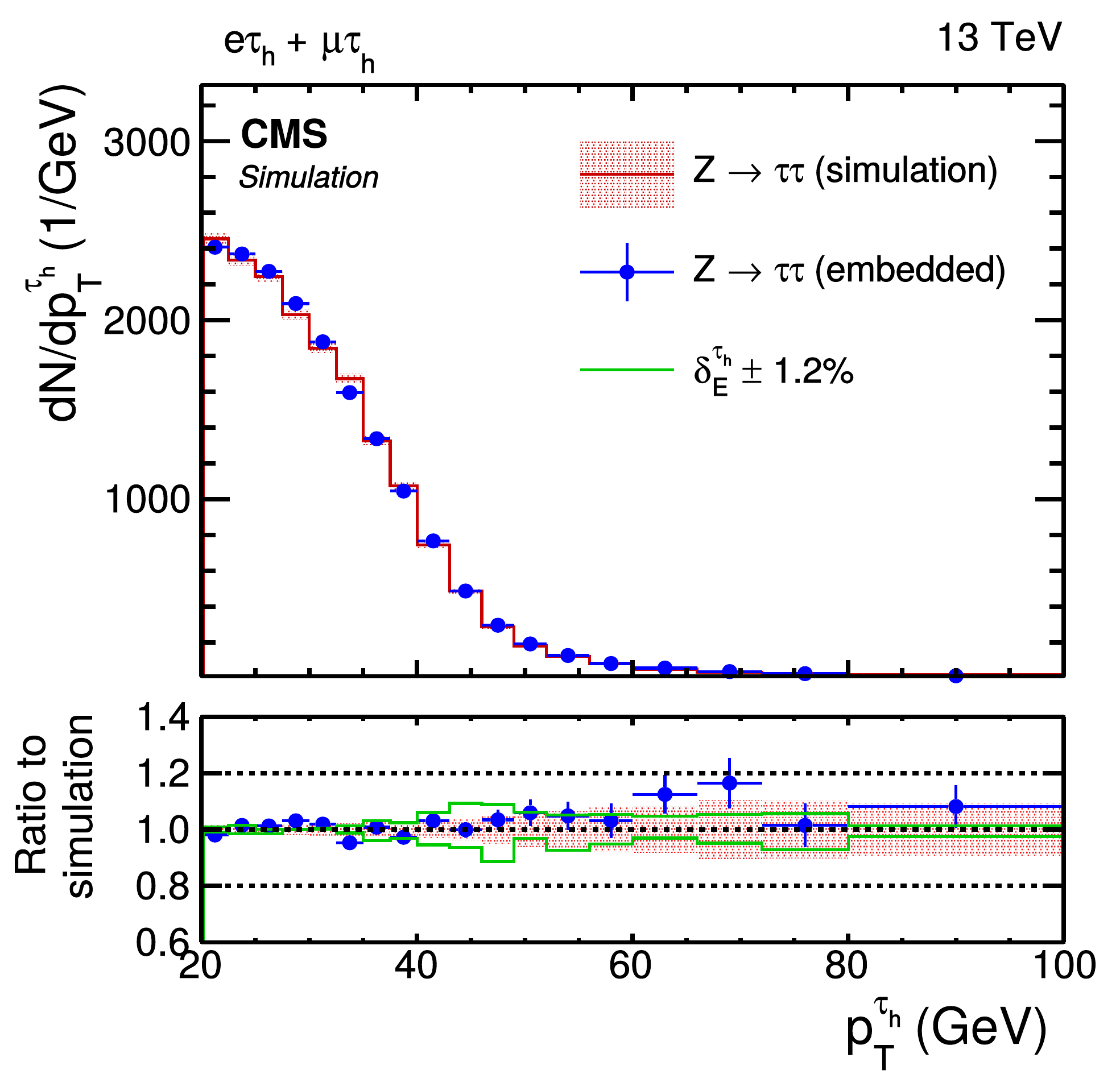}
		\caption {
			Comparison of $\Pgt$-embedded events with a statistically independent sample of simulated $\ZTT$
			events. Shown are the (left) $\eta$ and (right) $\pt$ distributions of the (upper row) electron
			in the $\emu{+}\etau$ final states, (middle row) muon in $\emu{+}\mutau$ final states, and (lower row)
			$\Pgth$ candidate in the $\etau{+}\mutau$ final states. The blue vertical bars and red-shaded bands
			correspond to the statistical uncertainty of each sample. The effect of a variation of the electron ($\Pgth$) energy scale of $\pm$1.0\%\,($\pm$1.2\%)
			is shown by the green lines.
		}
		\label{fig:tau-embedding-validation-lepton-kin}
	\end{figure}
	
	\begin{figure}[htbp]
		\centering
		\includegraphics[width=0.45\textwidth]{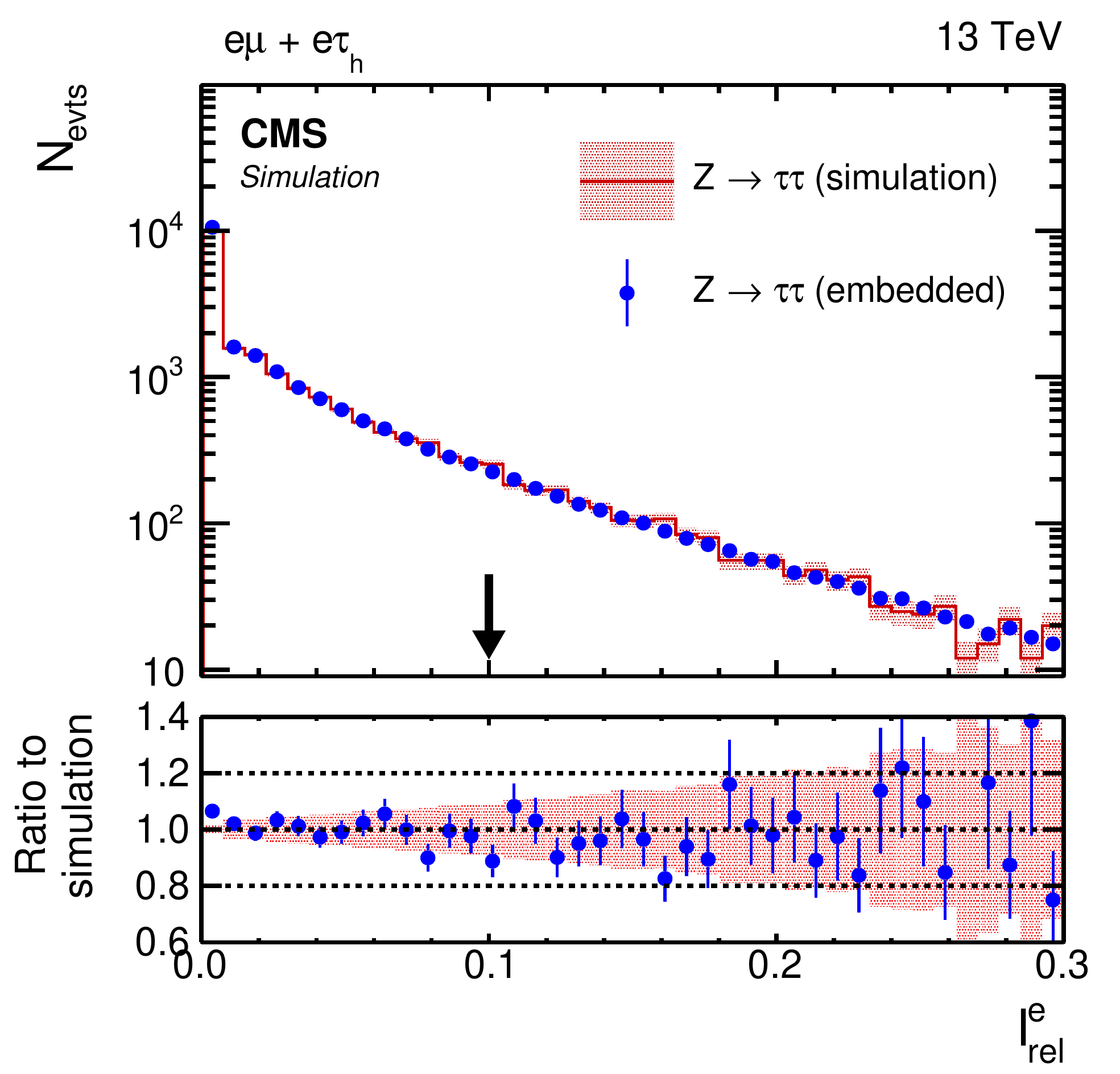}
		\includegraphics[width=0.45\textwidth]{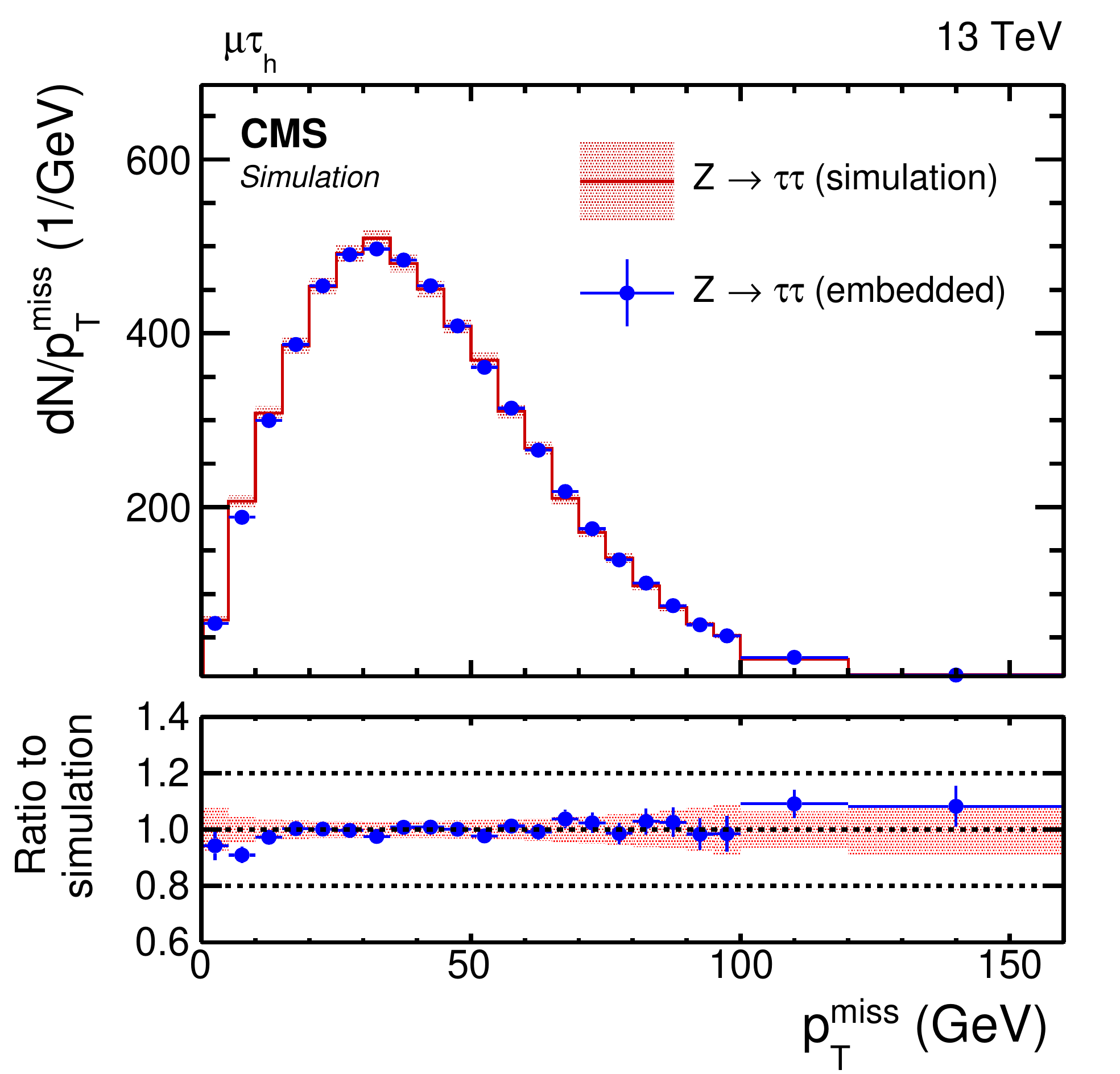}
		\includegraphics[width=0.45\textwidth]{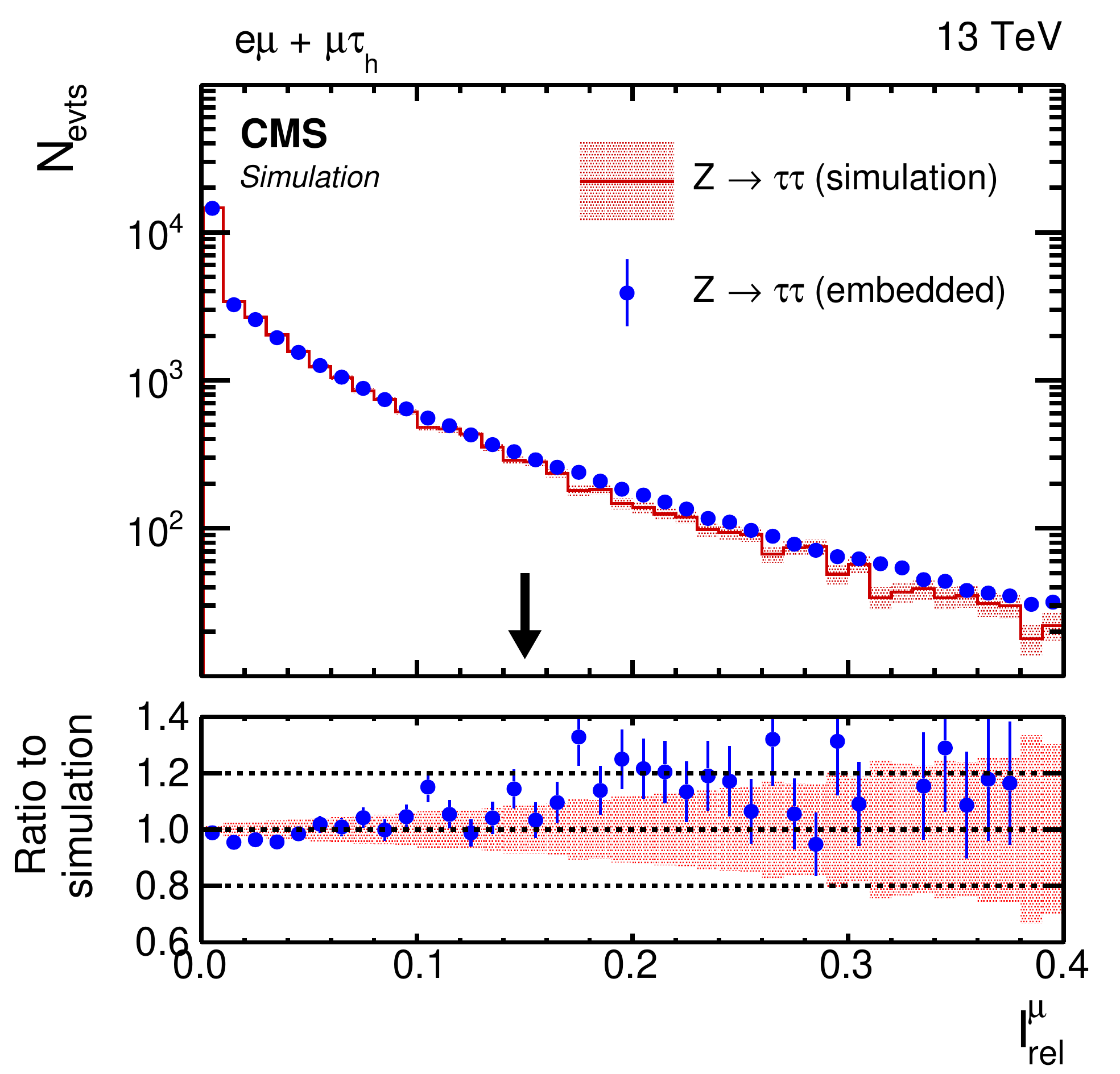}
		\includegraphics[width=0.45\textwidth]{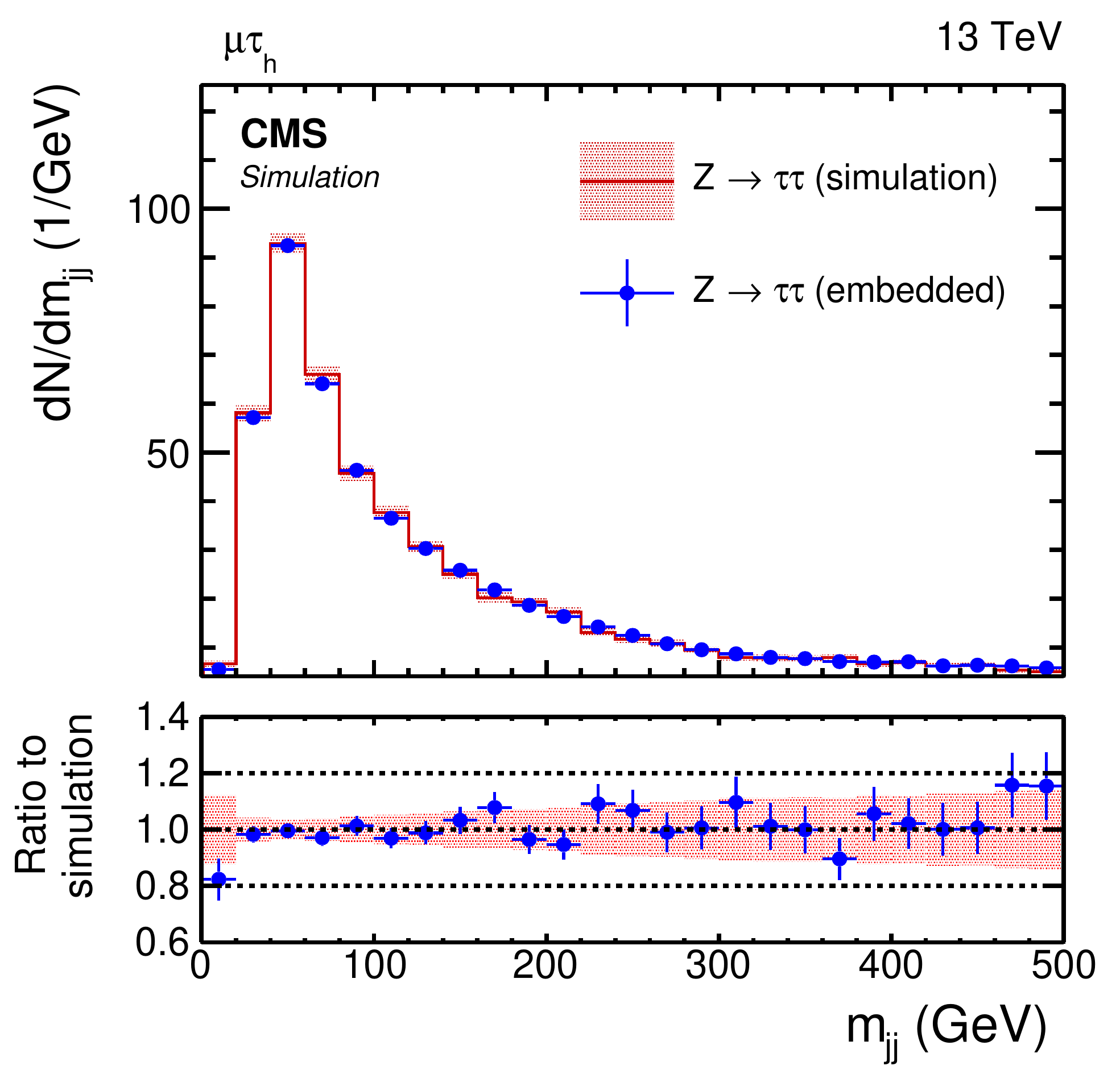}
		\includegraphics[width=0.45\textwidth]{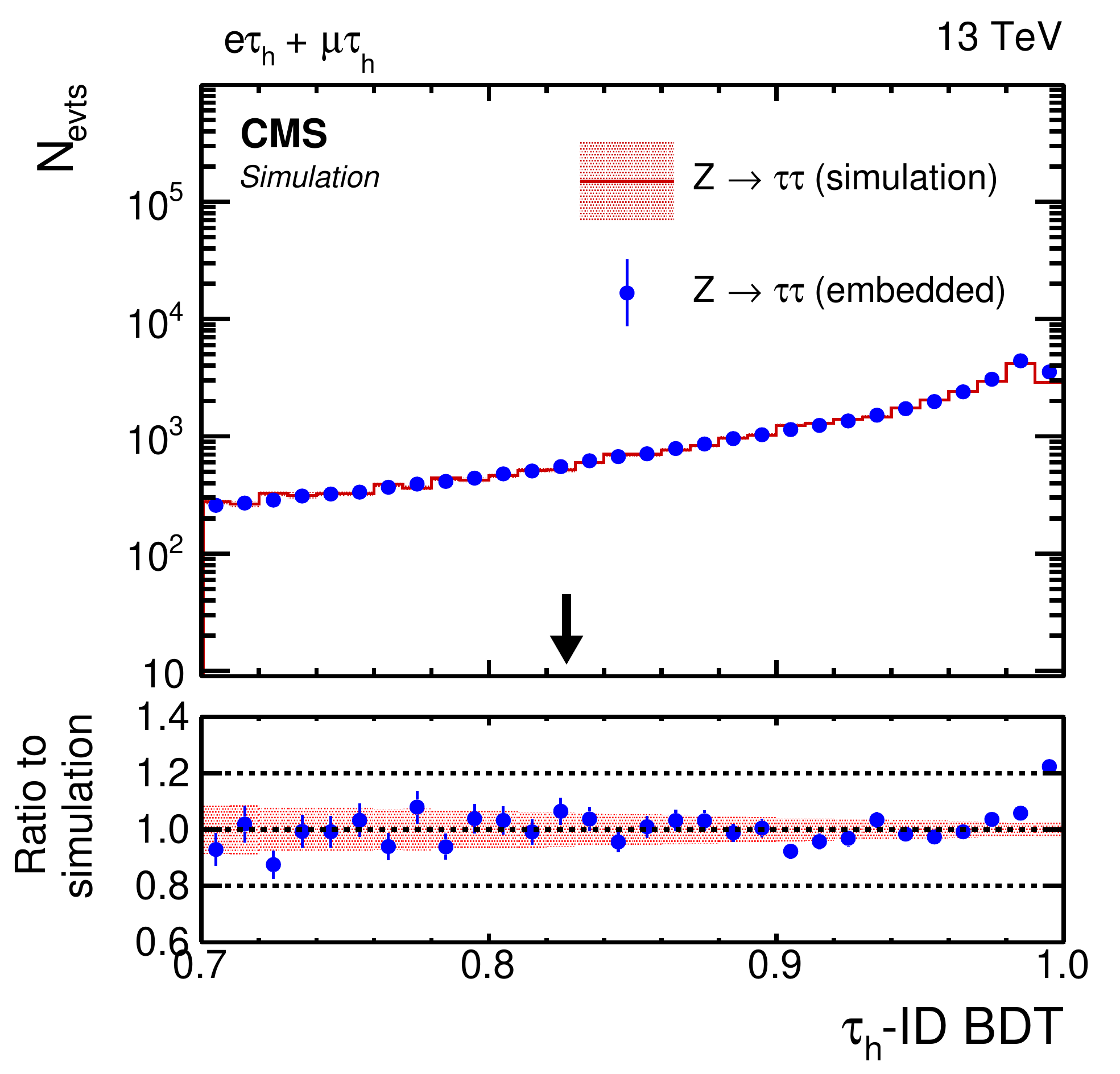}
		\includegraphics[width=0.45\textwidth]{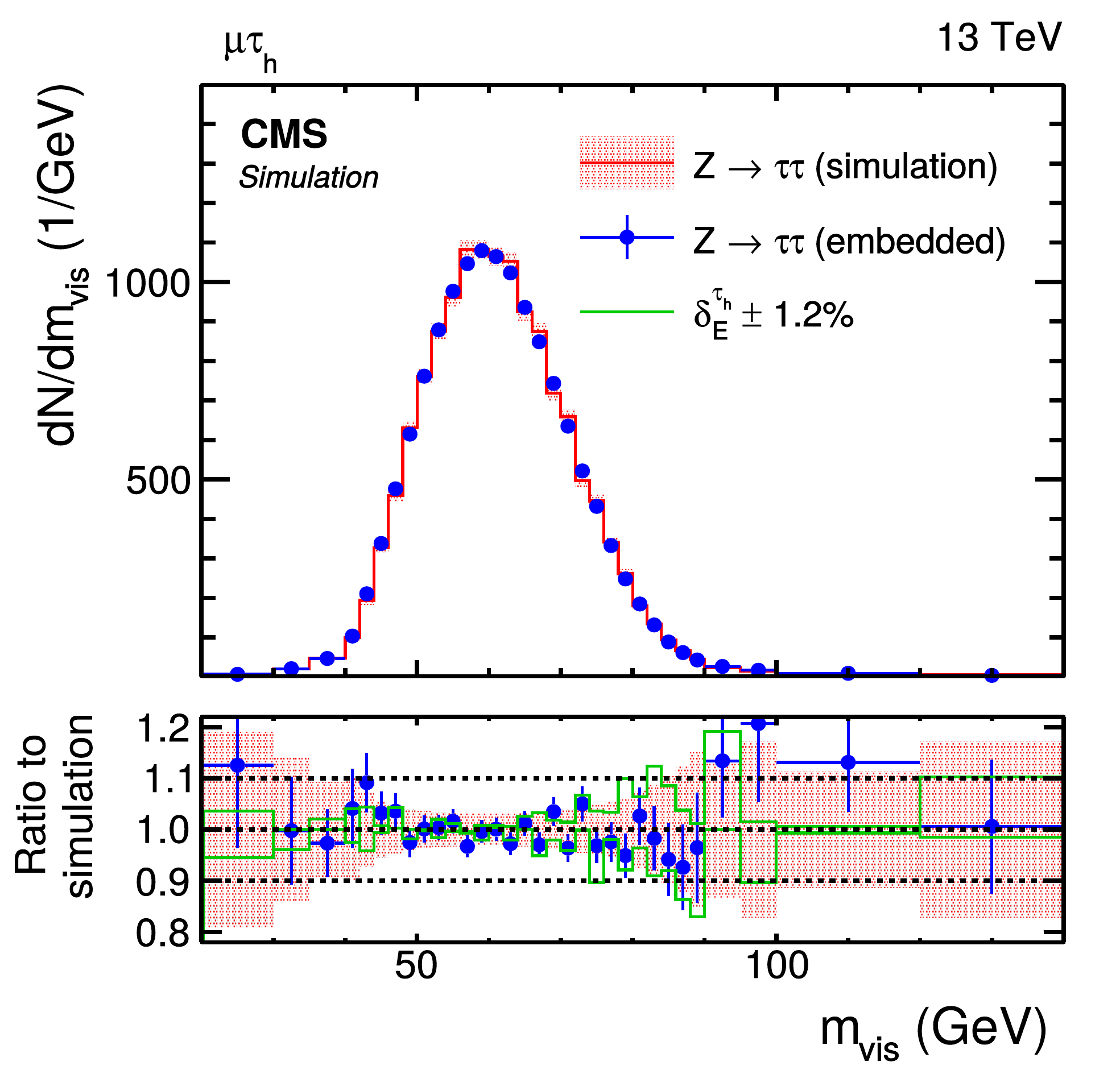}
		\caption {
			Comparison of $\Pgt$-embedded events with a statistically independent sample of simulated $\ZTT$
			events. Shown are distributions of (upper left) $\Irele$, (upper right) $\ptmiss$, (middle left)
			$\Irelm$, (middle right) $\mjj$, (lower left) $\Pgth$-ID BDT, and (lower right) $\mvis$, as discussed in the
			text. The black arrows indicate the working points usually used in the target analyses. The blue
                        vertical
			bars and red-shaded bands correspond to the statistical uncertainty of each sample. The effect
                        of a variation of the $\Pgth$ energy scale of $\pm$1.2\%
			is shown by the green lines.
		}
		\label{fig:tau-embedding-validation-event}
	\end{figure}
	
	\subsection{Validation using the \texorpdfstring{$\Pgt$}{tau}-embedding
		technique}
	\label{sec:validation-tautau}
	
	The main target of the embedding technique, the estimation of $\ZTT$ events is validated by comparing
	$\Pgt$-embedded events to a statistically independent sample of simulated $\ZTT$ events in each
	of the previously discussed $\Pgt\Pgt$ final states. In Fig.~\ref{fig:tau-embedding-validation-lepton-kin}
	the $\pt$ and $\eta$ distributions of the electron, muon, and $\Pgth$ candidate are shown using the
        $\emu$, $\etau$ and, $\mutau$ final states. To increase the statistical
	significance of the validation results, the distributions of the purely lepton related quantities are shown
	for the combination of multiple final states. Figure~\ref{fig:tau-embedding-validation-event} shows the
	distributions of the electron and muon isolation, $\Irelem$, the multivariate $\Pgth$ discriminant
	($\Pgth$-ID BDT), $\ptmiss$, $\mjj$, and the invariant mass of the visible decay products of the tau
	leptons, $\mvis$ in the $\mutau$ final state. The $\Pgt$-embedded event samples, by construction, have
        a larger size than the simulated validation
	sample and thus smaller statistical uncertainties, which becomes apparent from the smaller fluctuations,
	especially in the tails of the steeply falling distributions in the upper panels of the subfigures.
	
	In general, a good agreement is observed, within the statistical precision. Effects of FSR in the
	selection of the $\Pgm\Pgm$ event are not visible in
	the muon $\pt$ and $\mvis$ distributions. This is true for all $\Pgt\Pgt$ final states under
	investigation. Also shown for these distributions are the effects of a shift of the electron energy
	scale by $\pm$1\% and a shift of the tau lepton energy scale by $\pm$1.2\%, corresponding to the
	uncertainties usually applied to the target analyses. Differences in the electron and muon $\eta$
	are covered by the additional uncertainties in the correction for the geometrical $\Pgm\Pgm$ detector
	acceptance. Potential differences in the electron
	$\pt$ are small compared to the electron energy scale uncertainty usually applied to the target
	analyses, as discussed above. The effect of a corresponding shift in the electron energy scale is
	also shown in the corresponding subfigure. The same is true for the $\pt$ of the $\Pgth$ candidate.
	More pronounced deviations are visible in the $\Irelm$ distribution. These are explained by an
	incomplete removal of the energy deposits of the initially selected muons. Integrated over the full
        isolation cone, the expected difference in
	$\pt$ amounts to less than 200\MeV, corresponding to the excess in $\langle\Delta\pt\rangle$,
	as observed in the context of the discussion of Fig.~\ref{fig:mm-embedding-ptflow}. The fact that
	similar effects are not visible in $\Irele$ can be explained by the different reconstruction of
	electrons that may associate parts of the remaining energy deposits of the initially selected muons
	in the calorimeters to the electron clusters, thus removing them from the objects taken into account
	for the calculation of $\Irele$. A 20\% difference in the highest bin of the $\Pgth$-ID BDT
	distribution is explained by the reconstruction of tracks in the otherwise empty detector in
	the simulation step, for $\Pgth$ decays with one or three charged and no additional neutral hadrons.
	The overall effect on the identification efficiency is small and included in corresponding
	correction factors that are discussed in Section~\ref{sec:correction-factors}.
	
	In summary, in all investigated Drell--Yan final states, the agreement of the embedded event samples
	with the corresponding validation sample is observed to be compatible with the simulation. Most of
        the observed differences are within the statistical precision of the
	validation sample and smaller than the statistical precision of the target analyses in the $\Pgt\Pgt$
	final state. Residual systematic trends have been checked to have negligible effects on the target
	analyses. No further measures are taken to improve the agreement of the embedded event samples with
	the simulation. Instead, correction factors for the reconstruction and identification of the simulated
	electrons, muons and tau leptons are derived from $\Pe$-, $\Pgm$- and $\Pgt$-embedded events, in analogy
	to the correction factors usually provided for fully simulated events, as will be discussed in
	Section~\ref{sec:correction-factors}.
	
	\section{Application of the \texorpdfstring{$\Pgt$}{tau}-embedding
		technique to data}
	\label{sec:application}
	
	The $\Pgt$-embedded event samples used for the target analyses are obtained using the
	$\Pgm\Pgm$ data event selection.
	They replace the simulation of all $\ZTT$, $\ttbar(\Pgt\Pgt)$ and diboson$(\Pgt\Pgt)$ events in the
	$\Pgt\Pgt$ final states. To prevent double counting,
	$\ttbar(\Pgt\Pgt)$ and diboson$(\Pgt\Pgt)$ events are removed from background estimates that
        use simulation. Their
	selection must be performed on the undecayed tau leptons, at the stable particle level.

        The $\Pgt$-embedded event sample, except for the $\Pgt$ decays, provides a data description better than
        the $\ZTT$ simulation. The simulation
	can only reach an equivalent performance after a significant amount of tuning. This is true for the
	time-dependent PU profile of the data, the production of additional jets, especially in exclusive
	kinematic corners, like for multijet, multi \PQb jet, forward jet, or vector boson fusion topologies and
	the underlying event. Other event quantities which are typically difficult to model in the simulation
	are the number of reconstructed primary interaction vertices, or $\ptmiss$. All quantities
	referring to the part of the event that is obtained from the data may be used in the target analyses
	without any further corrections. The time needed to produce
	the $\Pgt$-embedded event sample is of the order of time necessary to reprocess
	the collected $\Pgm\Pgm$ data set. The size of the $\Pgt$-embedded event sample is 5 to 60 times the
        size of the data sample used for
	the target analyses. These are advantages over the simulation that will become even more
        important for the planned High-Luminosity LHC upgrade, where typically between 140 and 200 PU
        collisions are expected.

	The ability of the $\Pgt$-embedded event samples to describe the data is demonstrated below using a data set
	corresponding to an integrated luminosity of $41.5\fbinv$, collected with the CMS detector in 2017.
	
	\subsection{Correction factors}
	\label{sec:correction-factors}
	
	Residual differences between the $\Pgt$-embedded event samples and the data in individual
	control distributions, related to the simulated part of the event, can be adjusted by
	$\pt$- and $\eta$-dependent correction factors for the efficiencies of the selection and isolation
	requirements on each corresponding lepton. These correction factors map the efficiencies observed in the
        embedded event samples to the efficiencies observed in data. For electrons and
	muons they are obtained from a comparison of $\Pe\Pe$ ($\Pgm\Pgm$) selected events on the $\Pe$ ($\Pgm
	$)-embedded event samples with the same event selection on data, using the ``tag-and-probe''
	method~\cite{Khachatryan:2010xn}. They are provided as individual correction factors for the lepton
	identification and isolation efficiency, and the corresponding leg of the triggers used in the target
	analyses. The estimate of the reconstruction efficiency is included in the identification efficiency.
	
\begin{figure}[htbp]
		\centering
		\includegraphics[width=0.45\textwidth]{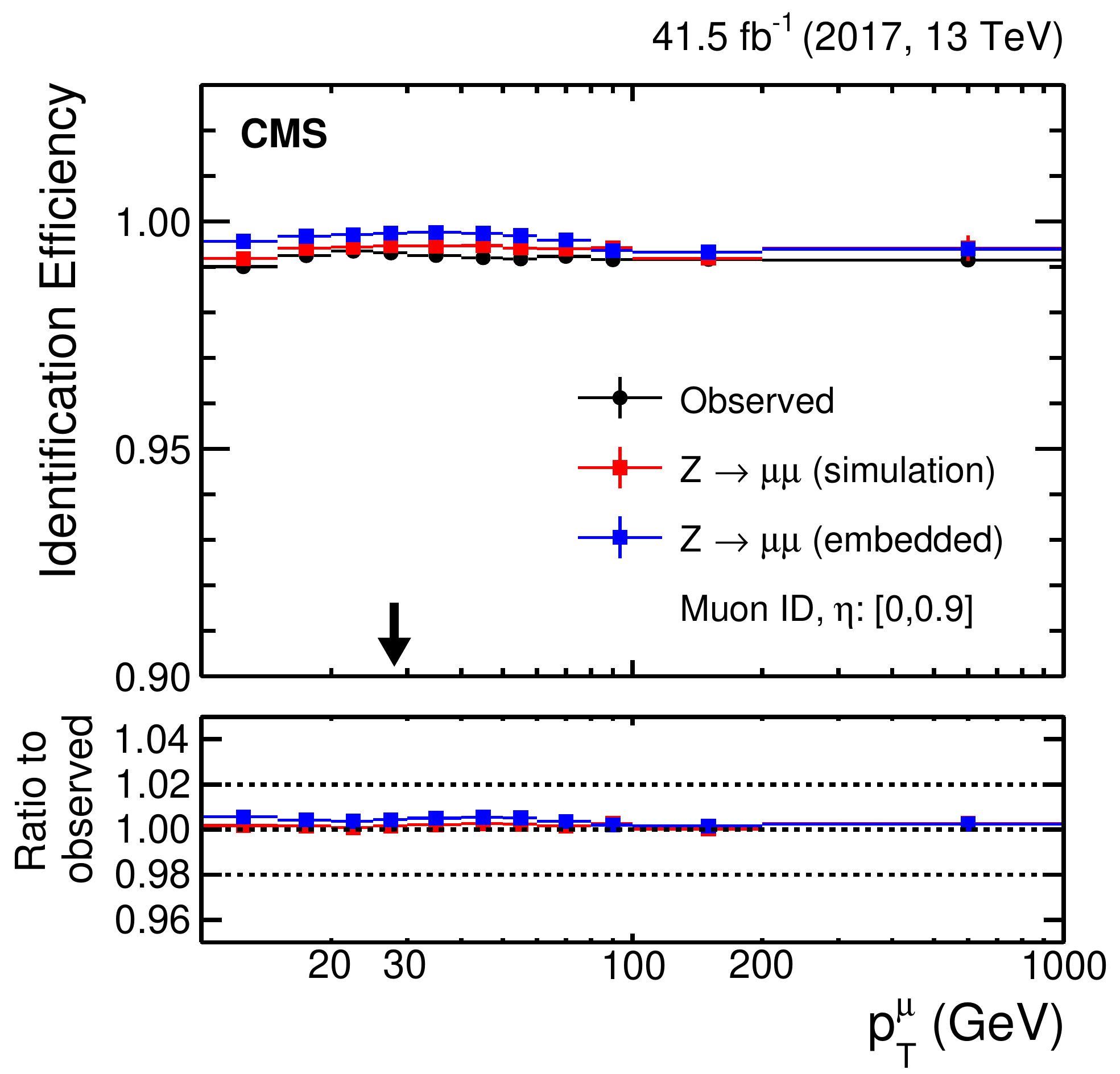}
		\includegraphics[width=0.45\textwidth]{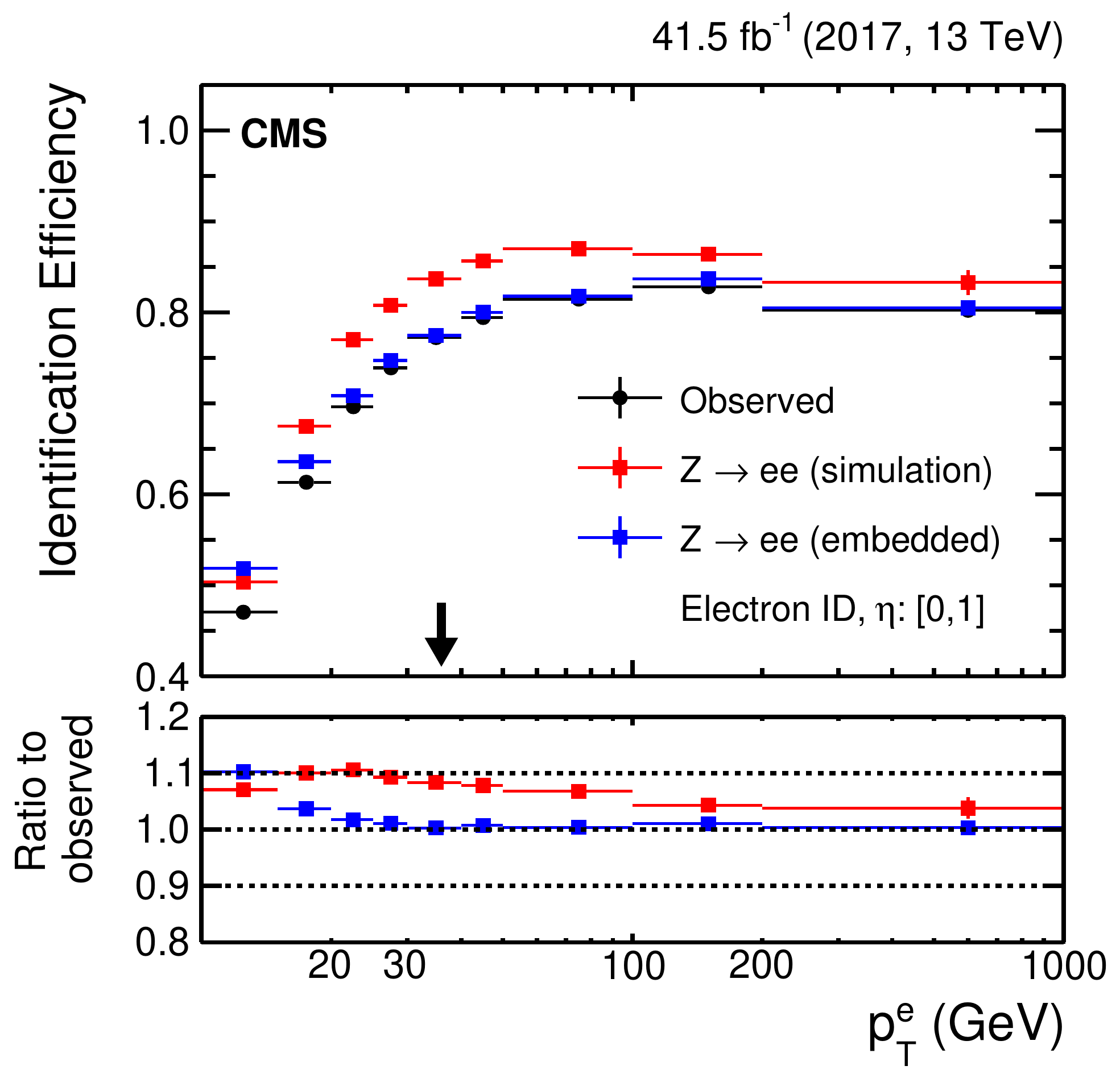}
		\includegraphics[width=0.45\textwidth]{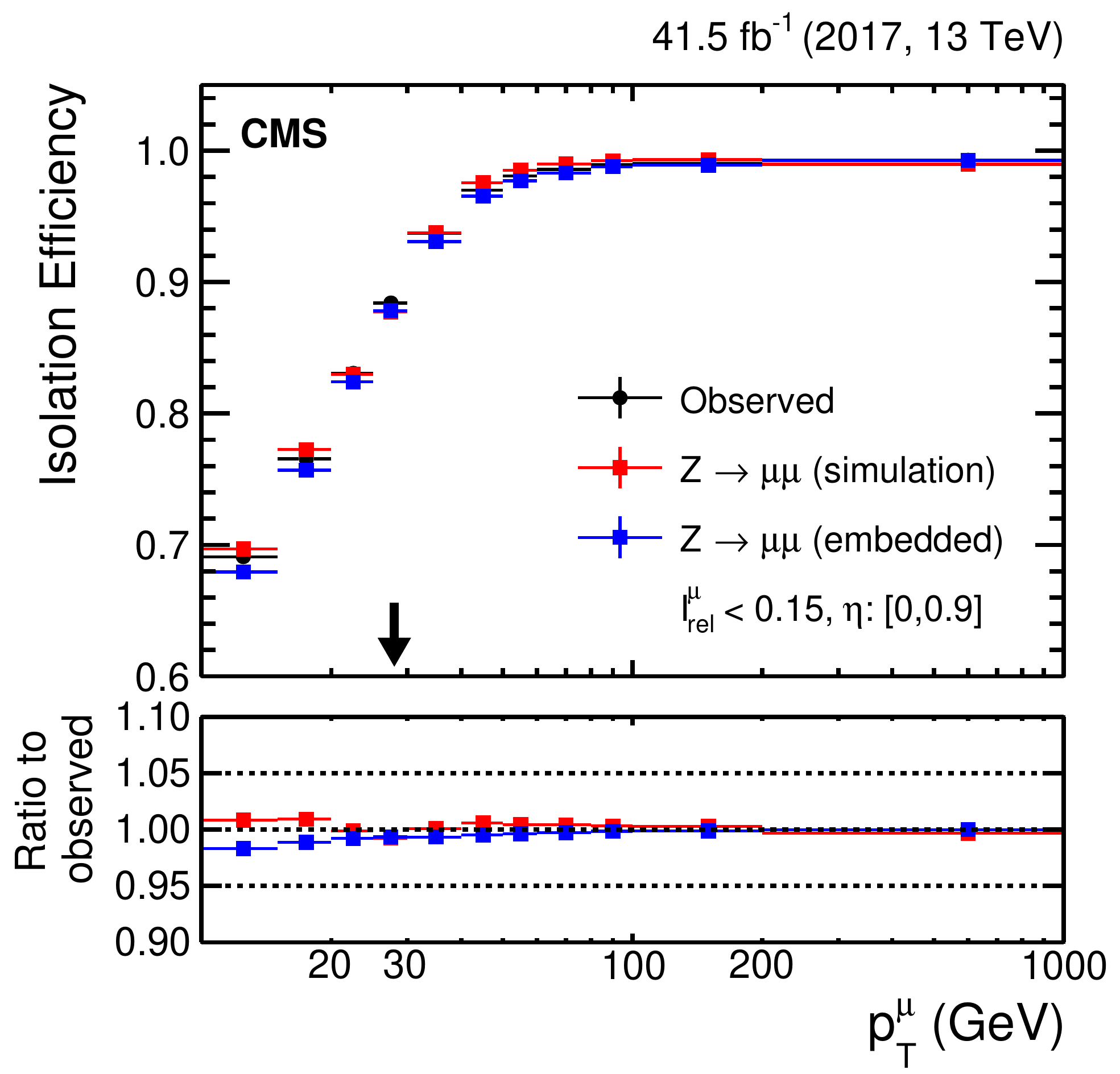}
		\includegraphics[width=0.45\textwidth]{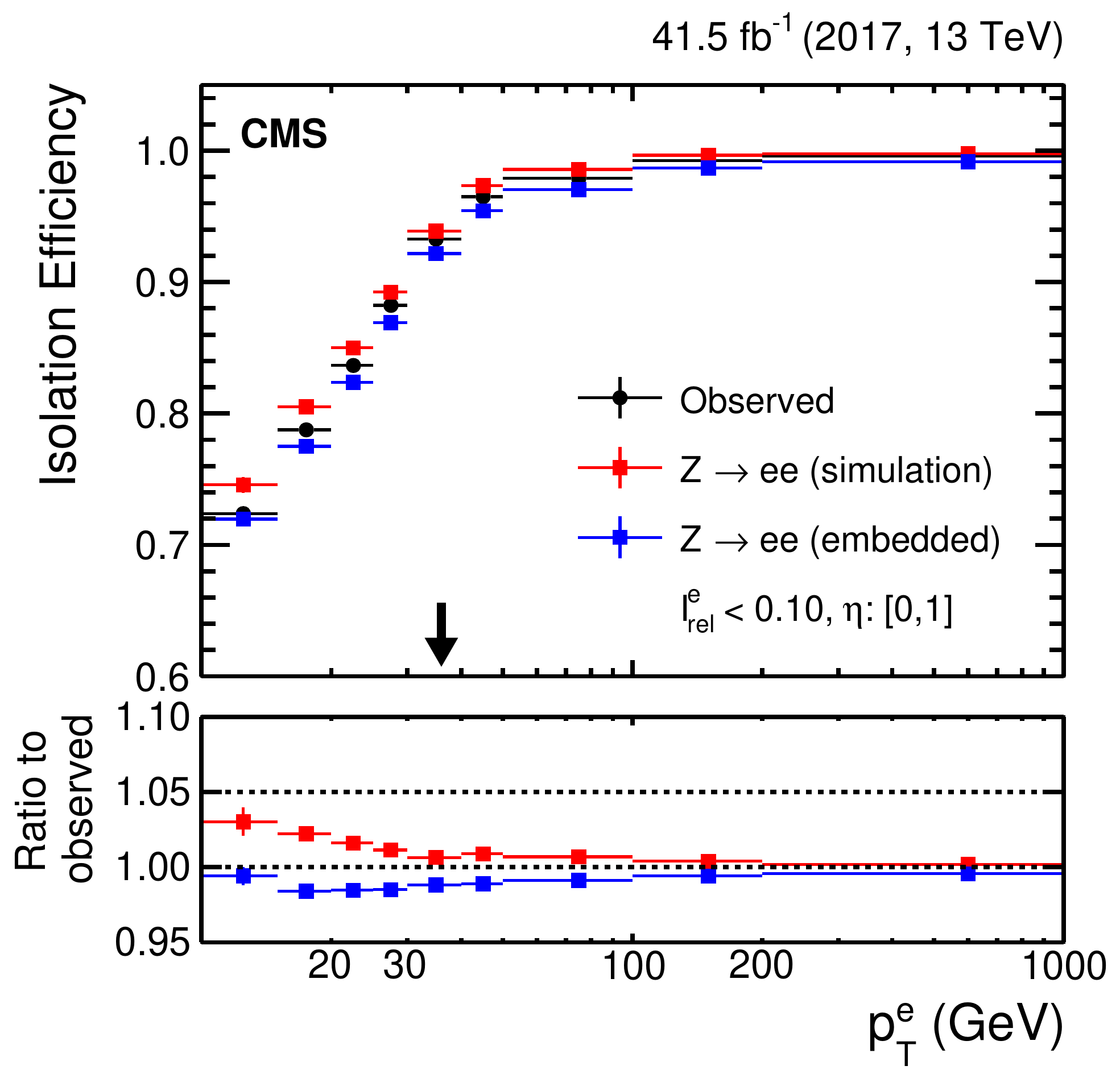}
		\caption {
			(Left column) muon and (right column) electron (upper row) identification and (lower row) isolation
			efficiencies as a function of the $\pt$ of the corresponding lepton in the central region of the
			detector. The black arrows indicate typical trigger thresholds of the target analyses. In the upper
			panel of each subfigure, the black dots correspond to the efficiencies obtained in data,
			the blue dots to the efficiencies obtained in the corresponding embedded event sample,
			and the red dots to the efficiencies obtained from the simulation. The lower panels show the
			ratios of the (blue) embedded event sample and (red) simulation, to the efficiency observed in data,
			which corresponds to the correction factors.
		}
		\label{fig:embedding-scale-factors}
	\end{figure}

        For the identification efficiency of the $\Pgth$ candidate, a global correction factor of $0.97\pm0.02$ 
        is obtained from a likelihood fit to the yield of $\ZTT$ events in the $\mutau$ final state in a control 
        region. Figure~\ref{fig:embedding-scale-factors} shows typical correction factors
	for the electron and muon identification and isolation efficiencies in the central region of the detector,
	as function of the $\pt$ of the corresponding lepton. Clear turn-on curves are visible for the muon
	isolation and the electron identification and isolation efficiencies. In each case, a plateau is
	reached for each efficiency above a $\pt$ threshold of about 30\GeV, which is close to the
	80\% efficiency working point discussed in Section~\ref{sec:reconstruction} for the electron
	identification, and close to unity otherwise. In general, the correction factors differ from
	the efficiencies observed in data by less than 5\% in the relevant kinematic regions, and
	they are smaller for the embedded event samples than for the simulated ones.
	
	\subsection{Uncertainties}
	\label{sec:uncertainties}
	
	When applied to the target analyses, the following uncertainties, which are specific to the method,
	should be applied:
	
	\begin{itemize}
		\item
		For the normalization of the $\Pgt$-embedded event samples, a global uncertainty of 2\% should
		be assumed due to the insufficient knowledge of the unfolding corrections of the initially
		selected muons, as described in Section~\ref{sec:Zmm-treatment}. The 2\% is chosen in accordance
                to the usual uncertainty in trigger leg efficiencies. This uncertainty should be applied
		per muon, resulting in an overall uncertainty in the normalization of 4\%.
		\item
		For the simulated leptons, a variation of 1.2\% in the $\Pgth$ energy scale, split by decay mode,
		as described in Ref.~\cite{Sirunyan:2018qio}, should be applied; a variation in the electron
		energy scale of 1\% in the central detector and 2.5\% in the endcaps of the ECAL should be applied.
		\item
		The uncertainty in the expected fraction of $\ttbar(\Pgt\Pgt)$ events in the embedded event samples
		is estimated from a 10\% up and down variation of the expected fraction in simulation. The
                estimate is based on a study in a $\ttbar$-enriched control region. It includes the uncertainty 
                in the number of $\ttbar$ events that do not contain muons in the final state (as given in 
                Table~\ref{tab:event-composition}) and a general uncertainty in the $\ttbar$ event yield in the
                selected kinematic regime.
		\item
		The uncertainties in the correction factors for the trigger leg, identification, and isolation
		efficiencies are usually of the order of 2\% in the kinematic regions relevant for the target
		analyses, which include the uncertainty in the removal of the energy deposits of the selected 
                muon that primarily affects the isolation efficiency for muons.
		\item
		The effects of the finite angular and $\pt$ resolutions of the detector are checked and have
		negligible influence on the $\Pgt$-embedded events. They are covered by the variation in the
		$\Pgth$ and electron energy scale given above. This is also true for a variation of $\ptmiss$ within
		the observed discrepancies visible in Fig.~\ref{fig:mm-embedding-validation} (middle left).
	\end{itemize}
	
	These uncertainties are usually a part of a more complex uncertainty model such as described in
	Ref.~\cite{Sirunyan:2018qio}. For the simulated processes that are replaced by the $\Pgt$-embedded
	event sample, they replace uncertainties in the integrated luminosity, production cross sections, jet
        energy scale,
	$\ptmiss$ scale and resolution, as well as in the tagging and mistag rates of \PQb jets.
	
	\subsection{Comparison to data}
	\label{sec:performance}
	
	To demonstrate how the embedding technique can help in a physics analysis on data, an inclusive event
	selection is performed for the $\Pgt\Pgt$ final states following typical selection requirements, as detailed
        in Ref.~\cite{Sirunyan:2018qio}.
	
	The online selection for the $\emu$ final state relies on a logical \emph{or} of two lower-threshold
	triggers that both require the presence of an electron and a muon in the event with $\pt>23\GeV$ for
	the higher-$\pt$ lepton and $\pt>12\,(8)\GeV$ for the lower-$\pt$ electron (muon).
	
	In the offline selection of the $\emu$ final state, an electron with $\pt>13\GeV$ and $\abs{\eta}<
	2.5$ and a muon with $\pt>9\GeV$ and $\abs{\eta}<2.4$ are required. If the event passed only one
	trigger, the lepton identified with the higher-$\pt$ trigger object is required to have a $\pt>24
	\GeV$, which guarantees a trigger acceptance well above the turn-on of at least one of the triggers
	used. Both leptons are required to pass identification criteria and to be isolated according to
        $\Irelem<0.15\,(0.20)$. Events with
	additional electrons or muons fulfilling looser selection requirements than these are rejected.
	
	The $\etau$ ($\mutau$) final state is based on the presence of at least one electron (muon) with  $\pt>35\,(27)
	\GeV$ and $\abs{\eta}<2.1$ at the trigger level.
	In these final states, an electron (muon) with $\pt>36\,(28)\GeV$ and $\abs{
		\eta}<2.1$ and a $\Pgth$ candidate with $\pt>30\GeV$ and $\abs{\eta}<2.3$ are required. The electron
	(muon) and the $\Pgth$ candidate must fulfill the identification requirements described in
	Section~\ref{sec:reconstruction}. The $\Pgth$ candidate is required to pass the \text{tight} working point
	of the $\Pgth$ identification discriminant, the \text{tight} (\text{very loose}) working point of
	the discriminant to suppress electrons and the \text{loose} (\text{tight}) working point of the
	discriminant to suppress muons in the $\etau$ ($\mutau$) case. In addition, the electron (muon)
	is required to be isolated, according to $\Irelem<0.10\,(0.15)$. Events with additional electrons or muons
	fulfilling looser selection requirements are rejected.

	In the $\tautau$ final state, a trigger decision
	based on the presence of two hadronically decaying tau leptons with $\pt>35\GeV$ and $\abs{\eta}
	<2.1$ is used. Furthermore, two $\Pgth$ candidates with $\pt>40\GeV$ and $\abs{\eta}<2.1$ are
	required. Both must pass the \text{tight} working point of the $\Pgth$ identification discriminant,
	the very loose working point of the discriminant against electrons and the \text{loose} working point
	of the discriminant against muons. Events with additional electrons or muons fulfilling looser
	requirements on identification, isolation, and \pt than described for the $\etau$ or $\mutau$ final
	state above are rejected.
	
	In all cases, the decay products of the two tau leptons are required to be oppositely charged,
	separated by more than 0.5 units in $\Delta R$, and associated with the PV within a distance of 0.045\unit{cm}
	in the transverse plane for electrons and muons and 0.2\unit{cm} along the beam axis for all
	final-state particles. The vetoing of additional electrons or muons ensures that no event is
	used for more than one $\Pgt\Pgt$ final state. At most 0.8\% of the selected events contain more
	$\Pgth$ candidates than required for the corresponding final state. In this case, the $\Pgt\Pgt$
	pair with the most isolated final state products is chosen. In the $\etau$ and $\mutau$ final state,
	the events are further selected according to the transverse mass,
	\begin{linenomath}
		\begin{equation}
		\label{eqn:mt_definition}
		\mTem = \sqrt{2\,\ptem\,\ptmiss\left(1 - \cos\Delta\phi\right)},
		\end{equation}
	\end{linenomath}
	where $\ptem$ refers to the $\pt$ of the electron (muon) and $\Delta\phi$ to the difference in the
	azimuthal angle between the electron (muon) momentum and $\ptvecmiss$. In the $\emu$ final state the events
	are further selected according to the event variable
	\begin{linenomath}
		\begin{equation}
		\label{eqn:Dzeta}
		\begin{split}
		\Dzeta &= p_{\zeta}^\text{miss} - 0.85\,p_{\zeta}^\text{vis} ; \qquad
		p_{\zeta}^\text{miss} = \ptvecmiss\cdot\hat{\zeta} ; \qquad
		p_{\zeta}^\text{vis} = \left(\vec{p}_{\text{T}}^{\kern1pt\Pe} + \vec{p}_{\text{T}}^{\kern1pt\Pgm}\right)
		\cdot\hat{\zeta}, \\
		\end{split}
		\end{equation}
	\end{linenomath}
	where $\vec{p}_{\text{T}}^{\,\Pe(\Pgm)}$ corresponds to the transverse momentum vector of the electron
	(muon) and $\hat{\zeta}$ to the bisectional direction between the electron and the muon momenta in the transverse
	plane~\cite{Abulencia:2005kq}. Events with $\mTem<40\GeV$ and $-10<\Dzeta<30\GeV$ are used for further
	consideration in each corresponding final state. Both $\mTem$ and $\Dzeta$ quantify the size of
	$\ptmiss$ and how aligned it is with the momenta of the selected leptons. They are typical
	event variables to distinguish genuine $\Pgt\Pgt$ events from $\Wjets$ and $\ttbar$ events.
	
	\begin{figure}[t]
		\centering
		\includegraphics[width=0.45\textwidth]{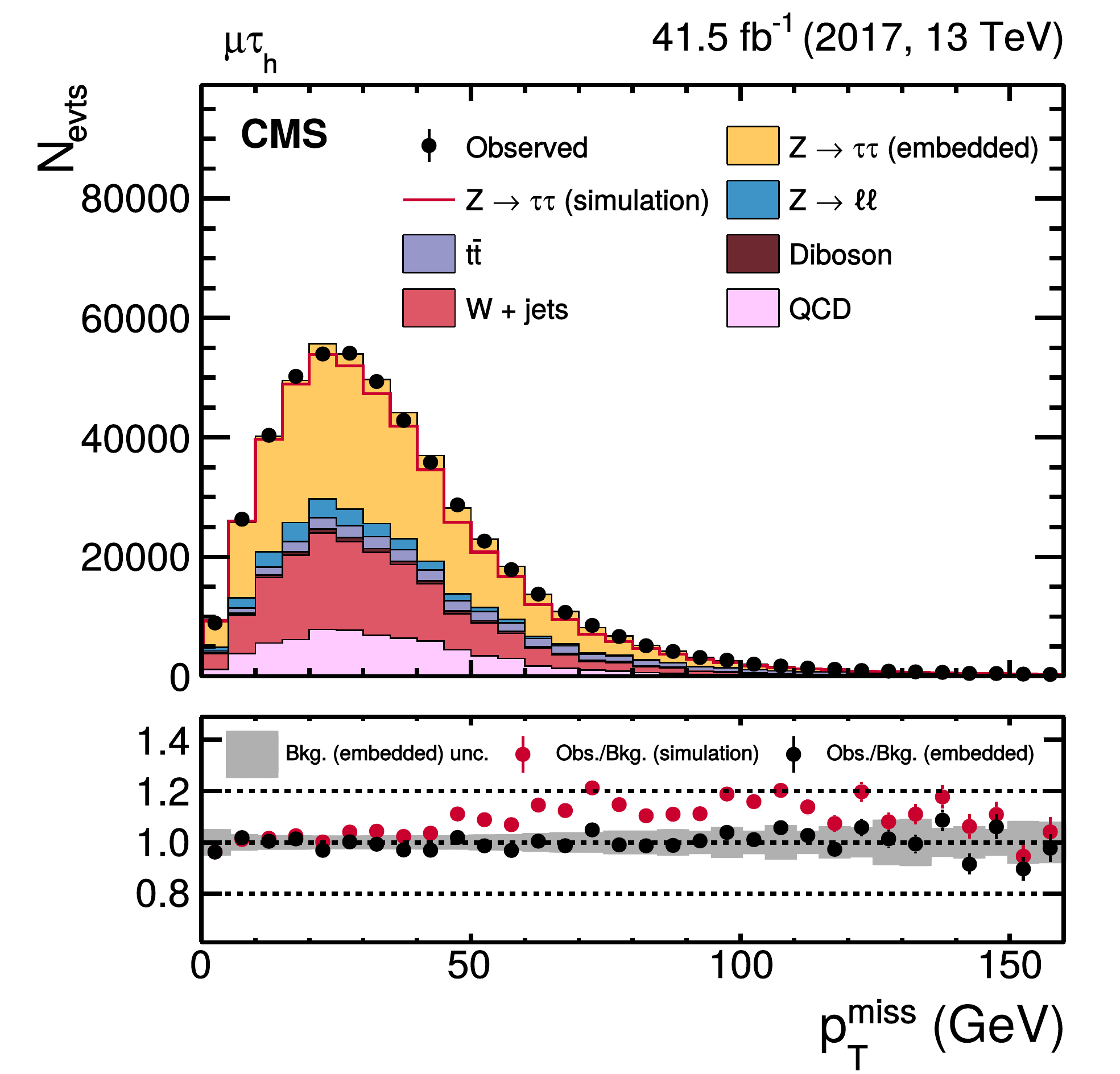}
		\includegraphics[width=0.45\textwidth]{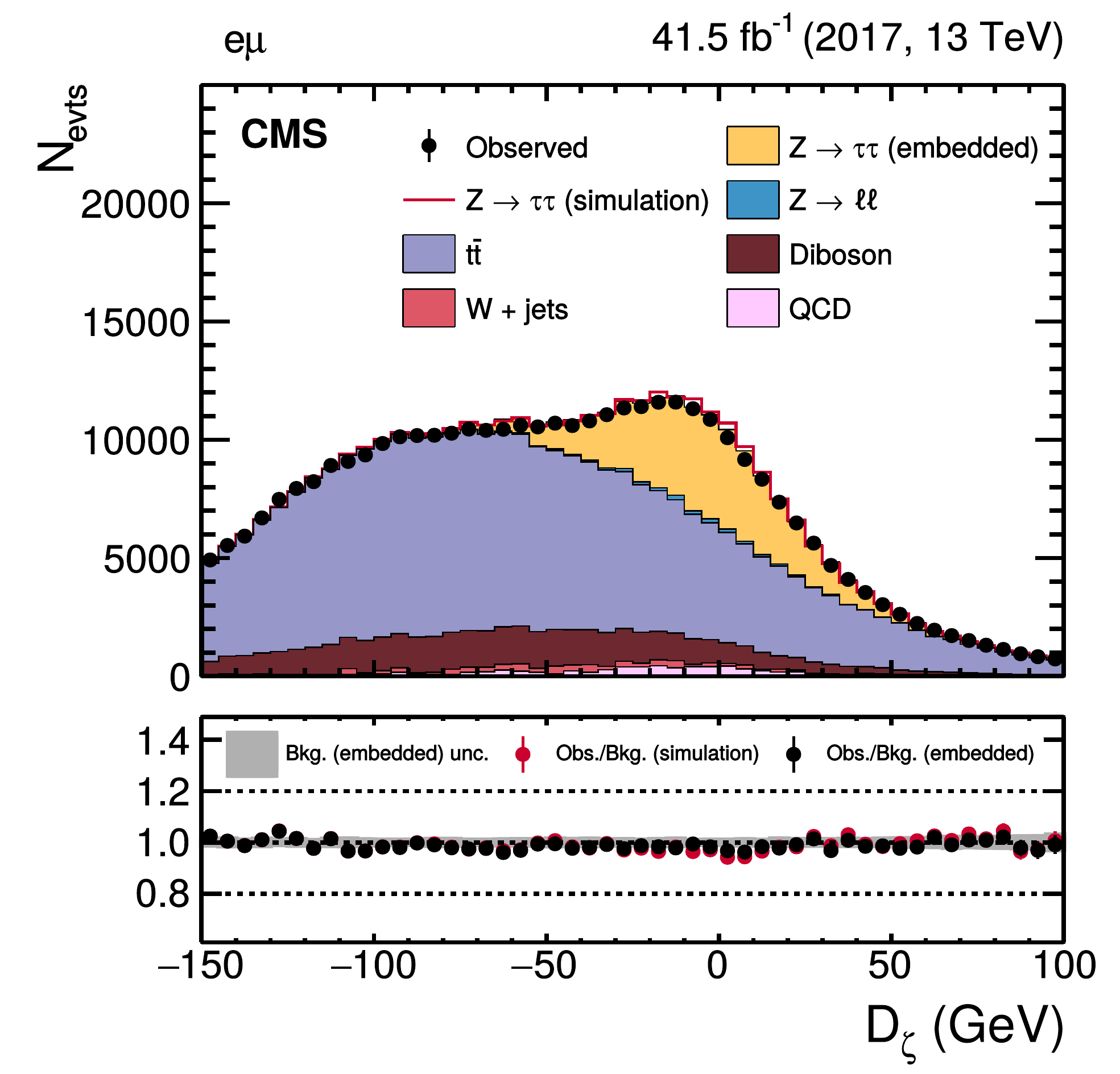}
		\includegraphics[width=0.45\textwidth]{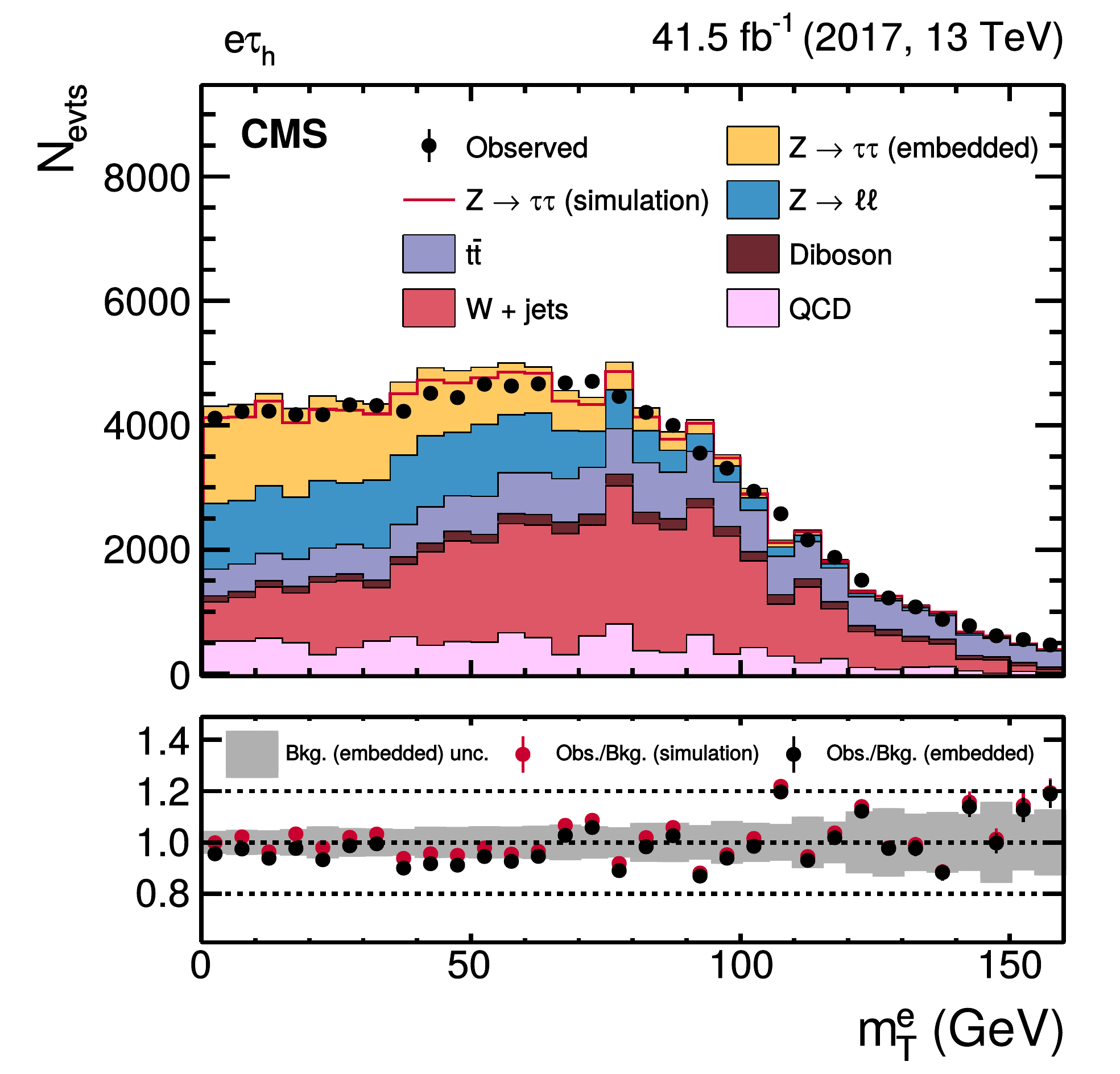}
		\includegraphics[width=0.45\textwidth]{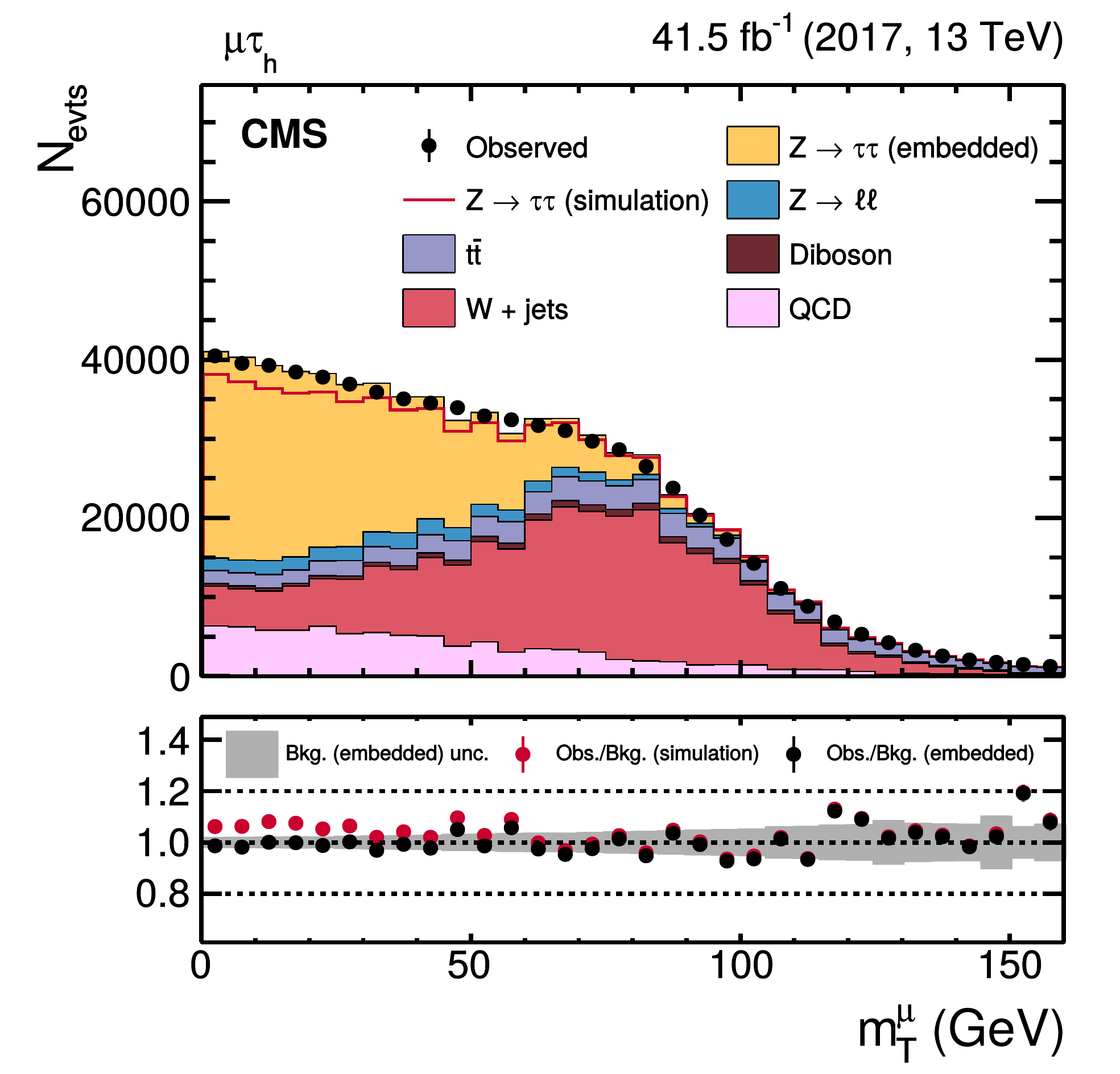}
		\caption {
			Distributions of (upper left) $\ptmiss$ in the $\mutau$ final state, (upper right) $\Dzeta$ in
			the $\emu$ final state, (lower left) $\mTe$ in the $\etau$ final state, and (lower right) $\mTm$
			in the $\mutau$ final state. The distributions are shown prior to the maximum likelihood fit
			described in the text. For these figures, no uncertainties that affect the shape of the
			distributions have been included in the uncertainty model. The background estimation purely from
			the CMS simulation is shown as an additional red line.
		}
		\label{fig:embedding-data-comparison_1}
	\end{figure}
	
	In Fig.\ref{fig:embedding-data-comparison_1}, the distributions of $\ptmiss$, $\Dzeta$, $\mTe$, and
	$\mTm$ are shown. In addition to the expectation using the $\Pgt$-embedded event samples, the overall
	expectation when using the simulation of $\ZTT$, $\ttbar(\Pgt\Pgt)$, and diboson($\Pgt\Pgt$) events is
        shown by an open
	histogram in the upper panel of the subfigures. For this comparison a series of corrections have been
	applied to the simulation, including a correction to match the pileup distribution in data, a reweighting
        of the $\PZ$ boson $\pt$ distribution of the LO
	simulation to that in $\ZMM$ events observed in data, corrections for the electron and muon legs of the
        corresponding
	trigger paths, and for the electron and muon identification and isolation, and corrections of the $\PZ$
	boson recoil, to mitigate differences in detector resolution, between the simulation and data, for
	the calculation of $\ptmiss$. For $\Pgt$-embedded events the corrections related to simulated
	leptons are applied, discussed in Section~\ref{sec:correction-factors}. A generally good
	agreement between the expectation and the data is observed, within the applied uncertainty model. A better
	agreement is found when using the $\Pgt$-embedded event samples instead of the simulation. Fluctuations
	in the distributions of $\mTe$ and $\mTm$ originate mostly from the limited size of the sample of
	simulated $\Wjets$ events. In the target analyses, a large fraction of $\Wjets$ and QCD multijet
	events are usually estimated from data, which implies that a fraction of up to 90\% of the typical
	background expectation for the target analyses can be estimated from data.
	
	\begin{figure}[b]
		\centering
		\includegraphics[width=0.45\textwidth]{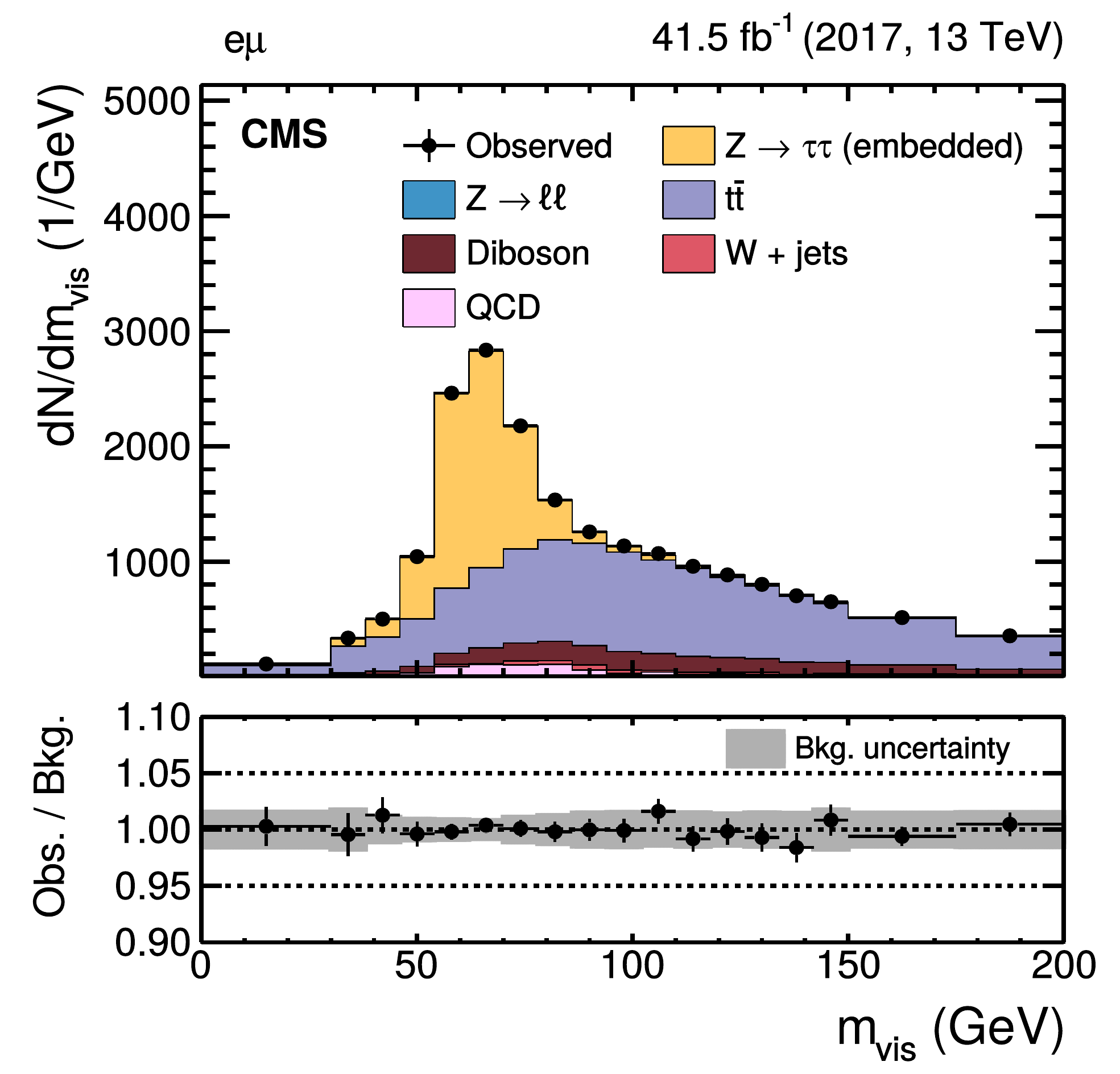}
		\includegraphics[width=0.45\textwidth]{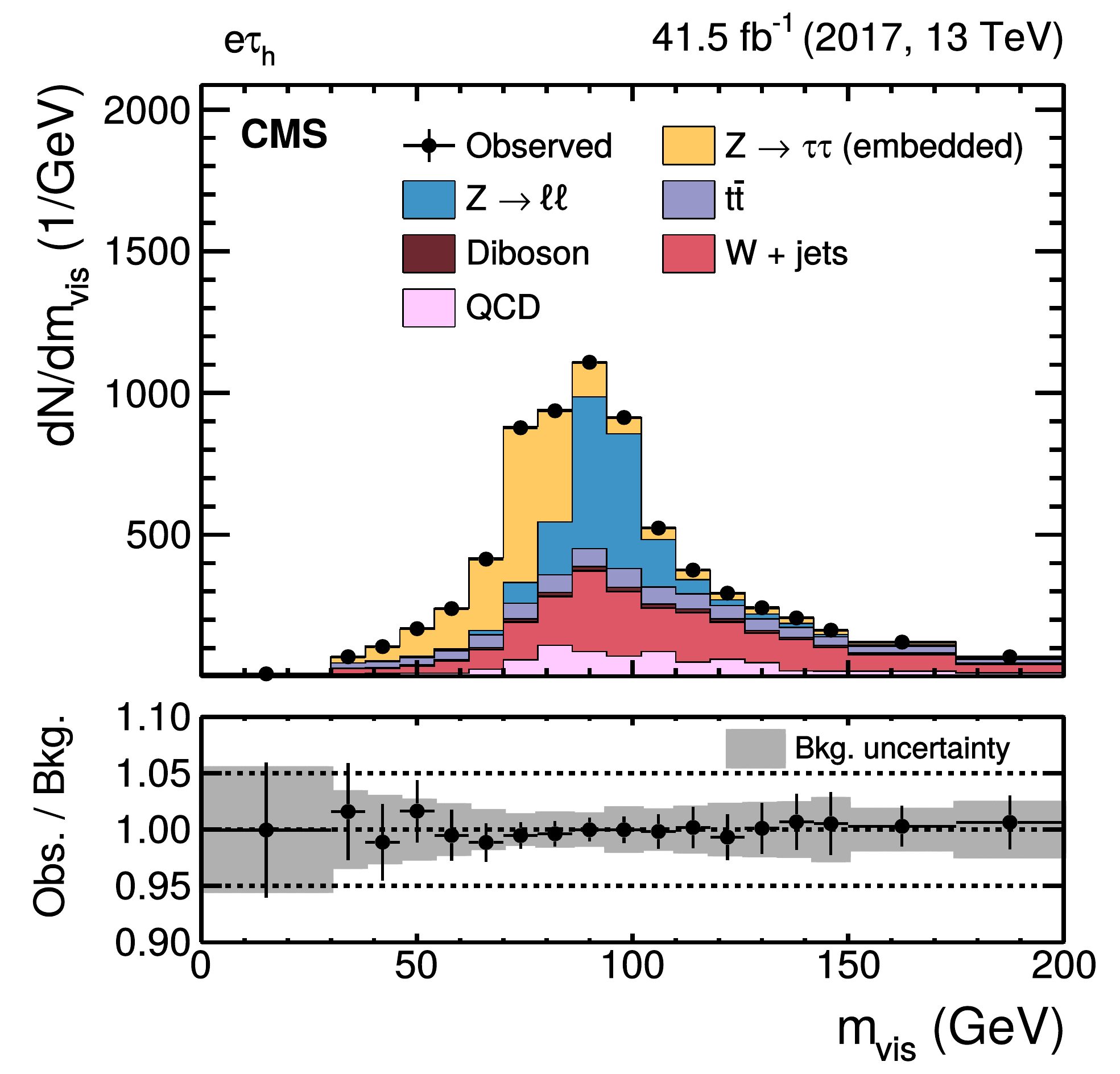}
		\includegraphics[width=0.45\textwidth]{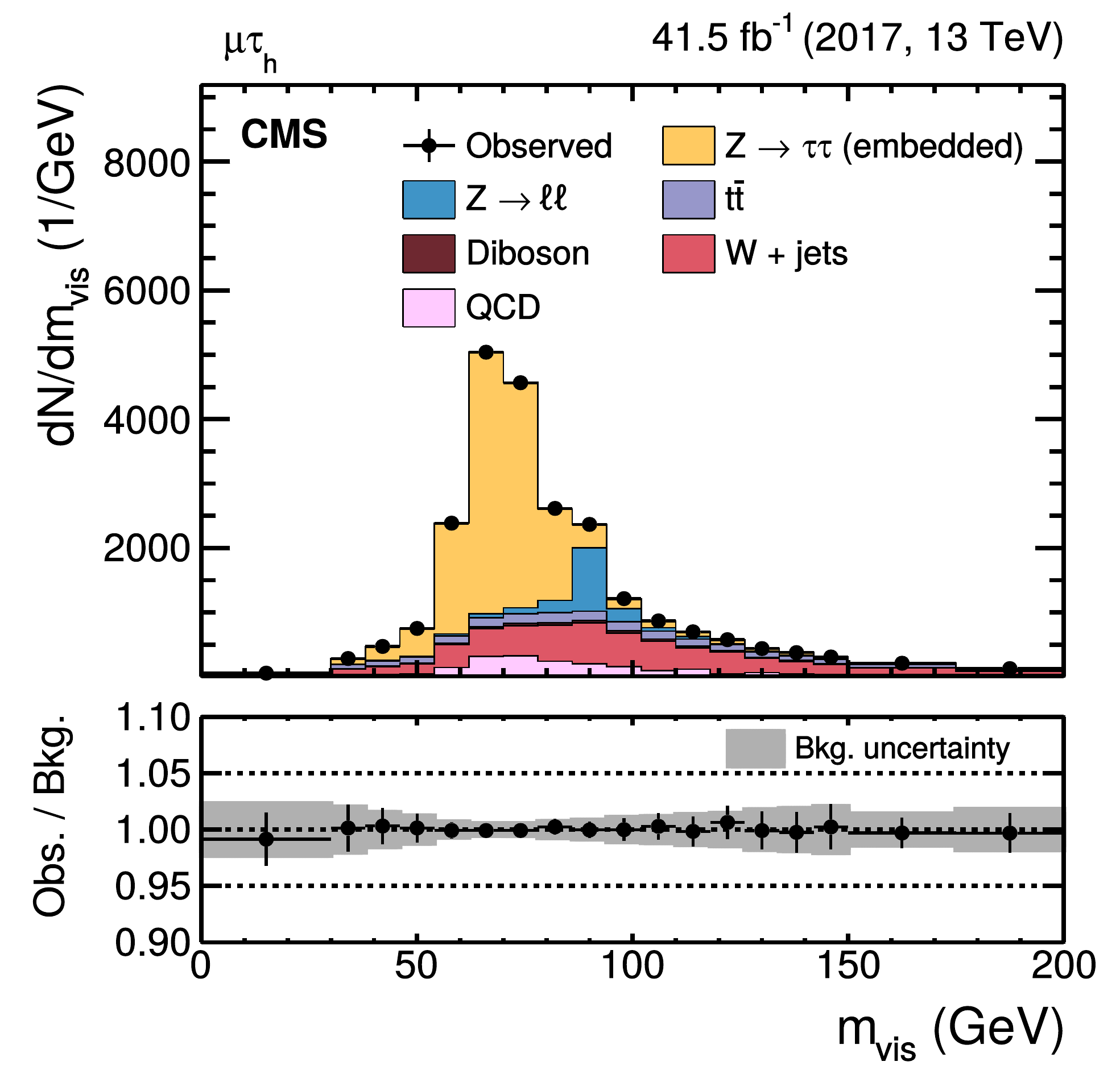}
		\includegraphics[width=0.45\textwidth]{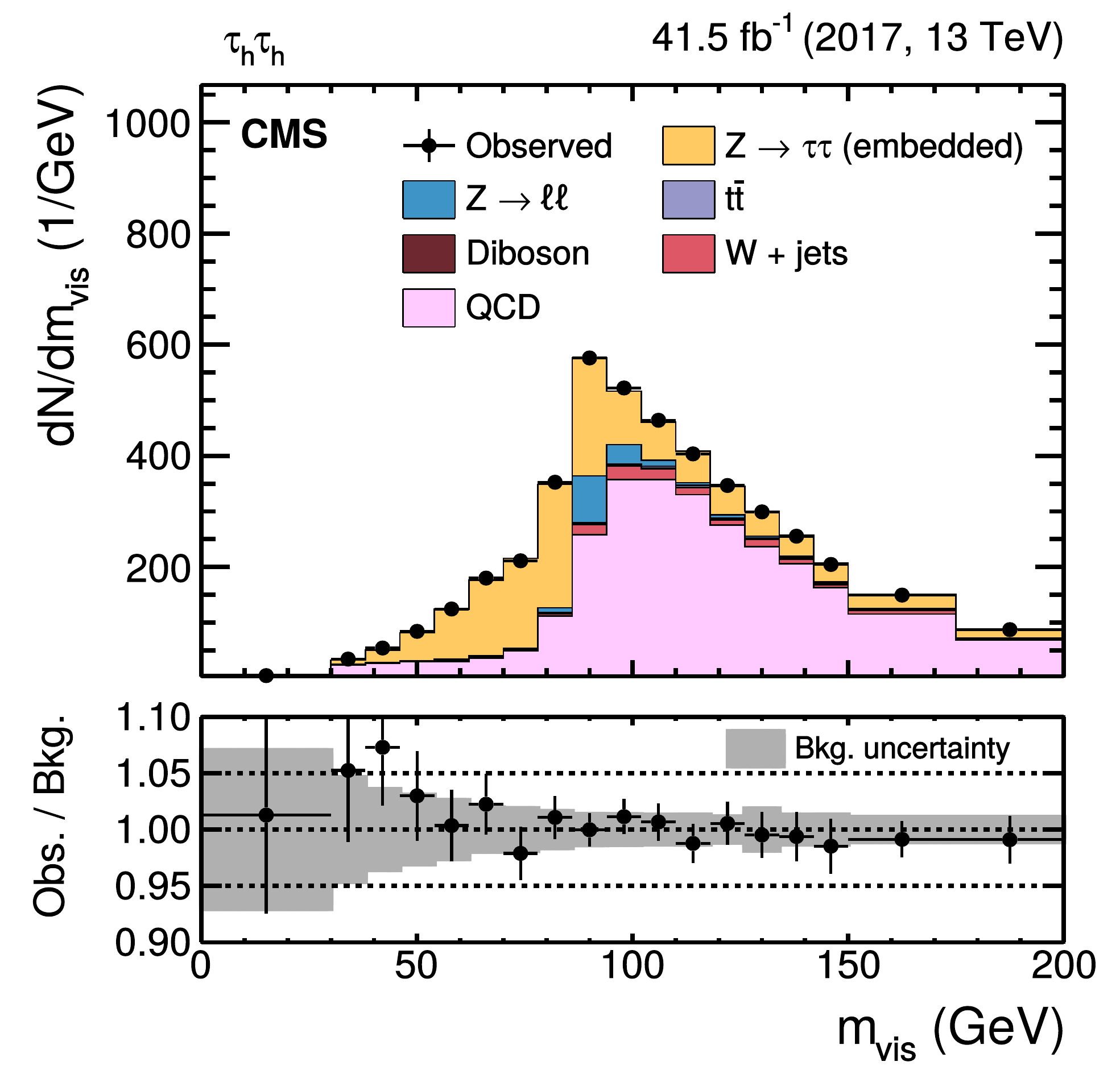}
		\caption {
			Invariant mass distribution of the visible $\Pgt\Pgt$ decay products, $\mvis$, in the (upper left)
			$\emu$, (upper right) $\etau$, (lower left) $\mutau$, and (lower right) $\tautau$  final states,
			after a fit to the data exploiting a typical uncertainty model as discussed
			in Ref.~\cite{Sirunyan:2018qio}. In the $\etau$ final state a significantly larger contribution 
                        of $\PZ\to\ell\ell$ events is visible compared to the $\mutau$ final state. The reason for this 
                        is that high-$\pt$ electrons have a higher probability to be misidentified as $\Pgth$ decays 
                        than muons.
		}
		\label{fig:embedding-data-comparison_2}
	\end{figure}
	
	In the target analyses the distributions of a variable related to the invariant mass of the $\Pgt\Pgt$
	system are usually used as input for a maximum likelihood fit to extract the actual signal. This
	signal can be an SM process, such as the SM $\PZ$ or Higgs boson production in the $\Pgt\Pgt$ final state,
	or any other process of physics beyond the SM. In Fig.~\ref{fig:embedding-data-comparison_2} the distributions
	of $\mvis$, as such a variable, in the $\emu$, $\etau$, $\mutau$, and $\tautau$ final states are shown,
	after the event selections, as described above, and after applying a maximum likelihood fit to the
	observation with the $\Pgt$-embedded event sample as signal. For this purpose, a likelihood model
	has been adapted from Ref.~\cite{Sirunyan:2018qio}. It incorporates $\mathcal{O}(100)$ uncertainties
	in form of nuisance parameters that may be correlated across the processes contributing to the input
	distributions, and across final states. Within a single input distribution the nuisance parameters
	may allow for correlated shifts across bins, such as process normalization or energy scale uncertainties,
	and for shifts of individual bins, within the statistical precision of the template distributions used
	in the model. The ability of the model to describe the data can be quantified using a goodness-of-fit
	test, based on a saturated likelihood model (SAT) described in Ref.~\cite{Baker:1983tu}, which
	corresponds to a generalization of a $\chi^{2}$ test including all systematic uncertainties of the
	model and their correlations. The SAT test indicates the overall statistical compatibility of the
	model with the observation, treating each bin of the input distributions as an independent measurement.
	Goodness-of-fit tests based on the empirical distribution function are usually more sensitive than a
	$\chi^{2}$-like test to small deviations that are correlated across several bins of a single histogram.
	A classical test of this kind that is mostly sensitive to deviations correlated across bins in the
	center of a given binned distribution is the Kolmogorov--Smirnov (KS) test~\cite{KolmogorovAN1933,smirnov1948}.
	A variant of this test that gives more emphasis to the edges of the given input distribution is the
	Anderson--Darling (AD) test~\cite{AndersonDarling}. The $p$-values for each of these tests, split by final
	state, are shown in Table~\ref{tab:p-values}. They have a one-to-one correspondence to the distributions
	shown in Fig.~\ref{fig:embedding-data-comparison_2}, with the small difference that these distributions
	are shown for a fit to the observation in all final states combined. The $p$-values are obtained
	from the comparison of the observed value for the corresponding test statistic with the outcome of
	pseudo-experiments based on the expectation. Their statistical precision is better than 0.5\%. The
	actual values range from 17\%, for the $p$-value of the AD test in the $\etau$ final state, to 82\%, for
	the $p$-value of the AD test in the $\mutau$ final state. All tests reveal good compatibility of the
	statistical model with the observation, which implies a successful description of the data with the given
	template distributions, especially with the $\Pgt$-embedded event samples. The fit to the observation
	in all final states combined reveals a $p$-value of 51\% and a normalization of $1.00\pm^{0.05}_{0.05
	}$ for the $\Pgt$-embedded event samples, which is in good agreement with the observations of
	Ref.~\cite{Sirunyan:2018qio} that have been made on an independent data set. Also a good compatibility
	of the normalization across all final states is observed. The normalization of the $\Pgt$-embedded
        samples is obtained from the data. Figures showing distributions of more quantities relevant for the
        analysis of $\Pgt\Pgt$ events are given in Appendix~\ref{app:performance}.
	
	\begin{table}[htbp]
		\topcaption{
			Normalization of the $\Pgt$-embedded event samples and $p$-values of the saturated model (SAT),
			Kolmogorov--Smirnov (KS) and Anderson--Darling (AD) test, as discussed in the text, separated by
			$\Pgt\Pgt$ final state, as introduced in Section~\ref{sec:embedding} and (where applicable)
			for all channels combined. The $p$-values have a statistical precision better than 0.5\%.
		}
		\label{tab:p-values}
		\centering
		\begin{tabular}{lcccc}
			& & \multicolumn{3}{c}{$p$-values} \\
			Final state
			& Normalization & SAT & KS & AD \\
			\hline
			$\hspace{0.3cm}\emu$     & $1.02\pm^{0.05}_{0.05}$ & $0.61$ & $0.29$  & $0.74$ \\[\cmsTabSkip]
			$\hspace{0.3cm}\etau$    & $0.87\pm^{0.08}_{0.07}$ & $0.69$ & $0.35$  & $0.17$ \\[\cmsTabSkip]
			$\hspace{0.25cm}\mutau$  & $0.96\pm^{0.07}_{0.06}$ & $0.76$ & $0.81$  & $0.82$ \\[\cmsTabSkip]
			$\hspace{0.20cm}\tautau$ & $1.10\pm^{0.12}_{0.11}$ & $0.71$ & $0.54$  & $0.30$ \\[\cmsTabSkip]
			$\hspace{0.20cm}$Combined& $1.00\pm^{0.05}_{0.05}$ & $0.51$ & \NA 	  & \NA	   \\
			\hline
		\end{tabular}
	\end{table}
	
\section{Summary}
\label{sec:summary}
	
The $\Pgt$-embedding technique developed for the CMS experiment is described and its validation and
relevant uncertainties are discussed. The 13\TeV proton-proton collisions collected by CMS in 2017
are used to demonstrate the performance of the technique with the data sample corresponding to an
integrated luminosity of 41.5\fbinv.

The main goal of the procedure is to estimate the background from $\ZTT$ events using recorded $\ZMM$
events. The estimate also includes events from $\ttbar$ and diboson production with two tau leptons
in the final state. Recorded $\Pgm\Pgm$ events are selected, the muons are removed from the
reconstructed event record, and replaced with simulated tau leptons with the same kinematic properties
as the removed muons. In that way hybrid events are obtained, which rely on the simulation only for
the decay of the tau leptons. Challenges in describing the underlying event or the production of
associated jets in the simulation, as well as the costly simulation of PU events thus are avoided.
The embedding technique decreases the uncertainties inherent in a typical simulation process, such as
the uncertainties in the missing transverse momentum, jet energy scale and resolution, \PQb tagging
efficiency, and misidentification probability.

A number of validation tests for $\Pgm$-, $\Pe$- , and $\Pgt$-embedding, as well as several
goodness-of-fit tests, show good agreement of embedded distributions with those obtained
using simulated and recorded data events. The embedding technique avoids time-consuming simulations
of events that becomes critical for the planned High-Luminosity LHC upgrade, where typical pileup of
140--200 collisions per bunch crossing is expected.

\begin{acknowledgments}
		\hyphenation{Bundes-ministerium Forschungs-gemeinschaft Forschungs-zentren Rachada-pisek} We congratulate our colleagues in the CERN accelerator departments for the excellent performance of the LHC and thank the technical and administrative staffs at CERN and at other CMS institutes for their contributions to the success of the CMS effort. In addition, we gratefully acknowledge the computing centers and personnel of the Worldwide LHC Computing Grid for delivering so effectively the computing infrastructure essential to our analyses. Finally, we acknowledge the enduring support for the construction and operation of the LHC and the CMS detector provided by the following funding agencies: the Austrian Federal Ministry of Education, Science and Research and the Austrian Science Fund; the Belgian Fonds de la Recherche Scientifique, and Fonds voor Wetenschappelijk Onderzoek; the Brazilian Funding Agencies (CNPq, CAPES, FAPERJ, FAPERGS, and FAPESP); the Bulgarian Ministry of Education and Science; CERN; the Chinese Academy of Sciences, Ministry of Science and Technology, and National Natural Science Foundation of China; the Colombian Funding Agency (COLCIENCIAS); the Croatian Ministry of Science, Education and Sport, and the Croatian Science Foundation; the Research Promotion Foundation, Cyprus; the Secretariat for Higher Education, Science, Technology and Innovation, Ecuador; the Ministry of Education and Research, Estonian Research Council via IUT23-4 and IUT23-6 and European Regional Development Fund, Estonia; the Academy of Finland, Finnish Ministry of Education and Culture, and Helsinki Institute of Physics; the Institut National de Physique Nucl\'eaire et de Physique des Particules~/~CNRS, and Commissariat \`a l'\'Energie Atomique et aux \'Energies Alternatives~/~CEA, France; the Bundesministerium f\"ur Bildung und Forschung, Deutsche Forschungsgemeinschaft, and Helmholtz-Gemeinschaft Deutscher Forschungszentren, Germany; the General Secretariat for Research and Technology, Greece; the National Research, Development and Innovation Fund, Hungary; the Department of Atomic Energy and the Department of Science and Technology, India; the Institute for Studies in Theoretical Physics and Mathematics, Iran; the Science Foundation, Ireland; the Istituto Nazionale di Fisica Nucleare, Italy; the Ministry of Science, ICT and Future Planning, and National Research Foundation (NRF), Republic of Korea; the Ministry of Education and Science of the Republic of Latvia; the Lithuanian Academy of Sciences; the Ministry of Education, and University of Malaya (Malaysia); the Ministry of Science of Montenegro; the Mexican Funding Agencies (BUAP, CINVESTAV, CONACYT, LNS, SEP, and UASLP-FAI); the Ministry of Business, Innovation and Employment, New Zealand; the Pakistan Atomic Energy Commission; the Ministry of Science and Higher Education and the National Science Center, Poland; the Funda\c{c}\~ao para a Ci\^encia e a Tecnologia, Portugal; JINR, Dubna; the Ministry of Education and Science of the Russian Federation, the Federal Agency of Atomic Energy of the Russian Federation, Russian Academy of Sciences, the Russian Foundation for Basic Research, and the National Research Center ``Kurchatov Institute''; the Ministry of Education, Science and Technological Development of Serbia; the Secretar\'{\i}a de Estado de Investigaci\'on, Desarrollo e Innovaci\'on, Programa Consolider-Ingenio 2010, Plan Estatal de Investigaci\'on Cient\'{\i}fica y T\'ecnica y de Innovaci\'on 2013-2016, Plan de Ciencia, Tecnolog\'{i}a e Innovaci\'on 2013-2017 del Principado de Asturias, and Fondo Europeo de Desarrollo Regional, Spain; the Ministry of Science, Technology and Research, Sri Lanka; the Swiss Funding Agencies (ETH Board, ETH Zurich, PSI, SNF, UniZH, Canton Zurich, and SER); the Ministry of Science and Technology, Taipei; the Thailand Center of Excellence in Physics, the Institute for the Promotion of Teaching Science and Technology of Thailand, Special Task Force for Activating Research and the National Science and Technology Development Agency of Thailand; the Scientific and Technical Research Council of Turkey, and Turkish Atomic Energy Authority; the National Academy of Sciences of Ukraine, and State Fund for Fundamental Researches, Ukraine; the Science and Technology Facilities Council, UK; the US Department of Energy, and the US National Science Foundation.
		
		Individuals have received support from the Marie-Curie program and the European Research Council and Horizon 2020 Grant, contract No. 675440 (European Union); the Leventis Foundation; the A. P. Sloan Foundation; the Alexander von Humboldt Foundation; the Belgian Federal Science Policy Office; the Fonds pour la Formation \`a la Recherche dans l'Industrie et dans l'Agriculture (FRIA-Belgium); the Agentschap voor Innovatie door Wetenschap en Technologie (IWT-Belgium); the F.R.S.-FNRS and FWO (Belgium) under the ``Excellence of Science - EOS'' - be.h project n. 30820817; the Ministry of Education, Youth and Sports (MEYS) of the Czech Republic; the Lend\"ulet (``Momentum'') Program and the J\'anos Bolyai Research Scholarship of the Hungarian Academy of Sciences, the New National Excellence Program \'UNKP, the NKFIA research grants 123842, 123959, 124845, 124850 and 125105 (Hungary); the Council of Scientific and Industrial Research, India; the HOMING PLUS program of the Foundation for Polish Science, cofinanced from European Union, Regional Development Fund, the Mobility Plus program of the Ministry of Science and Higher Education, the National Science Center (Poland), contracts Harmonia 2014/14/M/ST2/00428, Opus 2014/13/B/ST2/02543, 2014/15/B/ST2/03998, and 2015/19/B/ST2/02861, Sonata-bis 2012/07/E/ST2/01406; the National Priorities Research Program by Qatar National Research Fund; the Programa de Excelencia Mar\'{i}a de Maeztu, and the Programa Severo Ochoa del Principado de Asturias; the Thalis and Aristeia programs cofinanced by EU-ESF, and the Greek NSRF; the Rachadapisek Sompot Fund for Postdoctoral Fellowship, Chulalongkorn University, and the Chulalongkorn Academic into Its 2nd Century Project Advancement Project (Thailand); the Welch Foundation, contract C-1845; and the Weston Havens Foundation (USA).
	\end{acknowledgments}
	
	\clearpage
	
	\bibliography{auto_generated}
	
	\clearpage
	
	\appendix
	
	\section{Performance of the \texorpdfstring{$\Pgt$}{tau}-embedding
		method on data}
	\label{app:performance}
	
	Distributions of more quantities relevant for the analysis of $\Pgt\Pgt$ events. The distributions
	are shown prior to the maximum likelihood fit discussed in Section~\ref{sec:performance}. In addition
	to the expectation using the $\Pgt$-embedded event samples, the overall expectation when using the
	simulation of $\ZTT$, $\ttbar(\Pgt\Pgt)$, and diboson($\Pgt\Pgt$) events is shown by a red line in
        the upper panel of the
	subfigures.  For this comparison a series of corrections have been applied as discussed in
        Section~\ref{sec:performance}. For $\Pgt$-embedded events the
	corrections related to the simulated leptons, discussed in Section~\ref{sec:correction-factors},
	have been applied. For these figures, no uncertainties that affect the shape of the distributions
        have been included in the uncertainty model.
	
	\begin{figure}[htbp]
		\centering

		\includegraphics[width=0.45\textwidth]{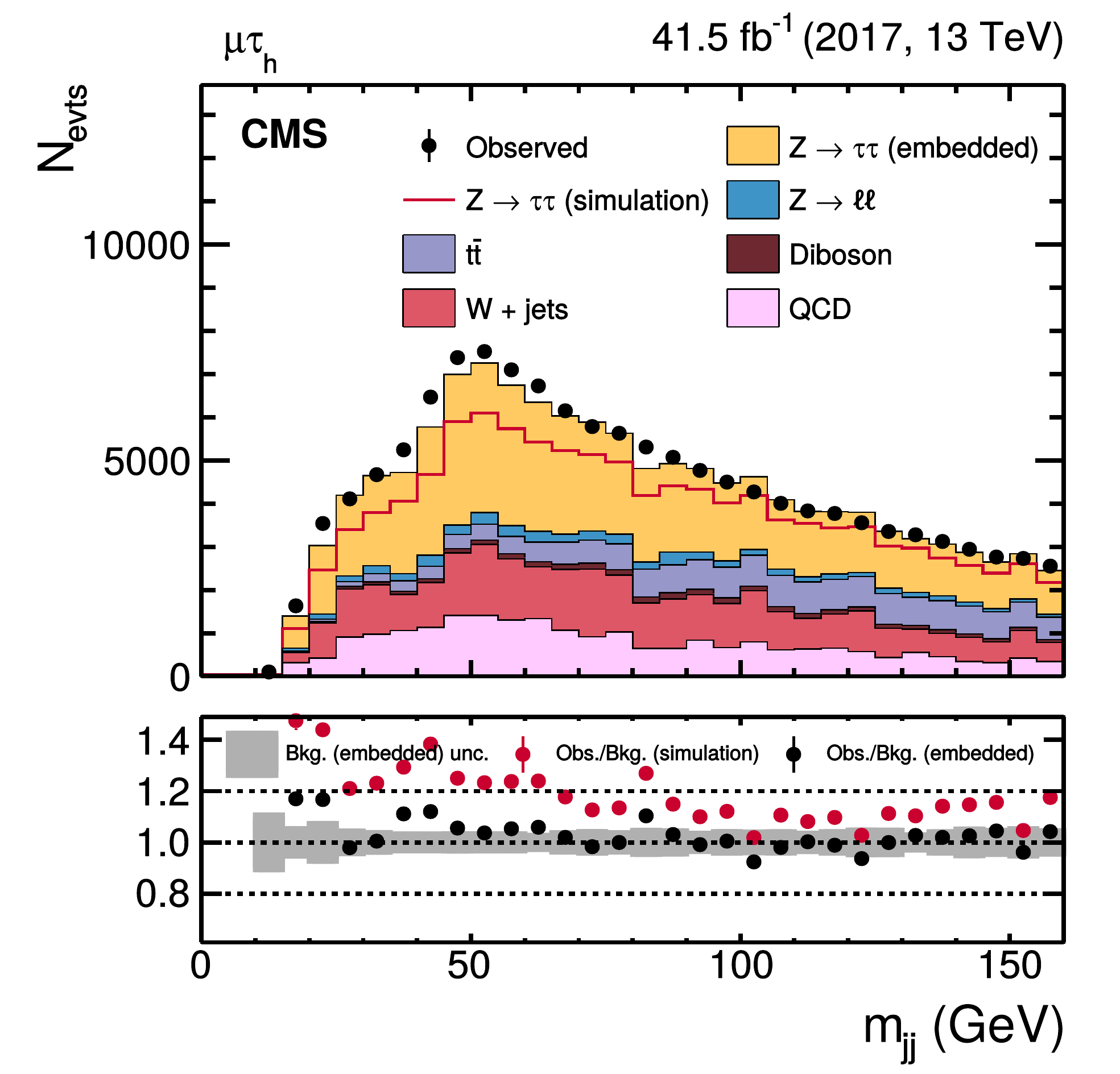}
		\includegraphics[width=0.45\textwidth]{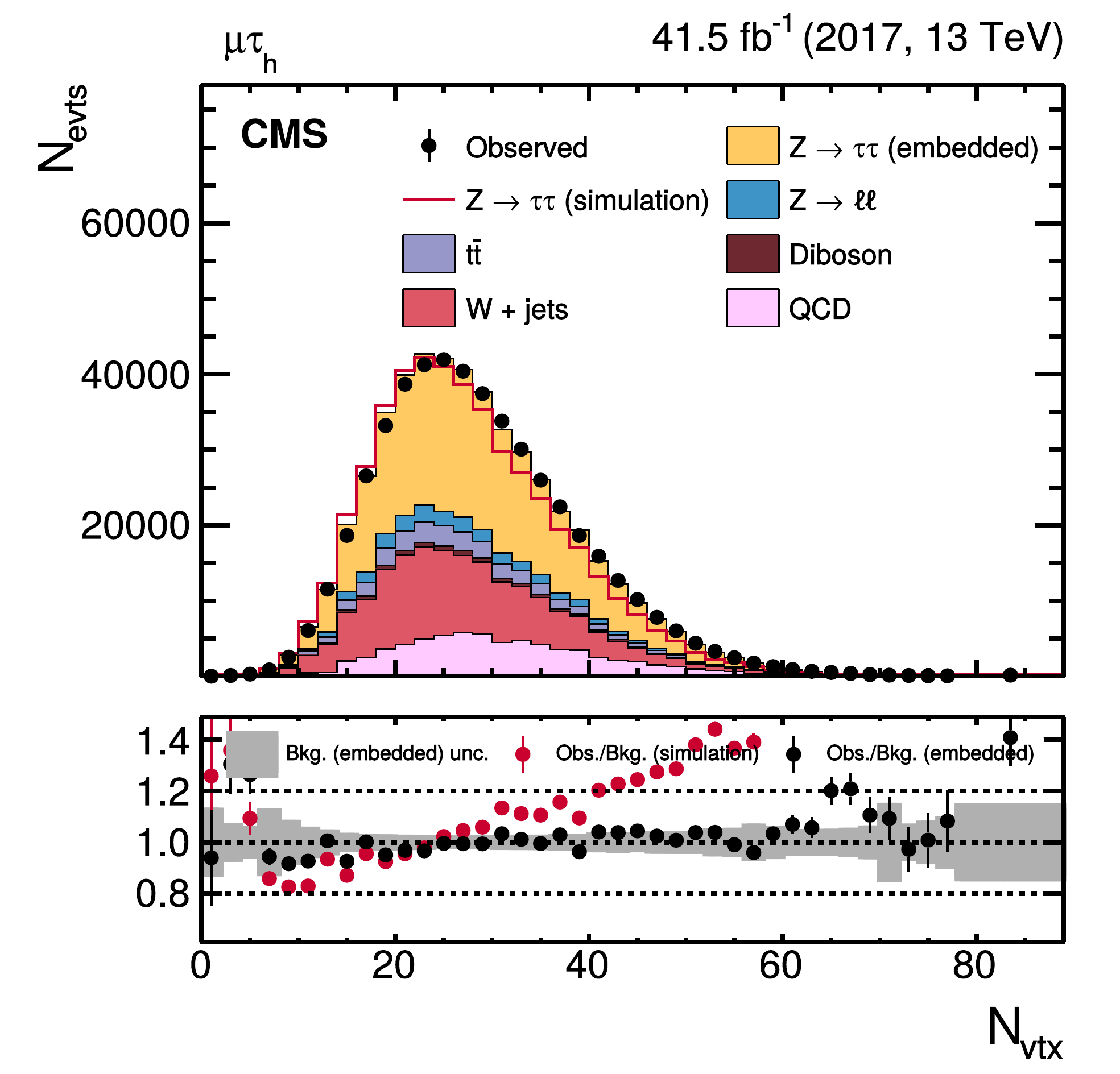}
		\caption {
			Distributions of (left) $\mjj$ and (right) the number of reconstructed primary vertices $N_{\text{vtx}}$
			in the $\mutau$ final state.
		}
		\label{fig:embedding-data-comparison_3}
	\end{figure}
	
	\begin{figure}[htbp]
		\centering
		\includegraphics[width=0.45\textwidth]{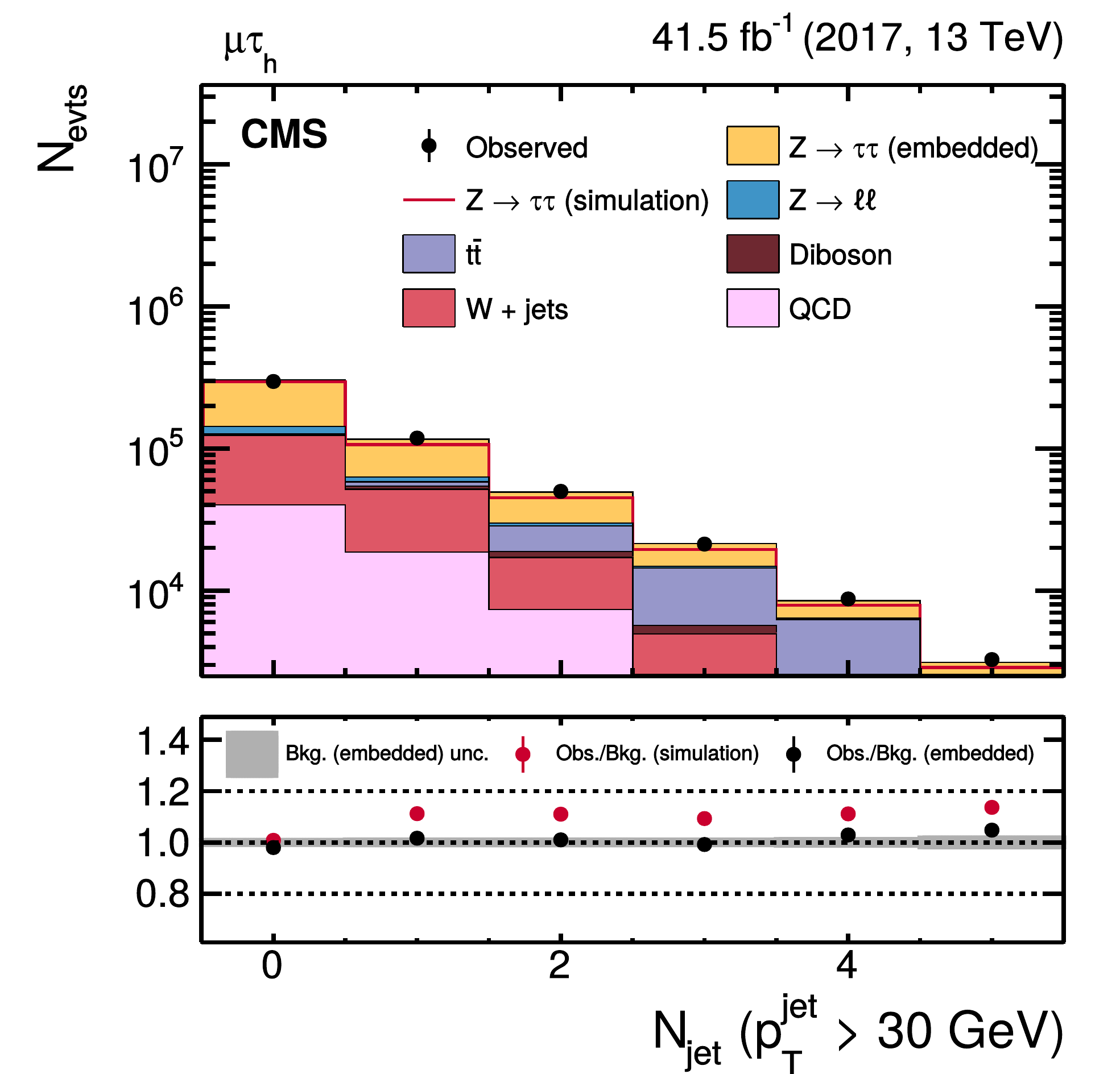}
		\includegraphics[width=0.45\textwidth]{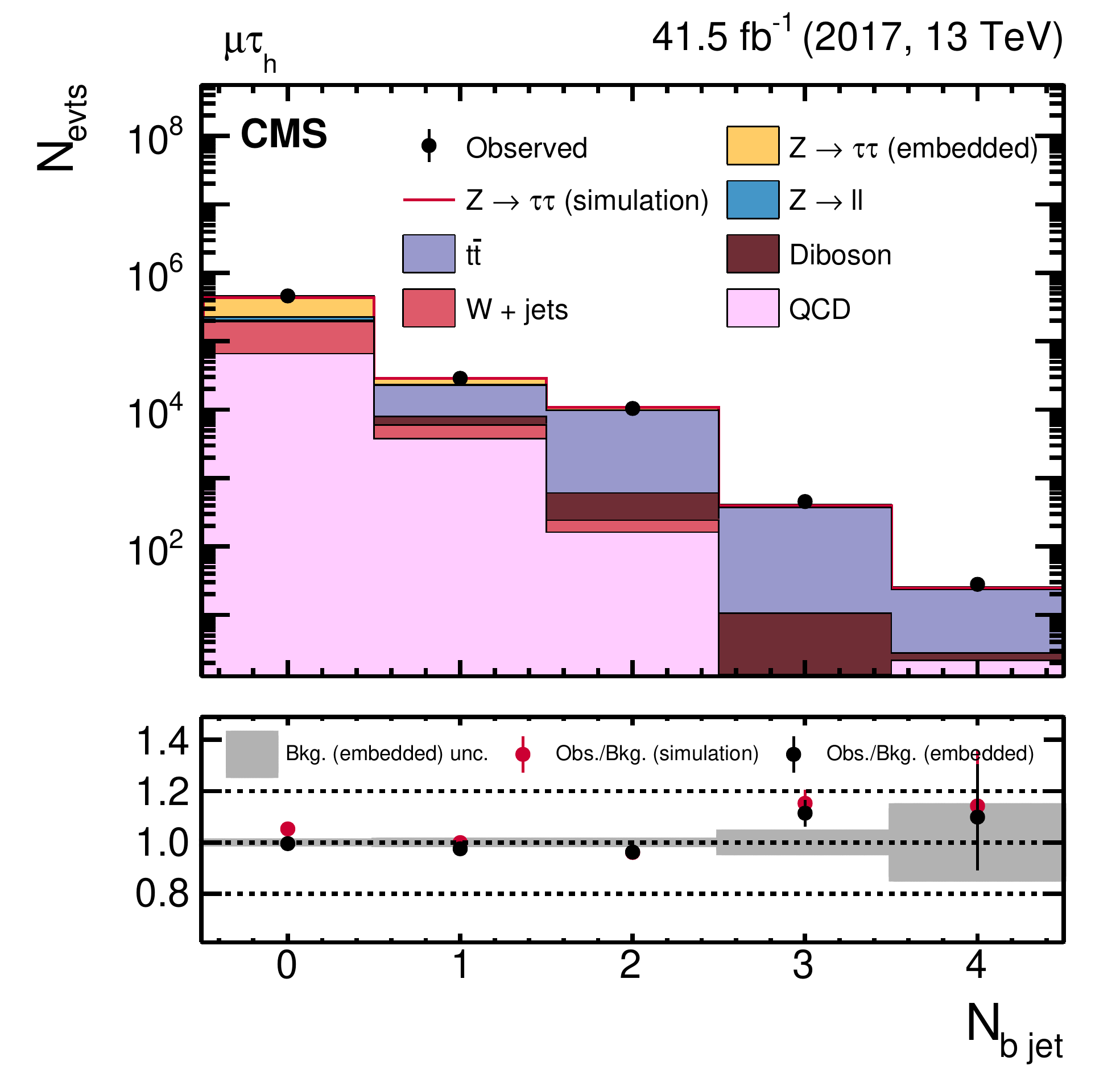}
		\caption {
			Distributions of the (left) jet and (right) \PQb jet multiplicity, as described in the text, in the
			$\mutau$ final state.
		}
		\label{fig:embedding-data-comparison_4}
	\end{figure}
	
	\begin{figure}[htbp]
		\centering
		\includegraphics[width=0.45\textwidth]{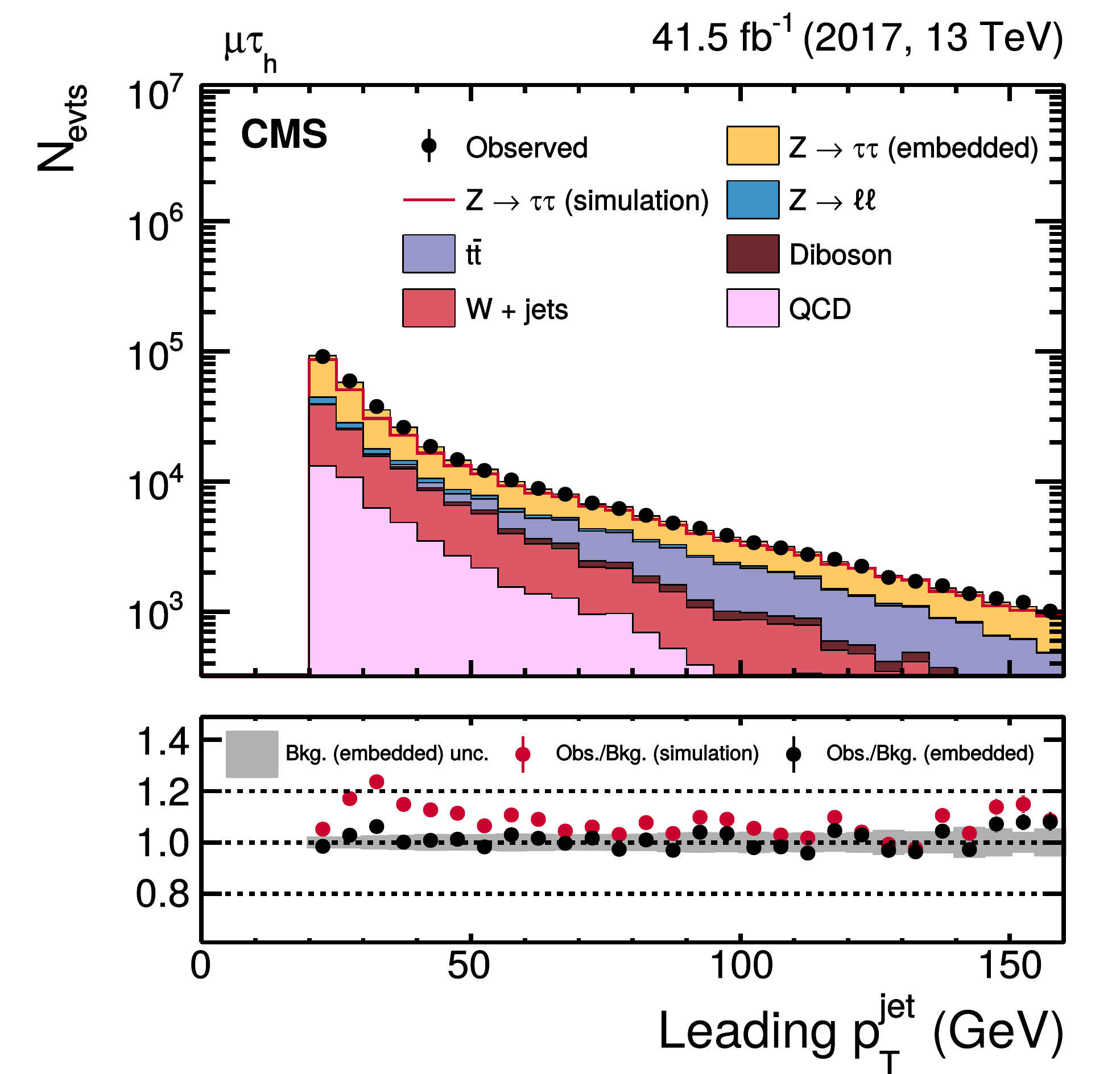}
		\includegraphics[width=0.45\textwidth]{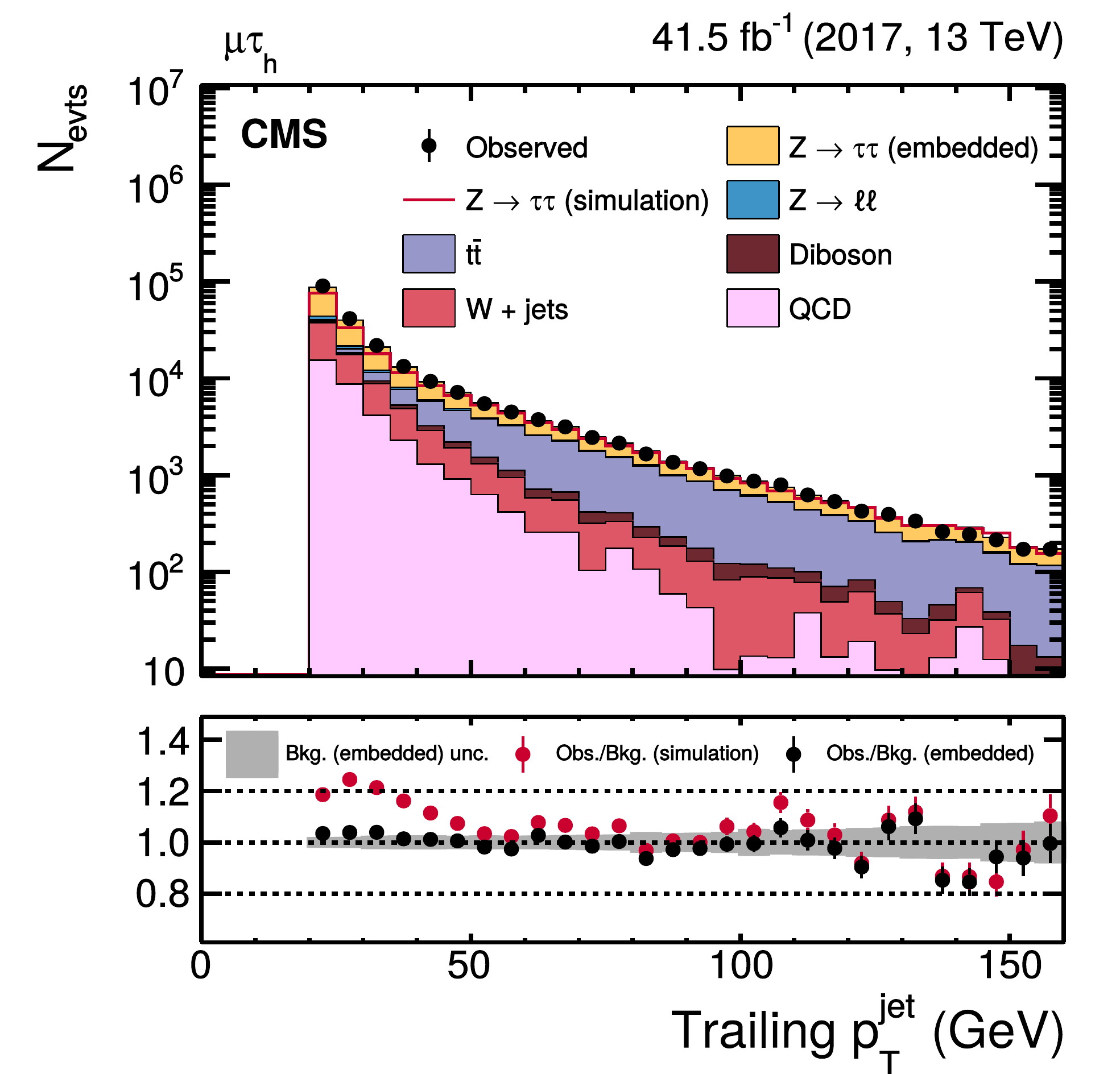}
		\caption {
			Distributions of the $\pt$ of the (left) leading and (right) trailing jet for events with more
			than one jet in the $\mutau$ final state.
		}
		\label{fig:embedding-data-comparison_5}
	\end{figure}
\cleardoublepage \section{The CMS Collaboration \label{app:collab}}\begin{sloppypar}\hyphenpenalty=5000\widowpenalty=500\clubpenalty=5000\vskip\cmsinstskip
\textbf{Yerevan Physics Institute, Yerevan, Armenia}\\*[0pt]
A.M.~Sirunyan, A.~Tumasyan
\vskip\cmsinstskip
\textbf{Institut f\"{u}r Hochenergiephysik, Wien, Austria}\\*[0pt]
W.~Adam, F.~Ambrogi, E.~Asilar, T.~Bergauer, J.~Brandstetter, M.~Dragicevic, J.~Er\"{o}, A.~Escalante~Del~Valle, M.~Flechl, R.~Fr\"{u}hwirth\cmsAuthorMark{1}, V.M.~Ghete, J.~Hrubec, M.~Jeitler\cmsAuthorMark{1}, N.~Krammer, I.~Kr\"{a}tschmer, D.~Liko, T.~Madlener, I.~Mikulec, N.~Rad, H.~Rohringer, J.~Schieck\cmsAuthorMark{1}, R.~Sch\"{o}fbeck, M.~Spanring, D.~Spitzbart, W.~Waltenberger, J.~Wittmann, C.-E.~Wulz\cmsAuthorMark{1}, M.~Zarucki
% ...
\vskip\cmsinstskip
% ...
1:  Also at Vienna University of Technology, Vienna, Austria\\
% ...
\end{sloppypar}
\end{document}